\documentclass[11pt]{JHEP}
\usepackage{amsmath}
\usepackage{amssymb}
\usepackage{amsmath,amsthm,epsfig,euscript,array,cancel}
\font\mybb=msbm10 at 10pt
\def\bb#1{\hbox{\mybb#1}}

\newcommand{\be}{\begin{equation}}
\newcommand{\ee}{\end{equation}}

\def\bea{\begin{eqnarray}}
\def\eea{\end{eqnarray}}
\preprint{ May 26, 2017. V2: 07/05/2018 
}
\renewcommand{\theequation}{\arabic{section}.\arabic{equation}}
\title{An analytic superfield formalism for tree superamplitudes  in D=10 and D=11}

\author{Igor Bandos
 \\ {\small\it Department of
Theoretical Physics, University of the Basque Country UPV/EHU, \\ P.O. Box 644, 48080 Bilbao, Spain} \\  {\small\it
and IKERBASQUE, Basque Foundation for Science, 48011, Bilbao, Spain} }

\date{12/10/16--26/05/17,
V2: June 14, 2017, V3: August 14- Oct. 18, V.4. Feb.15, 2018 Printed \today }

\abstract{Tree amplitudes of 10D supersymmetric Yang-Mills theory (SYM) and 11D  supergravity (SUGRA) are collected in multi-particle counterparts of analytic  on-shell superfields. These have essentially the same form as their chiral 4D counterparts describing ${\cal N}=4$ SYM and ${\cal N}=8$ SUGRA, but with components dependent on a
different set of bosonic variables. These are the D=10 and D=11 spinor helicity variables, the set of which includes the spinor frame variable (Lorentz harmonics) and a scalar density, and generalized homogeneous coordinates of the coset $\frac{SO(D-2)}{SO(D-4)\otimes U(1)}$  (internal harmonics).

We present an especially convenient parametrization of the spinor harmonics (Lorentz covariant gauge fixed with the use of an auxiliary gauge symmetry) and use this to find (a gauge fixed version of) the 3-point tree  superamplitudes of 10D SYM and 11D SUGRA which generalize the 4 dimensional anti-MHV superamplitudes.
}

\keywords{supergravity, supersymmetric gauge theory, amplitudes, superamplitudes, twistor approach, spinor moving frame, harmonic superspace}

\begin{document}

\section{Introduction}

An impressive recent  progress in calculation of multi-loop amplitudes of d=4 supersymmetric Yang-Mills (SYM)
and supergravity (SUGRA) theories, especially of their maximally supersymmetric versions ${\cal N}=4$ SYM and
${\cal N}=8$ SUGRA \cite{Bern:2011qn,Drummond:2008vq,Drummond:2009fd,Eden:2011ku,Kallosh:2012yy}, was reached in its significant part with the use of spinor helicity formalism and of its superfield generalization \cite{Witten:2003nn,Britto:2005fq,Bianchi:2008pu,Brandhuber:2008pf,ArkaniHamed:2008gz,Heslop:2016plj,Herrmann:2016qea}. This latter  works with superamplitudes depending on additional fermionic variables and unifying a number of different amplitudes of the bosonic and fermionic fields from the SYM or SUGRA supermultiplet.

The spinor helicity formalism for D=10 SYM was developed by Caron-Huot and O'Connel in \cite{CaronHuot:2010rj}
and for D=11 supergravity in \cite{Bandos:2016tsm} (more details can be found in \cite{Bandos:2017eof}). The progress in the  latter  was reached
due to the observation that the 10D spinor helicity variables of \cite{CaronHuot:2010rj} can be identified
with spinor Lorentz harmonics or spinor moving frame variables  used for the description of massless D=10
superparticles  in \cite{Galperin:1991gk,Delduc:1991ir,Bandos:1996ju}.
(Similar observation was made and used in D=5 context in \cite{Uvarov:2015rxa}).
The spinor helicity formalism of \cite{Bandos:2016tsm} uses 11D  spinor harmonics  of \cite{Galperin:1992pz,Bandos:2006nr,Bandos:2007mi,Bandos:2007wm}.

As far as the generalization of D=4 superamplitudes is concerned, in \cite{CaronHuot:2010rj}
a kind of Clifford superfield representation  of the amplitudes of 10D SYM was constructed.
However, this later happened to be quite nonminimal and difficult to apply. Then the subsequent papers
\cite{Boels:2012ie,Boels:2012zr,Wang:2015jna,Wang:2015aua} used the D=10 spinor helicity formalism of
\cite{CaronHuot:2010rj} in the context of type II supergravity where the natural complex structure helped
to avoid the use of the above mentioned Clifford superfields\footnote{An interesting  recent
analysis of the divergences of higher dimensional maximal SYM theory \cite{Bork:2015zaa,Borlakov:2016mwp} avoids an explicit use of the
10D spinor helicity formalism but assumes some generic properties of the amplitudes in this formalism.}.
An alternative, constrained superfield formalism was proposed  for 11D SUGRA amplitudes in \cite{Bandos:2016tsm};
its  10D SYM cousin will be briefly described here (see also \cite{Bandos:2017kdq} and \cite{Bandos:2017eof} for details).
In it the superamplitudes carry the indices of 'little groups' $SO(D-2)_i$ of the light-like momenta $k_{a(i)}$ of $i$-th scattered particles and obey a
set of differential equations involving fermionic covariant derivatives $D^+_{q(i)}$.
This formalism is quite different from the 4D superamplitude approach; some efforts on development of
the necessary technique and on deeper understanding of its structure are still required to be accomplished
to make possible its efficient application  to physically interesting problems.

In this paper we develop a simpler {\it analytic superfield formalism} for the description of 11D SUGRA and 10D SYM amplitudes. In it the superamplitudes are multiparticle counterparts of an on-shell analytic superfields, which depend on the fermionic variable in exactly  the same manner as the chiral superfields describing ${\cal N}=8$ SUGRA and ${\cal N}=4$ SYM. However, the  component fields in these analytic superfields depend on another  set of bosonic variables including some internal harmonic variables (see \cite{Galperin:1984av,Galperin:1984bu,Galperin:2001uw})  $w_q^A, \bar{w}_{qA}$ parametrizing the coset $\frac{Spin(D-2)}{Spin(D-4)\otimes U(1)}$. These are used to split the set of $(2{\cal N})$ real  spinor fermionic coordinates $\theta^-_q$ of the natural  on-shell superspaces of 11D SUGRA and 10D SYM  on the set of ${\cal N}$ complex spinor coordinates $\eta^-_A$ and its complex conjugate $\bar{\eta}{}^{-A}$. The {\it analytic} on--shell superfields describing 11D SUGRA and 10D SYM depend on $\eta^-_A$ but not on $\bar{\eta}{}^{-A}$ and, in this sense, are similar to the {\it chiral} on-shell superfields describing ${\cal N}=8$ SUGRA and ${\cal N}=4$ SYM. However, as in higher dimensional case  $\eta^-_A= \theta^-_q \bar{w}_{qA} $ is formed with the use of harmonic variable $\bar{w}_{qA}$, we call these superfields {\it analytic} rather than chiral.

We show how the analytic superamplitudes are constructed from the basic constrained superamplitudes of
10D SYM and 11D SUGRA and  the set of complex $(D-2)$ component null-vectors $U_{I\, i}$ related to the internal frame associated to $i$-th scattered particle. We describe the properties of analytic superamplitudes  and present a convenient parametrization of the spinor harmonics (gauge fixing with respect to a set of auxiliary symmetries acting on spinor frame variables), which allows to establish relations between D=10, 11 superamplitudes and their 4d counterparts. Using such relation we have found a gauge fixed
expressions for the on-shell 3-point tree superamplitudes.
These can be used as basic elements of the  analytic superamplitude formalism based on a generalization of the BCFW recurrent relations \cite{Britto:2005fq}. The derivation and application of these latter, as well as the use of analytic superamplitudes to gain new insight for further development of the constrained superamplitude formalism will be the subject of future papers.

The rest of this paper has the following structure.

In the remaining part of the Introduction, after a resume  of our notation, we briefly review  $D=4$ spinor
helicity and on-shell superfield description of ${\cal N}=4$ SYM and ${\cal N}=8$  SUGRA.  In sec. 2 we  describe the D=10  spinor helicity
formalism.
In sec. 3 we review briefly the on-shell superfield description of 10D SYM \cite{Galperin:1992pz}. Analytic on-shell superfield approach is developed in sec. 4. The spinor helicity formalism, constrained on-shell superfield and analytic on-shell superfield descriptions of  D=11 SUGRA are presented in sec. 5.
 In sec. 6 we introduce the analytic D=10 and D=11 superamplitudes and describe their properties and their relation with constrained superamplitudes.
A real supermomentum, which is supersymmetric invariant due to the momentum conservation, is introduced there.

A convenient parametrization of the spinor harmonics is described in sec. 7. Its study  indicated the necessity  to  impose a relation between  internal harmonics corresponding to different scattered particles, which then allowed to associate a complex spinor frames to each of them.  In sec. 7 we also present a  convenient gauge fixing of the auxiliary gauge symmetries which leads to a simple gauge fixed form of both real and complex spinor harmonics.
This has been used to obtain gauge fixed expressions for 3-point analytic superamplitudes of 10D SYM and 11D SUGRA, which can be found  in sec. 8. We conclude in sec. 9.

Appendix A is devoted to spinor frame re-formulation of 4D spinor helicity formalism, which is useful for comparison of 4D and 10/11D (super)amplitudes. Appendix B shows how to obtain the BCFW--like deformation of the 10/11D spinor helicity and complex fermionic variables from the deformation of real spinor frame and real fermionic variables found in \cite{CaronHuot:2010rj,Bandos:2016tsm}.

\bigskip

\subsection{Notation}

As we will use many different types of indices, for reader convenience we resume the index notation here.

The equations in D=10 and D=11 cases often have similar structure and we use similar notations in these two cases. To describe these in a universal manner and also to stress this similarity,
it is convenient to introduce parameters ${\cal N}$ and $s$, which take values ${\cal N}=4,8$ and $s=1,2$ for the case 10D SYM and 11D SUGRA, respectively,
$$ 10D \; SYM \; : \quad  {\cal N}=4\; , \qquad   s=1\; ,  $$
$$ 11D \; SUGRA \; : \qquad  {\cal N}=8\; , \qquad   s=2\; .  $$
These characterize the number of supersymmetries and maximal spin of the quanta of the dimensionally reduced theories, $ {\cal N}=4$ SYM and $ {\cal N}=8$ SUGRA. Clearly, $s={\cal N}/4$.

The symbols from the beginning of the Greek alphabet denote $Spin (1,D-1)$ indices (this is to say, indices of the minimal spinor representation of $SO(1,D-1)$)
$$ \alpha,\beta, \gamma , \delta =1,..., 4{\cal N}\; . $$
Notice that, when we consider D=4 SYM and SUGRA, we use the complex Weyl spinor indices
$\alpha,\beta=1,2$ and $\dot{\alpha}, \dot{\beta}=1,2$ so that the above equations do not apply.

The spinor indices of the small group $SO(D-2)$ (indices of $Spin(D-2)$)  are denoted by
$$ q,p=1,..., 2{\cal N} \qquad and \qquad \dot{q}, \dot{p}=1,..., 2{\cal N} \; . \qquad $$
In the case of D=11 the dotted $Spin (9)$ indices are identical to undotted,
$\dot{q}=q$, while for D=10 they are transformed by different (although equivalent) 8s and 8c representations of $SO(8)$.

The vector indices of $SO(D-2)$ are denoted by
$$ I,J,K,L= 1,..., (D-2)$$
while
$$ \check{I}, \check{J}, \check{K}, \check{L}= 1,..., (D-4)\;  $$
are vector indices of 'tiny' group $SO(D-4)$. Spinor indices of $SO(D-4)$ ($Spin(D-4)$ indices) are denoted by $$ A,B,C,D =1, ..., {\cal N} \; . $$
The latter notation  also applies to the 4D dimensional reduction of 11D and 10D theories, where $A,B,C,D$  denote the indices of the fundamental representation of $SU({\cal N})$ R-symmetry group.

Finally, $a,b,c,d=0,1, ... , (D-1)$ are D-vector indices. In D=4 we also use $\mu,\nu,\rho=0,1,2,3$ to stress the difference from $D=10$ and $D=11$.

The symbols $i,j=1,..,n$ are used to enumerate the scattered particles described by $n$-point (super)amplitude.

\subsection{D=4 spinor helicity formalism }

In spinor helicity formalism the scattering  amplitudes of $n$ massless particles
${\cal A}(1,...,n):= {\cal A}(p_{(1)},\varepsilon_{(1)}; ...,p_{(n)},\varepsilon_{(n)})$ are considered to be homogeneous functions of $n$ pairs of 2-component bosonic Weyl spinors $\lambda^{\alpha}_{(i)}= (\bar{\lambda}{}^{\dot{\alpha}}_{(i)})^*$ (${\alpha}=1,2$; $\dot{\alpha}=1,2$),
\begin{eqnarray}\label{cAn=cAnl}
{\cal A}(1,..,n):= {\cal A}(p_{(1)},\varepsilon_{(1)}; ...;p_{(n)},\varepsilon_{(n)}) =
{\cal A}(\lambda_{(1)}, \bar{\lambda}_{(1)}; \ldots ;\lambda_{(n)}, \bar{\lambda}_{(n)})
\;  . \qquad \end{eqnarray}
The spinor  $\lambda^{\alpha}_{(i)}$ carries the information about  momentum and polarization of $i$-th particle. In particular, $i$-th light-like 4--momentum $p_{(i)}^\mu$ is  determined in terms of $\lambda^{\alpha}_{(i)}= (\bar{\lambda}{}^{\dot{\alpha}}_{(i)})^*$ by Cartan-Penrose relation (${\alpha}=1,2$, $\;\dot{\alpha}=1,2$, $\;   \mu=0,...,3$)  \cite{Penrose:1967wn,Penrose:1972ia}
\begin{eqnarray}\label{p=ll=4D}
p_{_{A\dot{A}(i)}}:= p_{\mu (i)} \sigma^\mu_{{\alpha}\dot{\alpha}}=
2\lambda_{{\alpha}(i)}  \bar{\lambda}{}_{{\dot{\alpha}(i)}}   \qquad \Leftrightarrow \qquad p_{\mu(i)}= \lambda_{(i)} \sigma_\mu \bar{\lambda}_{(i)} . \qquad
\end{eqnarray}
Here  $\sigma^\mu_{{\alpha}\dot{\alpha}}$ are relativistic Pauli matrices obeying
 $\sigma^\mu{}_{{\alpha}\dot{\alpha}} \sigma{_{\mu \beta\dot{\beta}}}= 2\epsilon_{{\alpha}{\beta}}\epsilon_{\dot{\alpha}\dot{\beta}}$
with $\epsilon_{{\alpha}{\beta}}=\left(\begin{matrix} 0 & -1\cr 1 & 0 \end{matrix}\right) = \epsilon_{\dot{\alpha}\dot{\beta}}$. This identity explains equivalence of two forms of the Cartan-Penrose representation  (\ref{p=ll=4D}) and also allows to show that $p_{a(i)}p_{(i)}^a=0$.

The  $n$-particle amplitude is restricted by  $n$ helicity constraints
\begin{eqnarray}\label{hhiA=hiA}
\hat{h}_{(i)} {\cal A}(1,...,n) = h_i {\cal A}(1,...,n)\; ,\qquad
\end{eqnarray}
where the operator
\begin{eqnarray}
 \label{Ui=b}
\hat{h}_{(i)}:=
 \frac{1}{2} \left(\lambda^\alpha_{(i)} \frac \partial  {\partial \lambda^\alpha_{(i)} }-
 \bar{\lambda}{}^{\dot{\alpha}}_{(i)} \frac \partial  {\partial \bar{\lambda}{}^{\dot{\alpha}}_{(i)}}\right)   \qquad \end{eqnarray}
 counts the difference between degrees of homogeneity in $\lambda^\alpha_{(i)}$ and $ \bar{\lambda}{}^{\dot{\alpha}}_{(i)}$.
 Its eigenvalue $h_i$, the helicity of $i$-th particle, defines the amplitude homogeneity property with respect to the phase transformations of $\lambda^\alpha_{(i)}$ and $ \bar{\lambda}{}^{\dot{\alpha}}_{(i)}$,
 \begin{eqnarray}\label{cA=e2ihicA}
 {\cal A} (..., e^{i\beta_i} \lambda^\alpha_{(i)},  e^{-i\beta_i}  \bar{\lambda}{}^{\dot{\alpha}}_{(i)}, ...) = e^{2ih_i \beta_i } {\cal A}(..., \lambda^\alpha_{(i)},   \bar{\lambda}{}^{\dot{\alpha}}_{(i)}, ...)\; . \qquad
\end{eqnarray}
 It is quantized: the amplitude is a well defined function of complex variable $\lambda^\alpha_{(i)}$ if and only if $\beta_i $ is equivalent to $\beta_i +2\pi$, and this happens when $2h_i\in {\bb Z}$.  In the case of gluons $h_i=\pm 1$  and in the case of gravitons  $h_i=\pm 2$.

\subsection{D=4 superamplitudes and on-shell superfields}

A superamplitude  of ${\cal N}=4$ SYM or ${\cal N}=8$ supergravity  depends, besides $n$ sets of complex bosonic spinors, on $n$ sets of complex fermionic variables $\eta_{(i)}^A$ ($(\eta_{(i)}^A)^*=\bar{\eta}{}_{A(i)}$)  carrying the index of fundamental representation of the $SU({\cal N})$ R-symmetry group  $A, B=1, ..., {\cal N}$,
  \begin{eqnarray}\label{cAn=cAnlf}
{\cal A}(1; ...;n) = {\cal A}(\lambda_{(1)}, \bar{\lambda}_{(1)}, \eta_{(1)}; ...;\lambda_{(n)}, \bar{\lambda}_{(n)},  \eta_{(n)})\; , \qquad \eta_{(i)}^A\eta_{(j)}^B=- \eta_{(j)}^B\eta_{(i)}^A
\; .    \qquad \end{eqnarray}
It obeys $n$ super-helicity constraints,
  \begin{eqnarray}\label{UcAn=2cAn}
  \hat{h}_{(i)} {\cal A}(\{\lambda_{(i)}, \bar{\lambda}_{(i)}, \eta^A_{(i)}\}) = \frac {\cal N} 4 {\cal A}(\{\lambda_{(i)}, \bar{\lambda}_{(i)}, \eta_{(i)}^A\})\; ,\qquad A =1,..., {\cal N}\,  \qquad
  \end{eqnarray}
with
 \begin{eqnarray} \label{Ui=}
 2\hat{h}_{(i)}=
 \lambda^\alpha_{(i)} \frac \partial  {\partial \lambda^\alpha_{(i)} }-
 \bar{\lambda}{}^{\dot{\alpha}}_{(i)} \frac \partial  {\partial \bar{\lambda}{}^{\dot{\alpha}}_{(i)}}  +
 \eta_i^A \frac \partial  {\partial  \eta^A_{i} } \, .    \qquad
 \end{eqnarray}
It is important that the dependence of amplitude on fermionic variables is holomorphic:  it depends on $\eta_i^A$ but is independent of $\bar{\eta}{}_{A(i)}=(\eta_{(i)}^A) ^*$. Furthermore, according to (\ref{Ui=}),  the degrees of homogeneity in these fermionic variables is related to the helicity $h_i$ characterizing dependence on bosonic spinors. Hence, decomposition of superamplitude on the fermionic variables involves amplitudes of different helicities.

These superamplitudes can be regarded as multiparticle generalizations of the so-called on-shell superfields
 \begin{eqnarray}\label{Phi=3,4}
 \Phi (\lambda ,\bar{\lambda}, \eta^A) = f^{(+s)} +  \eta^A \chi_A + \frac 1 2 \eta^B \eta^A s_{AB}+
 \ldots +  \eta^{\wedge ({\cal N}-1)}{}_A \bar{\chi}^A + \eta^{\wedge {\cal N}}  f^{(-s)} \; , \qquad
 \\
 \label{etaN:=}
 \eta^{\wedge {\cal N}}=  \frac 1 {{\cal N}!} \eta^{A_1} \ldots \eta^{A_{{\cal N}}}
 \epsilon_{{A_1}  \ldots A_{{\cal N}} }\; , \qquad  \eta^{\wedge ({\cal N}-1)}{}_A= \frac 1 {({\cal N}-1)!} \eta^{B_2} \ldots \eta^{B_{{\cal N}}}
 \epsilon_{A{B_2}\ldots B_{{\cal N}}} \; , \qquad
\end{eqnarray}
which  obey the  super-helicity constraint
 \begin{eqnarray}\label{UPhi=1}
 \hat{h}  \Phi (\lambda ,\bar{\lambda}, \eta) &=&
 s \Phi (\lambda ,\bar{\lambda}, \eta) \; , \quad  s= \frac {\cal N} 4 \, , \quad  2\hat{h}=
 -\lambda^\alpha \frac \partial  {\partial \lambda^\alpha }+
 \bar{\lambda}{}^{\dot{\alpha}}\frac \partial  {\partial \bar{\lambda}{}^{\dot{\alpha}}}  +
 \eta^A \frac {\partial }  {\partial  \eta^A} \, ,  \quad A=1,...,{\cal N}\, .  \qquad
 \end{eqnarray}
The chiral superfields on a real  superspace $\Sigma^{(4|2{\cal N})}=\{\lambda ,\bar{\lambda}, \eta , \bar{\eta}\}$ obeying Eq. (\ref{UPhi=1}) describe the on-shell states of ${\cal N}=4$ SYM and ${\cal N}=8$ SUGRA.
They can be considered as  homogeneous  superfields on  {\it chiral on-shell superspace}
\begin{eqnarray}\label{on-sh=4Dssp}
\Sigma^{(4|{\cal N})}=\{\lambda ,\bar{\lambda}, \eta \}\;  \qquad
\end{eqnarray}
satisfying  Eq. (\ref{UPhi=1}), which just fixes the charge of superfield with respect to a phase transformations of its arguments
\footnote{The relative charges of bosonic and fermionic coordinates of this phase transformations can be restored from the relation between supertwistors and standard superspace coordinates
\cite{Ferber:1977qx}. In superamplitude context these relations can be found {\it e.g.} in \cite{Drummond:2008vq}. }.

Such on-shell superfields can be obtained by quantization of $D=4$ Brink-Schwarz superparticle with ${\cal N}$-extended supersymmetry in its Ferber-Shirafuji formulation \cite{Ferber:1977qx,Shirafuji:1983zd} (see also \cite{Gumenchuk:1990db,Sorokin:1989jj} as well as  \cite{Bandos:1990ji} and \cite{Bandos:1991my}).
This observation has served us as an important guide:
in   \cite{Bandos:2017eof} we show how to obtain the 10D and 11D on-shell superfield formalism  from  D=10 and D=11  superparticle quantization. Here we will not consider superparticle quantization but describe briefly the resulting constrained on-shell superfields and constrained superamplitude formalism of  \cite{Bandos:2016tsm,Bandos:2017eof} and use these as a basis to search for the analytic on-shell superfields and analytic  superamplitude formalism.

To conclude our brief review,  let us present the expressions for basic building blocks of the 4D superamplitude formalism, the 3-point superamplitudes of D=4 ${\cal N}=4$ SYM theory. These are two:
the anti-MHV ($\overline{{\rm MHV}}$)
   \begin{eqnarray}\label{cA3g=bMHV}
 {\cal A}^{\overline{{\rm MHV}}}(1,2, 3)= \frac {1} {<12> <23><31>}\; \delta^4 \left( \eta_{A(1)}<23> +  \eta_{A(2)}<31>+  \eta_{A(3)} <12> \right)   \qquad
  \end{eqnarray}
  and  the MHV superamplitude
  \begin{eqnarray}\label{cA3g=MHV}
 {\cal A}^{MHV}(1,2, 3)=  \frac {1} {[12] \,[23]\, [31]}\; \delta^8 \left( \bar{\lambda}_{ \dot{\alpha}1}\eta_{A1}+  \bar{\lambda}_{ \dot{\alpha}2}\eta_{A2}+ \bar{\lambda}_{ \dot{\alpha}3} {\eta}_{A3} \right) .  \qquad
  \end{eqnarray}
Here we set the SYM coupling constant to unity and use the standard notation for the contraction of 4D Weyl spinors  \begin{eqnarray}\label{ij=lili}
  <ij> &=&<\lambda_i\lambda_j>=\lambda^\alpha_{i}\lambda_{\alpha j} = \epsilon^{\alpha\beta}\lambda_{\beta i}\lambda_{\alpha j} \; ,  \qquad \nonumber \\ {} [\; ij\; ]\; &=& <ij>^*=
  [\bar{\lambda}_i\bar{\lambda}_j]=\bar{\lambda}{}^{\dot\alpha}_{i}\bar{\lambda}_{\dot{\alpha} j} = \epsilon^{\dot{\alpha}\dot{\beta}}\bar{\lambda}{}_{\dot{\beta} i}\bar{\lambda}_{\dot{\alpha} j} \; .  \qquad\end{eqnarray}

\section{Spinor helicity formalism in D=10}

As we have already mentioned in the Introduction, the D=10 spinor helicity formalism \cite{CaronHuot:2010rj} can be constructed using the spinor (moving) frame or Lorentz harmonic variables. To describe these it is convenient  to start with introducing the vector frame variables or vector harmonics (called light-cone harmonics in \cite{Sokatchev:1985tc,Sokatchev:1987nk}).

\subsection{Vector harmonics}
\label{vectorFrame}

The property of vector harmonic variables are universal so that, instead of specifying  ourselves to D=10 dimensional case, we write the equations of this section for arbitrary number $D$ of spacetime dimensions. This will allow us to refer on these equations when considering spinor helicity formalism for 11D supergravity.

Let us consider a vector frame
\begin{eqnarray}\label{uaib=}
u_{ai}^{(b)}= \left( \frac 1 2 \left(u_{a (i)}^\#+ u_{a i}^=\right) , u_{a i}^I\, , \frac 1 2 \left(u_{a i}^\# - u_{a i}^=\right)\right) \quad \in \quad SO(1, D-1)\; . \qquad
\end{eqnarray}

It can be associated with D-dimensional light-like momentum $k_{a(i)}$,  $k_{a(i)}k^a_{(i)}=0$, by the condition that  one of the light-like vectors of the frame, say
$u_{a i}^= = u_{a i}^0- u_{a i}^{(D-1)}$, is proportional to this  $k_{a(i)}$,
\begin{eqnarray}\label{kia=ru=}
k_{(i)}^a =  \rho^{\#}_{(i)} u_{(i)}^{a=}\, . \qquad
\end{eqnarray}
The additional index  $i$ will enumerate particles scattered  in the process described by an  on-shell amplitude.
Below in this section, to lighten the equations,  we will omit this index when this does not lead to a confusion.

The condition (\ref{uaib=}) implies $u_a^{(c)}\eta^{ab}u_b^{(d)}=\eta^{(c)(d)}$, which can be split into \cite{Sokatchev:1985tc,Sokatchev:1987nk}
\begin{eqnarray}\label{uu=0}
&& u_a^{=}u^{a=}=0 \; , \qquad \\ \label{uu=2}
 && u_a^{\#}u^{a\#}=0 \; , \qquad  u_a^{=}u^{a\#}=2 \; , \qquad \\ \label{uui=0} && u_a^{I}u^{a=}=0 \; , \qquad u_a^{I}u^{a\#}=0 \; , \qquad  u_a^{I}u^{aJ}=-\delta^{IJ} \; , \qquad
\end{eqnarray}
and also $u_a^{(c)}\eta_{(c)(d)}u_b^{(d)}=\eta_{ab}$, which can be written in the form of
\begin{eqnarray}\label{I=UU}
\delta_a{}^b= {1\over 2}u_a^{=}u^{b\#}+ {1\over 2}u_a^{\#}u^{b=}-  u_a^{I}u^{bI} \; . \qquad
\end{eqnarray}

Notice that the sign indices $^{=}$ and $^{\#}$ of two light--like elements  of the vector frame (see (\ref{uu=0}) and (\ref{uu=2})) indicate their weights under the
transformations of $SO(1,1)$ subgroup of the Lorentz group $SO(1,D-1)$,
\begin{eqnarray}\label{SO1-1=}
 u_a^{=}\mapsto \; e^{-2\alpha } u^{a=} \; , \qquad u_a^{\#} \mapsto \; e^{+2\alpha }  u^{a\#} \; , \qquad   u_a^{I} \mapsto \; u^{aI} \; . \qquad
\end{eqnarray}

It is convenient to  change the basis and  to consider the  splitting of the vector frame matrix (\ref{uaib=}) on two light-like and $(D-2)$ orthogonal vectors  in the form \cite{Sokatchev:1985tc}
\begin{eqnarray}\label{harm=++--I}
u_{a}^{(b)}= \left( u_{a }^=,  u_{a }^\# ,  u_{a}^I\right)\; ,  \qquad u_{c}^{(a)}u^{c(b)}=\eta^{(a)(b)}= \left(\begin{matrix}   0\; &\;\; 2 & 0 \cr
2\; &\;\; 0 & 0 \cr 0\; &\;\; 0 & -\delta^{IJ} \end{matrix}\right)\; . \qquad
\end{eqnarray}
This is manifestly invariant under the direct product $SO(1,1)\otimes SO(D-2)$ of the above scaling symmetry (\ref{SO1-1=}) and the rotation group $SO(D-2)$ mixing the spacelike vectors $ u_{a}^I$,
\begin{eqnarray}\label{SOD-2=}
SO{(D-2)}\; : \qquad u_{a }^= &\mapsto& \; u_{a }^= \; , \qquad
  u_{a}^\# \mapsto  \; u_{a}^\#  \; , \qquad
 u_{a}^I \mapsto \; u_{a}^J {\cal O}^{JI}  \; , \qquad  {\cal O}{\cal O}^T={I} \; . \qquad
\end{eqnarray}

If only one light-like  vector $u_{a}^=$ of the frame is  relevant, as it will be the case in our discussion below,
the transformations mixing $u_{a}^\#$ and $ u_{a}^I$ can be also considered as a symmetry.
These are so--called  $K_{(D-2)}$  transformations
\begin{eqnarray}\label{KD-2=}
K_{(D-2)}\; : \qquad u_{a }^= &\mapsto& \; u_{a }^= \; , \qquad \nonumber \\
  u_{a}^\# &\mapsto & \; u_{a}^\#  +  u_{a }^I \; K^{\#I}  + \frac 1 4 u_{a }^=(K^{\#I}K^{\#I})\; , \qquad
 u_{a}^I \mapsto \; u_{a}^I + \frac 1 2 u_{a }^=K^{\#I} \;  \qquad
\end{eqnarray}
(identified in \cite{Galperin:1991gk,Delduc:1991ir} as conformal boosts of the conformal group of Euclidean space).

To make  the associated  momentum (\ref{kia=ru=})  invariant under  $SO(1,1)$ transformations (\ref{SO1-1=}),  we have to require that
\begin{eqnarray}\label{SO1-1=r}
 \rho^{\#} \mapsto \; e^{+2\alpha } \rho^{\#} \; ,  \qquad
\end{eqnarray}
and this explains the index $^{\#}$ of  $\rho$ multiplier in  (\ref{kia=ru=}). Of course, we can use (\ref{SO1-1=r}) to set $\rho^{\#} =1$. However, it happens to be much more convenient
to keep  $SO(1,1)$ unfixed  and to use it as identification relation (gauge symmetry acting on) vector harmonics  (\ref{uaib=}).

The complete expression for light-like momentum (\ref{kia=ru=})
is invariant under
$H_B= [SO(1,1)\otimes SO(D\!-\!2)]\subset\!\!\!\!\!\!\times  K_{(D-2)}$ transformations (\ref{SO1-1=}), (\ref{SOD-2=}), (\ref{KD-2=}). This is the Borel subgroup of $SO(1,D-1)$ so that  $SO(1,D-1)/H_B$ coset is compact;  actually it is isomorphic to the  sphere ${\bb S}^{(D-2)}$. If we use $H$ transformations as identification relation on the set of vector harmonics, these can be considered as a kind of  homogeneous coordinates of such a sphere \cite{Galperin:1991gk,Delduc:1991ir}
\begin{eqnarray}\label{++--I=SD-2}
\left\{ \left( u_{a }^=,  u_{a }^\# ,  u_{a}^I\right)\right\}\; = \frac {SO(1,D-1)} {[SO(1,1)\otimes SO(D-2)]\otimes K_{(D-2)}}= {\bb S}^{(D-2)} \; .
\end{eqnarray}
Such a treatment as constrained homogeneous coordinates of the coset makes the vector frame variable similar to the internal coordinate of harmonic superspaces introduced in \cite{Galperin:1984av,Galperin:1984bu}, and
stays beyond the name  {\it vector harmonics} or {\it vector  Lorentz harmonics},  which we mainly use for them.

In the context of (\ref{kia=ru=}),  ${\bb S}^{(D-2)}$ in (\ref{++--I=SD-2}) can be identified with the celestial sphere of a $D$-dimensional observer.
Notice that this is in agreement with the fact that a light-like $D$--vector defined up to a scale factor can be considered as providing  homogeneous coordinates for the  ${\bb S}^{(D-2)}$ sphere
\begin{eqnarray}\label{u--=SD-2}
\left\{  u_{a }^= \right\} = {\bb S}^{(D-2)} \; .
\end{eqnarray}
The usefulness of seemingly superficial construction with the complete frame (\ref{++--I=SD-2}) becomes clear when we consider spinor frame variables, which provide a kind of square roots  of the light-like vectors of the Lorentz frame.

\subsection{Spinor frame in D=10 }

To each vector frame $u_b^{(a)}$ we can associate a spinor frame described by $Spin(1,D-1)$ valued matrix $V_{\alpha}^{(\beta)} \in Spin(1,D-1)$ related to $u_b^{(a)}$ by the condition of the preservation of $D$-dimensional Dirac matrices
\begin{eqnarray}\label{VGVt=G} V\Gamma_b V^T =  u_b^{(a)} {\Gamma}_{(a)}\, , \qquad V^T \tilde{\Gamma}^{(a)}  V = \tilde{\Gamma}^{b} u_b^{(a)}\;
 \, , \qquad \end{eqnarray}
 and also of the charge conjugation matrix if such exists in the minimal spinor representation of $D$-dimensional Lorentz group,
 \begin{eqnarray}\label{VCVt=C}
 VCV^T=C \; , \qquad if \quad  C\; exists\; for\; given\; D\; .  \qquad
\end{eqnarray}

In the case of D=10,  where the minimal Majorana-Weyl (MW) spinor representation  is 16-dimensional,
the $SO(1,1)\times SO(8)$ invariant splitting of vector frame in (\ref{uaib=}) is reflected by splitting the spinor frame matrix on two rectangular blocks, $ v_{\alpha \dot{q}}^{\; +}$ and $v_{\alpha q}^{\; -}$,
\begin{eqnarray}\label{harmV=D}
V_{\alpha}^{(\beta)}= \left(\begin{matrix} v_{\alpha \dot{q}}^{\; +} , & v_{\alpha q}^{\; -}
  \end{matrix}\right) \in Spin(1,D-1)\; , \qquad
\end{eqnarray}
which are called  {\it spinor  frame variables} or {\it Lorentz harmonic} ({\it spinor Lorentz harmonic}).
Their sign indices $^\pm$  indicate their scaling properties with respect to the $SO(1,1)$ transformations, and their  columns  are enumerated by indices of  different, $c$-spinor and $s$-spinor representations of $SO(8)$ group,
\begin{eqnarray}\label{q=D10}
 D=10:\qquad \alpha=1,...,16\; , \quad \dot{q}=1,...,8 \; , \quad q=1,...,8
 \; .   \qquad
\end{eqnarray}
The set of constraints on 10D Lorentz harmonics are given by Eqs. (\ref{VGVt=G}) in which
$\Gamma^a_{\alpha\beta}= \sigma^a_{\alpha\beta}=\sigma^a_{\beta\alpha}$ and $\tilde{\Gamma}^{a \; \alpha\beta}= \tilde{\sigma}^{a \; \alpha\beta}=\tilde{\sigma}^{a \;\beta\alpha}$ are 16$\times 16$ generalized Pauli matrices, which obey
$\sigma^a\tilde{\sigma}^b +\sigma^b\tilde{\sigma}^a=
2\eta^{ab}{\bb I}_{16\times 16}$.
We prefer to write this relation in the universal form
\begin{eqnarray}\label{GtG=I}
 \Gamma^a_{\alpha\gamma}\tilde{\Gamma}^{b \; \gamma\beta} + \Gamma^b_{\alpha\gamma}\tilde{\Gamma}^{a \; \gamma\beta}= 2 \delta_{\alpha}{}^{\beta}\; ,  \qquad
\end{eqnarray}
which  also describes the properties of symmetric 32$\times$32 11D  Dirac matrices introduced below (see sec. \ref{sec=11Dharm}).

The charge conjugation matrix does not exist in 10D Majorana -Weyl spinor representation so that there is no way to rise or to lower the spinor indices. The elements of the inverse of the spinor  frame matrix
\begin{eqnarray}\label{harmV-1=D}
V_{(\beta)}^{\;\;\; \alpha}= \left(\begin{matrix}    v^{-\alpha}_{{\dot{q}}} \cr v^{+\alpha}_{{q}}
 \end{matrix} \right) \in Spin(1,D-1)
 \;  \qquad
\end{eqnarray}
are introduced as additional variables, which obey the constraints
\begin{eqnarray}\label{VV-1=I}
V_{\alpha}^{(\beta)}V_{(\beta)}{}^\gamma :=  v_{\alpha\dot{q}}^{\; +} v^{-\gamma}_{\dot{q}}
+ v^{\; -}_{\alpha q} v^{+\gamma}_q
=
\delta_{\alpha}{}^\gamma
\;   \qquad
\end{eqnarray}
 and
\begin{eqnarray}\label{v-qv+p=}
&
v^{-\alpha}_{\dot{q}}   v_{\alpha \dot{p}}^{\; +}=\delta_{\dot{q}\dot{p}}
 \; ,  \qquad & v^{-\alpha}_{\dot{q}}   v_{\alpha q}^{\; -}=0 \;  , \qquad
 \nonumber  \\
 & v^{+\alpha}_{{q}}  v_{\alpha \dot{p}}^{\; +}=0
 \;  , \qquad & v^{+\alpha}_{{q}} v_{\alpha {p}}^{\; -} = \delta_{qp} \;  . \qquad
\end{eqnarray}
For brevity, we will call  $v^{-\alpha}_{{\dot{q}}}$ and $v^{+\alpha}_{{q}}$ {\it inverse harmonics}.

The constraints (\ref{VGVt=G}) can be split on the following
set of $SO(1,1)\otimes SO(8)$ covariant relations
\begin{eqnarray}\label{u==v-v-}
 & u_a^= \Gamma^a_{\alpha\beta}= 2v_{\alpha q}{}^- v_{\beta q}{}^-  \; , & \qquad
 u_a^= \delta_{{q}{p}} = v^-_{{q}} \tilde{\Gamma}_{a}v^-_{{p}}  \qquad  \\
\label{v+v+=u++}
& v_{\dot{q}}^+ \tilde{\Gamma}_{ {a}} v_{\dot{p}}^+ = \; u_{ {a}}^{\# } \delta_{\dot{q}\dot{p}}\; , & \qquad 2 v_{{\alpha}\dot{q}}{}^{+}v_{{\beta}\dot{q}}{}^{+}= {\Gamma}^{ {a}}_{ {\alpha} {\beta}} u_{ {a}}^{\# }\; , \qquad \\
 \label{uIs=v+v-}
& v_{{q}}^- \tilde {\Gamma}_{ {a}} v_{\dot{p}}^+=u_{ {a}}^{I} \gamma^I_{q\dot{p}}\; , &\qquad
 2 v_{( {\alpha}|{q} }{}^- \gamma^I_{q\dot{q}}v_{|{\beta})\dot{q}}{}^{+}= {\Gamma}^{a}_{\alpha\beta} u_{ {a}}^{I}\; , \quad  \end{eqnarray}
where  $\gamma^I_{q\dot{p}}= \tilde{\gamma}{}^I_{\dot{p}q}$ with  $I=1,...,8$ are  $SO(8)$ Clebsh-Gordan coefficients obeying
\begin{eqnarray}\label{gIgJ+=I}
\gamma^I\tilde{\gamma}{}^J+ \gamma^J\tilde{\gamma}{}^I= \delta^{IJ}I_{8\times 8}\; , \qquad \tilde{\gamma}{}^I\gamma^J+ \tilde{\gamma}{}^J\gamma^I= \delta^{IJ}I_{8\times 8}
\; . \quad  \end{eqnarray}

Although the constraints for the inverse harmonics (\ref{harmV-1=D})
 \begin{eqnarray}
 \label{u==v-v-1}
  & u_a^= \tilde{\Gamma}{}^{a\, \alpha\beta}= 2v^{-\alpha}_{\dot q}   v^{-\beta}_{\dot q} \; , &\qquad  u_a^= \delta_{\dot{q}\dot{p}}= v^-_{\dot{q}} {\Gamma}_{a} v^-_{\dot{p}}
 \; , \qquad \\
\label{v+v+=u++1}
& v_{{q}}^+ {\Gamma}_{ {a}} v_{{p}}^+ = \; u_{ {a}}^{\# } \delta_{{q}{p}}\; , & \qquad 2 v_{{q}}^{+ {\alpha}}v_{{q}}^{+}{}^{ {\beta}}= \tilde{\Gamma}^{ {a} {\alpha} {\beta}} u_{ {a}}^{\# }\; , \qquad \\ \label{uIs=v-v+1}
&
 v_{\dot q}^- {\Gamma}_{ {a}} v_{{p}}^+ = - u_{ {a}}^{I} \gamma^I_{p\dot{q}}\; , &\qquad
 2 v_{\dot q}^{-( {\alpha}}\gamma^I_{q\dot{q}}v_{{q}}^{+}{}^{ {\beta})}=-  \tilde{\Gamma}^{ {a} {\alpha} {\beta}} u_{ {a}}^{I}\; , \quad
\end{eqnarray}
can be obtained from (\ref{u==v-v-})--(\ref{uIs=v+v-}) and (\ref{v-qv+p=}), it is convenient to keep their form in mind.

The constraints (\ref{u==v-v-}) allow us to treat harmonic $v_{\alpha q}^{\; -}$ as a kind of square root of
the light-like vector $u_a^=$ of the vector frame. Similar to this latter, $v_{\alpha q}^{\; -}$ can be also treated as a constrained homogeneous coordinates of the coset isomorphic to the celestial sphere
\begin{eqnarray}\label{v-qi=G-H}
\{ v_{\alpha q }^{\; -} \}  \; \in \; {\bb S}^{8}\; .   \qquad
\end{eqnarray}
Actually, Eq. (\ref{v-qi=G-H}) abbreviates  the spinorial counterparts of (\ref{++--I=SD-2}) and (\ref{u--=SD-2}); the complete form of the first of these is
\begin{eqnarray}\label{v-v+=G-H}
\{ (v_{\alpha \dot{q} }^{\; +}, v_{\alpha q }^{\; -}) \}  \; =  \; \frac {Spin(1,D-1)} {[SO(1,1)\otimes Spin(8)]\subset\!\!\!\!\!\!\times  K_{8}} = \; {\bb S}^{8}\; ,   \qquad
\end{eqnarray} where $K_{D-2}$ ($K_{8}$ in our 10D case) leaves $v_{\alpha {q} }^{\; -}$ invariant and acts on the complementary harmonics $v_{\alpha \dot{q} }^{\; +}$ by
\begin{eqnarray}\label{KD-2=v}
K_{D-2}\; :\qquad v_{\alpha \dot{q} }^{\; +} \mapsto v_{\alpha \dot{q} }^{\; +} + \frac 1 2 K^{\# I}v_{\alpha p }^{\; -}\gamma^I_{p\dot{q}}\; .   \qquad
\end{eqnarray}
In a model with $[SO(1,1)\otimes Spin(D-2)]\subset\!\!\!\!\!\!\times  K_{D-2}\;$ gauge symmetry
$v_{\alpha \dot{q} }^{\; +}$ does not carry degrees of freedom: any $v_{\alpha \dot{q} }^{\; +}$  forming $Spin(1,D-1)$ matrix with given $v_{\alpha q }^{\; -}$ can be obtained from some reference solution of this condition, $ v_{\alpha \dot{q} 0}^{\; +}$, by $K_{D-2}$ transformations (\ref{KD-2=v}).
This justifies the simplified form of (\ref{v-qi=G-H}) where only $v_{\alpha p }^{\; -}$ are presented as the constrained homogeneous coordinates of the  sphere.

\subsection{D=10 spinor helicity formalism}

\label{spinorHelicity10D}

When the vector frame is attached  to a light-like momentum as in
 (\ref{kia=ru=}),
\begin{eqnarray}\label{k=pu--}
 & k_a= \rho^\# u_a^= \; ,  \quad  \end{eqnarray}
 the  constraints (\ref{u==v-v-}) for the associated spinor frame  imply that the  following D=10  counterparts of the D=4 Cartan-Penrose relations (\ref{p=ll=4D}) hold:
\begin{eqnarray}\label{k=pv-v-}
k_a
\Gamma^a_{\alpha\beta}= 2\rho^{\#} v_{\alpha q}^{\; - } v_{\beta q}^{\; - } \; , \qquad
 \rho^{\#} v^-_{{q}} \tilde{\Gamma}_{a}v^-_{{p}}= k_{a} \delta_{{q}{p}}.  \qquad  \qquad  \end{eqnarray}
 In D=10 we should also mention the existence of the similar relations for the inverse harmonics (\ref{harmV-1=D}),
 \begin{eqnarray}\label{k=pv-v-1}  && k_{a} \tilde{\Gamma}{}^{a\, \alpha\beta}= 2\rho^{\#}v^{-\alpha}_{\dot q}   v^{-\beta}_{\dot q} \; , \qquad \rho^{\#}  v^-_{\dot{q}} {\Gamma}_{a}v^-_{\dot{p}}=  k_{a}  \delta_{\dot{q}\dot{p}}
 \; . \qquad
\end{eqnarray}

Contracting the first equations in  (\ref{k=pv-v-})  and in (\ref{k=pv-v-1})  with   $v^{-\beta}_{\dot{q}}$ and $v_{\alpha {q}}^{\; -}$, and using (\ref{v-qv+p=}) we easily find that these obey the massless Dirac equations (or, better to say, $D=10$   Weyl equations)
\begin{eqnarray}\label{Dirac}
k_a\Gamma^a_{\alpha\beta}  v^{-\beta}_{\dot q}=0\; , \qquad  k_{a} \tilde{\Gamma}{}^{a\, \alpha\beta}v_{\beta  q}^{\; -} =0
\; . \qquad
\end{eqnarray}
Thus, they  can be identified, up to a scaling factor,  with D=10 spinor helicity variables of \cite{CaronHuot:2010rj}:
\begin{eqnarray}\label{shel=l=}
\lambda_{\alpha q}= \sqrt{\rho^{\#}} v_{\alpha  q}^{\; -} . \qquad
\end{eqnarray}
The polarization spinor of the D=10 fermionic fields \cite{CaronHuot:2010rj} can be associated with the
inverse harmonics $v^{-\alpha}_{\dot{q}}$:
\begin{eqnarray}\label{spol=l=} \lambda^{\alpha}_{\dot{q}}= \sqrt{\rho^{\#}} v^{-\alpha}_{\dot{q}}\; .  \qquad
\end{eqnarray}

\subsection{D=10 SYM  multiplet in the Lorentz harmonic spinor helicity formalism  }

The polarization vector of the vector field can be identified with spacelike vectors   $u_a^I$ of the frame adapted to the light-like momentum of the particle by (\ref{k=pu--}) ({\it cf.} \cite{CaronHuot:2010rj}) so that the on-shell field strength of the D=10 gauge field can expressed by
\begin{eqnarray}\label{Fab=kuwI}
D=10:\quad F_{ab} = k_{[a}u_{b]}{}^I\, w^I =
\rho^{\#} u^=_{[a}u_{b]}{}^I\, w^I
\; , \qquad a=0,1,...,9\; , \qquad I=1,...,8 \;  \qquad
\end{eqnarray}
in terms of one SO(8) vector $w^I$. It is easy to check that both Bianchi identities and Maxwell equations in momentum representations are satisfied, $k_{[a}F_{bc]}=0=k_aF^{ab}$.

As we have already said, the polarization spinor can be identified with the spinor frame variable  $v^{-\alpha}_{\dot q }$. Hence, in the  linear approximation, the on-shell states of spinor superpartner of the gauge field  can be described by
\begin{eqnarray}\label{chi10D=vpsi}
\chi^\alpha=v^{-\alpha}_{\dot q } \psi_{\dot q }
\;  \qquad
\end{eqnarray}
in terms of  a fermionic SO(8) c-spinor $\psi_{\dot q }$. Indeed,   due to (\ref{Dirac}), the field (\ref{chi10D=vpsi}) solves the free  Dirac equation.

When the formalism is applied to external particles of scattering amplitudes,
the bosonic  $w^I$ and fermionic  $\psi_{\dot{q}}$ are considered to be dependent on $\rho^{\#}$ and on spinors harmonics  $v_{\alpha q}^{\; -}$
 related to the momentum of the particle through (\ref{k=pv-v-}),
\begin{eqnarray}\label{w=w-rv}
w^I=w^I(\rho^{\#}, v_{q}^{ -})\qquad  and \qquad
\psi_q=\psi_q(\rho^{\#}, v_{q}^{ -})\;  \qquad
\end{eqnarray}
 When describing the on-shell states of the SYM multiplet, it is suggestive to replace $\rho^{\#}$ by  its conjugate coordinate and consider the field
on the nine-dimensional space ${\bb R}\otimes {\bb S}^{8}$:
\begin{eqnarray}\label{w=w-xv}
w^I=w^I(x^=, v_{q}^{ -})\qquad  and \qquad
\psi_q=\psi_q(x^=, v_{q}^{ -})\; . \qquad
\end{eqnarray}
The supersymmetry acts on these 9d fields by
\begin{eqnarray}\label{susy=8+8a}
\delta_{\epsilon}\psi_{\dot q }(x^=, v_{q}^{ -}) = \epsilon^{-{q}}  \gamma^I_{q\dot{q}}  \; w^I(x^=, v_{q}^{ -}) \; , \qquad \delta_{\epsilon} w^I(x^=, v_{q}^{ -}) = 2i \epsilon^{-{q}} \gamma^I_{q\dot{q}} \partial_{=} \psi_{\dot q }(x^=, v_{q}^{ -})\; \; , \qquad
\end{eqnarray}
where 8 component fermionic  $\epsilon^{-{q}}$ is the contraction of the  constant fermionic spinor $\epsilon^\alpha$ with the spinor frame variable,
\begin{eqnarray}\label{susy8=16v}
\epsilon^{-{q}}  =\epsilon^\alpha v_{\alpha q}^{\; -}  \; .  \qquad
\end{eqnarray}

\section{Constrained on-shell superfield description of 10D SYM}

The above described fields of the spinor helicity formalism for 10D SYM   can be collected in  on-shell superfields, which can be considered as one-particle prototypes of tree superamplitudes.
A constrained on-shell superfield formalism for linearized 10D SYM was proposed in
\cite{Galperin:1992pz}.  We briefly describe that in this Section and, in the next Sec. 4, use it as a starting point to obtain a new analytic superfield description of 10D SYM.

\subsection{ On-shell superspace for 10D SYM }

In \cite{Galperin:1992pz}  the constrained superfields describing 10D SYM are defined  on the real {\it on-shell superspace} with bosonic coordinates  $x^=$ and $v_{\alpha q}^{\; -}$, and fermionic  coordinates $\theta^-_q$
\begin{eqnarray}\label{On-shellSSP}
\Sigma^{(9|8)} : \qquad &&  \{ (x^= , \theta^-_q, v_{\alpha q}^{\; -})\} \; ,  \qquad \{ v_{\alpha q}^{\; -}\} = {\bb S}^{8}\; , \qquad \\ \nonumber &&
q=1,..., 8\; , \qquad \alpha =1,..., 16 \; .
\qquad \nonumber
\end{eqnarray}
The 10D  supersymmetry acts on the coordinates of $\Sigma^{(9|8)} $  by
\begin{eqnarray}\label{susy=S}
\delta_\epsilon x^== 2i \theta^-_q \; \epsilon^\alpha v_{\alpha q}^{\; -} \; , \qquad
\delta_\epsilon \theta^-_q =  \epsilon^\alpha v_{\alpha q}^{\; -}
\; , \qquad \delta_\epsilon v_{\alpha q}^{\; -}=0\; .  \qquad
\qquad
\end{eqnarray}
This specific form indicates that our on-shell superspace $\Sigma^{(9|8)}$  can be regarded as invariant subspace of the D=10 {\it Lorentz harmonic superspace}, i.e. of the direct product of standard 10D and 11D superspaces and of the internal sector parametrized by Lorentz harmonics
$(v_{\alpha \dot q}^{\; +}, v_{\alpha q}^{\; -} )\in \; Spin(1,9)$ considered as homogeneous coordinates of the coset $\frac {Spin(1,9)}{Spin(1,1)\otimes Spin(8)}$.

The generic unconstrained superfield on $\Sigma^{(9|8)}$  (\ref{On-shellSSP})  contains too many component fields so that  on-shell superfield describing linearized D=10 SYM  should obey some superfield equations.
Such equations have been proposed in \cite{Galperin:1992pz}. To write
them  in a compact form we will need the fermionic derivatives covariant under
(\ref{susy=S})
\begin{eqnarray}\label{D+dq:=}
D^+_{{q}}={\partial}^+_{{q}} + 2i  \theta^-_{{q}}\partial_{=} \; , \qquad \partial_{=}
:={\partial\over \partial x^=}\; , \qquad \partial^+_{{q}}
:={\partial\over \partial \theta^-_{{q}}}\; , \qquad q=1,..., 8\; .
\end{eqnarray}
These carry the s-spinor indices of $Spin(8)$ group and obey  $d=1$  $\frak{N}=8$ extended supersymmetry algebra
\begin{eqnarray}\label{D+qD+p=I}
\{ D^+_{{q}},  D^+_{{p}}\} = 4i \delta_{qp}\partial_=\,  .
\end{eqnarray}

A one particle counterpart of a superamplitude is actually given by Fourier images of the superfield on  (\ref{On-shellSSP}) with respect to $x^=$. These will depend on the set of coordinates
$(\rho^\# , \theta^-_q, v_{\alpha q}^{\; -})$, where $\rho^\#$ is a momentum conjugate to $x^=$. The fermionic covariant derivative acting on such Fourier-transformed  on-shell superfields reads
\begin{eqnarray}\label{D+dq:=rho}
D^+_{{q}}={\partial}^+_{{q}} + 2 \rho^\#  \theta^-_{{q}} \; , \qquad
\end{eqnarray} and obeys
\begin{eqnarray}\label{D+qD+p=rho}
\{ D^+_{{q}},  D^+_{{p}}\} = 4\rho^\# \delta_{qp}\,  .
\end{eqnarray}

\subsection{ On-shell superfields and superfield equations  of 10D SYM }

The basic superfield  equations of D=10 SYM  \cite{Galperin:1992pz}
\begin{eqnarray}\label{D+Psi=gV}
D=10\; : \qquad D^+_{{q}}\Psi_{\dot{q}} =\gamma^I_{q\dot{q}}\, V^I\; , \qquad
q= 1,...,8\; , \quad \dot{q}=1,..,8 \; , \quad I =1,..,8 \;
\end{eqnarray}
are imposed on the fermionic superfield  $\Psi_{\dot{q}}=\Psi_{\dot{q}}(x^=, \theta^-_{\dot{q}}, v_{\alpha \dot{q}}{}^-)$ carrying c-spinor  index of SO(8).
The superfield $V^I$ is defined by Eq. (\ref{D+Psi=gV}) itself, which also imply that it obeys
\begin{eqnarray}\label{D+V=gdPsi}
D^+_{{q}}V^I= 2i \gamma^I_{q\dot{q}}\partial_{=}\Psi_{\dot{q}}
\; . \qquad
\end{eqnarray}
This equation shows that there are no other independent components in the constrained on-shell superfield
$\Psi_{\dot{q}}$.

\section{An analytic on-shell superfield description of  10D  SYM }

In this section we present an {\it analytic} superfield formalism for the on-shell D=10 SYM, which is  alternative to both the Clifford superfield approach of \cite{CaronHuot:2010rj} and to the constrained superfield formalism, which we have described above (more details can be found  in \cite{Bandos:2017eof}).
 We begin by  solving the equations of the constrained on-shell superfields of 10D SYM from \cite{Galperin:1992pz} in terms of one analytic on-shell superfield. In sec. 6 we generalize this for the case of superamplitudes and describe an analytic superamplitude formalism.

\subsection{From constrained to unconstrained on-shell superfield formalism  }

To arrive at our unconstrained superfield formalism it is convenient to write the superspace equations  (\ref{D+V=gdPsi}) and (\ref{D+Psi=gV})  for on-shell superfields describing 10D SYM \cite{Galperin:1992pz} in the form of
\begin{eqnarray}\label{DWI=gPsi}
 && D^+_{{q}}W^I= 2i \gamma^I_{q\dot{q}}\Psi_{\dot{q}} \; , \qquad
\qquad \\
\label{DPsi=gW}
 && D^+_{{q}}\Psi_{\dot{q}} =\gamma^I_{q\dot{q}}\, \partial_{=} W^I\; , \qquad q= 1,...,8\; , \quad \dot{q}=1,..,8 \; , \quad I =1,..,8 \; .
\end{eqnarray}
The superfield $V^I$ in (\ref{D+Psi=gV}) and (\ref{D+V=gdPsi})  is related to $W^I$ by
$V^I=\partial_{=} W^I$. After such a  redefinition, we can discuss the bosonic superfield $W^I$ as
fundamental and state that $\Psi_{\dot{q}}$ is defined by the $\gamma$-trace part of (\ref{DWI=gPsi}).
The first terms in its decomposition on fermionic coordinates are
$$
W^I = w^I+2i  \theta^-\gamma^I\psi + i  \theta^-\gamma^{IJ} \theta^-  \partial_=w^I - \frac 2 3
\theta^-\gamma^{IJ} \theta^- \; \theta^- \gamma^I \partial_=\psi + \ldots \; . \qquad
$$

We are going to show that, after breaking SO(8) symmetry down to its SO(6)=SU(4) subgroup, Eq.
 (\ref{DWI=gPsi}) splits into a chirality condition for a single complex superfield ($\Phi=W^7+iW^8$) and other parts which, together with (\ref{DPsi=gW}), allow to determine $\Psi_{\dot{q}} $ and all the remaining components of $W^I$ in terms of this single chiral superfield.

\subsection{SU(4) invariant solution of the constrained superfield equations }

Breaking $SO(8)\mapsto SO(6)\otimes SO(2)\approx SU(4)\otimes U(1)$, we can split the vector representation {\bf 8}$_v$ of SO(8) on {\bf 6+1+1} of $SO(6)$,
\begin{eqnarray}\label{WI=6+1+1}
W^I= (W^{\check{I}}, W^7, W^8), \qquad  {\check{I}}=1,...,6\; .
\end{eqnarray}
Then introducing
\begin{eqnarray}\label{Phi=W7-iW8}
\Phi= \frac {W^7 -i W^8} 2, \qquad  \bar{\Phi}= \frac {W^7 +i W^8} 2, \qquad  \Psi_q = \gamma^8_{q\dot{q}}\Psi_{\dot{q}} \; ,
\end{eqnarray}
we find that (\ref{DWI=gPsi}) implies
\begin{eqnarray}\label{DPhi=pPsi}
 D^+_{{q}}\Phi &=& \left(\delta_{qp} + i (\gamma^7\tilde{\gamma}{}^8)_{qp}\right)\Psi_p \; , \qquad  \nonumber \\
  D^+_{{q}}\bar{\Phi} &=& -\left(\delta_{qp} - i (\gamma^7\tilde{\gamma}{}^8)_{qp}\right)\Psi_p
  . \qquad
\end{eqnarray}
It is important to notice that the matrices  \begin{eqnarray}\label{SO6proj}
{\cal P}^{\pm}_{qp}= \frac 1 {{2}}\left(\delta_{qp} \pm  i (\gamma^7\tilde{\gamma}{}^8)_{qp}\right)\; , \qquad
\end{eqnarray}
 are orthogonal projectors
\begin{eqnarray} \label{SO6proj*} {\cal P}^{+}{\cal P}^{+}={\cal P}^{+}\; , \qquad {\cal P}^{-}{\cal P}^{-}={\cal P}^{-}\; ,  \qquad
{\cal P}^{+}{\cal P}^{-}=0  \qquad  \\  \label{P+P=1}
{\cal P}^{+}+ {\cal P}^{-}= {\bb I}\; , \qquad ({\cal P}^{+})^*= {\cal P}^{-}\; , \qquad
\end{eqnarray}
and hance that (\ref{DPhi=pPsi}) implies
\begin{eqnarray}\label{I-ig7g8=}
 \left(\delta_{qp} - i (\gamma^7\tilde{\gamma}{}^8)_{qp}\right) D^+_{{p}}\Phi = 0 \; , \qquad
 \left(\delta_{qp} + i (\gamma^7\tilde{\gamma}{}^8)_{qp}\right)
  D^+_{{p}}\bar{\Phi} =0
  . \qquad
\end{eqnarray}
As, according to (\ref{P+P=1}), the projectors ${\cal P}^{+}$ and $ {\cal P}^{-}$ are complementary and complex conjugate, we can introduce complex 8$\times$4 matrix $w_q{}^A$ and its complex conjugate $\bar{w}_{qA}$ such that
\begin{eqnarray}\label{p+=wbw}
 \left(\delta_{qp} + i (\gamma^7\tilde{\gamma}){}^8_{qp}\right) = 2w_{q}{}^A \bar{w}_{pA}
  \; , \qquad
 \left(\delta_{qp} - i (\gamma^7\tilde{\gamma}){}^8_{qp}\right)
= 2\bar{w}_{qA}w_p{}^A
  .  \qquad
\end{eqnarray}
In terms of these rectangular blocks Eqs. (\ref{I-ig7g8=}) can be written as chirality (analyticity) conditions
\begin{eqnarray}\label{bDPhi=0}
  \bar{D}{}^+_A
\Phi =0 \; , \qquad  D^{+A}\bar{\Phi}=0 \; ,
 \qquad
\end{eqnarray}
with
\begin{eqnarray}\label{D+A:=}
 \bar{D}^+_A = \bar{w}_{pA} D^+_q \; , \qquad  D^+_A  =  w_{q}{}^A D^+_q\; . \qquad
\end{eqnarray}
The remaining parts of Eqs. (\ref{DPhi=pPsi}) determine the fermionic superfield $\Psi_{\dot q}$,
\begin{eqnarray}\label{Psi=DPhi}
\Psi_{\dot q} =  w_{\dot q}{}^A \bar{\Psi}{}^{+A} + \bar{w}_{\dot q A} \Psi^+_A\; , \qquad  \Psi^+_A= - \frac i 4   D^+_A {\Phi}\; ,
  \qquad \bar{\Psi}{}^{+A}= - \frac i 4 \bar{D}{}^+_{A}\bar{\Phi}\; . \qquad
\end{eqnarray}
Eq.  (\ref{DPsi=gW}) allows us to find also the derivatives of the remaining 6 components $W^{\check{I}}$ of the SO(8) vector superfield $W^I$,
\begin{eqnarray}\label{W6=DDPhi}
  \partial_= W^{\check{I}} =  \frac 1 8  (\gamma^{\check{I}} \tilde{\gamma}{}^8)_{qp} D^+_qD^+_p (\Phi-\bar{\Phi})  \; . \qquad
\end{eqnarray}
To conclude, we have solved the equations for constrained on shell superfields of 10D SYM
\cite{Galperin:1992pz} in terms of one chiral (analytic) on-shell superfield $\Phi$ and its c.c. $\bar{\Phi}$ (\ref{Phi=W7-iW8}).

\subsection{The on-shell superfields are analytic rather  than chiral}
\label{sec=U-harm}

Our solution breaks explicitly the manifest $SO(D-2)=SO(8)$ 'little group' invariance of the constrained superfield formalism down to  $SO(D-4)=SO(6)$ (called 'tiny group' in \cite{Boels:2012zr}).
Actually, one can avoid this explicit  $SO(8)\mapsto SO(6)\otimes SO(2) \approx SU(4)\otimes U(1)$ symmetry breaking by using the method of harmonic superspaces \cite{Galperin:1984av,Galperin:2001uw}. To this end we must write the general solution of the constrained superfield equations in a formally  SO(8) invariant form  by introducing a `bridge' coordinates parametrizing $SO(8)/[SU(4)\otimes U(1)]$ coset: the  $SO(8)$ valued matrix
\begin{eqnarray}\label{UinSO8}
&  U_I^{(J)}= \left(U_I{}^{\check{J}}, U_I{}^{(7)}, U_I{}^{(8)}\right)= \left(U_I{}^{\check{J}}, \frac 1 2 \left( U_I+ \bar{U}_I\right), \frac 1 {2i} \left( U_I- \bar{U}_I \right)\right) \; \in \; SO(8)
 \; . \qquad
\end{eqnarray}
This  is transformed by multiplication on $SO(8)$ matrix from the left and by multiplication by $SO(6)\times SO(2)\subset SO(8)$ matrix from the right. The conditions of orthogonality   of the $ U_I^{(J)}$ matrix (\ref{UinSO8}), $  U_I^{(J)} U_I^{(K)}=\delta^{(J)(K)}$, imply that the complex vector $U_I$ is null and has the norm equal to 2,
\begin{eqnarray}\label{U2=0}
U_IU_I=0\; , \qquad \bar{U}_I\bar{U}_I=0
\; , \qquad U_I\bar{U}_I=2\; ,
\end{eqnarray}
as well as that it is orthogonal to six mutually orthogonal real vectors $U_I{}^{\check{I}}$
\begin{eqnarray}\label{UU6=0}
U_IU_I{}^{\check{J}}=0\; , \qquad \bar{U}_IU_I{}^{\check{J}}=0
\; , \qquad U_I{}^{\check{J}}U_I{}^{\check{K}}=\delta ^{\check{J}\check{K}} \; .
\end{eqnarray}
Now we can easily define $SO(8)$ covariant counterparts of the projectors in (\ref{SO6proj})
\begin{eqnarray}\label{SO6proj=}
{\cal P}^{+}_{qp}= \frac 1 {{2}}\left(\delta_{qp} +  i (\gamma^{I}\tilde{\gamma}{}^{J})_{qp} U_I^{(7)}U_J^{(8)}\right)
= \frac 1 4 \gamma^I\tilde{\gamma}{}^J \bar{U}_I U_J \; , \qquad  \nonumber \\ {\cal P}^{-}_{qp}= \frac 1 {{2}}\left(\delta_{qp} -   i (\gamma^{I}\tilde{\gamma}{}^{J})_{qp} U_I^{(7)}U_J^{(8)}\right)
= \frac 1 4 \gamma^I\tilde{\gamma}{}^J  U_I\bar{U}_J \; . \qquad
\end{eqnarray}

Furthermore, we  can define the 8$\times$8 SO(8) valued matrices $w_q^{(p)}$ and $w_{\dot{q}}^{(\dot{p})}$,
which are related to (\ref{UU6=0}) by
\begin{eqnarray}\label{Ug8=wg8w}
 \gamma^I_{q \dot{p}} U_I^{(J)}= w_q^{(p)}\gamma^{(J)}_{(p) (\dot{q})}  w_{\dot{p}}^{(\dot{q})} \; , \qquad
 w_{q'}^{(p)}w_{q'}^{(q)}=\delta^{(p)(q)}\; , \qquad  w_{\dot{p}'}^{(\dot{q})}  w_{\dot{p}'}^{(\dot{p})} = \delta^{(\dot{q})(\dot{p})}\; .
\end{eqnarray}
The elements of these real matrices can be combined in two rectangular $8\times 4$ complex conjugate blocks
\begin{eqnarray}\label{wqA=}
w_q^A = (\bar{w}_{qA})^* \; , \qquad w_{\dot{q}}^{\, A} = (\bar{w}_{\dot{q}A})^* \; , \qquad A=1,2,3,4\; . \qquad
\end{eqnarray}
These obey
\begin{eqnarray}\label{wbw+cc=1}
&& w_{q}{}^A \bar{w}_{pA}+ \bar{w}_{qA}w_p{}^A =\delta_{qp}\; , \qquad \\
\label{bww=1}
&& \bar{w}_{qB}w_{q}{}^A =\delta_B{}^A\; , \qquad w_{q}{}^A w_{q}{}^B =0 \; , \qquad \bar{w}_{qA} \bar{w}_{qB} =0\; . \qquad
\end{eqnarray}
and factorize the orthogonal projectors (\ref{SO6proj=})
\begin{eqnarray}\label{SO6proj=ww}
{\cal P}^{+}_{qp}=
\frac 1 4 \gamma^I\tilde{\gamma}{}^J \bar{U}_I U_J =  w_{q}{}^A \bar{w}_{pA}
  \; , \qquad {\cal P}^{-}_{qp}= \frac 1 4 \gamma^I\tilde{\gamma}{}^J  U_I\bar{U}_J = \bar{w}_{qA}w_p{}^A  \; \qquad
\end{eqnarray}
({\it cf.} (\ref{p+=wbw})).

With a suitable choice of representation of 8d Clebsch-Gordan coefficients $\gamma^I_{q\dot{q}}= \tilde{\gamma}{}^I_{\dot{q}q}$,
the first equation in  (\ref{Ug8=wg8w}) can be split into
\begin{eqnarray}\label{Ug86=wg6w+cc}
U\!\!\!\!/{}^{\check{J}}_{q \dot{p}}:= \gamma^I_{q \dot{p}} U_I^{\check{J}}= i w_q^{A}\sigma^{\check{J}}_{AB}  w_{\dot{p}}^{B} + i \bar{w}_{qA}\tilde{\sigma}{}^{\check{J}AB } \bar{w}_{\dot{p}B} \; , \qquad \\
 \label{Ug8=bww}
U\!\!\!\!/{}_{q \dot{p}}:=  \gamma^I_{q \dot{p}} U_I = 2  \bar{w}_{{q}A} w_{\dot{p}}^A  \; , \qquad
 \bar{U}\!\!\!\!/{}_{q \dot{p}}:= \gamma^I_{q \dot{p}} \bar{U}_I = 2   w_q^{A} \bar{w}_{\dot{p}A} \; . \qquad
\end{eqnarray}
 In (\ref{Ug86=wg6w+cc})
\begin{eqnarray}\label{sigma6d=}
{\sigma}^{\check{I}}_{AB}= - {\sigma}^{\check{I}}_{BA}= - (\tilde{\sigma}^{\check{I} AB})^*=  {1\over 2}\epsilon_{ABCD}
\tilde{\sigma}^{\check{I}\, CD} \; ,  \qquad \check{I}=1,\ldots, 6\; , \quad  A,B,C,D=1,\ldots , 4\;
\qquad
\end{eqnarray}
are 6d Clebsch-Gordan coefficients which obey
\begin{eqnarray}\label{Cliff6d} {\sigma}^{\check{I}}\tilde{\sigma}^{\check{J}}+ {\sigma}^{\check{J}}\tilde{\sigma}^{\check{I}}= 2{\delta}^{\check{I}\check{J}}
\delta_A{}^B \; , \qquad  \label{so(6)id} {\sigma}^{\check{I}}_{AB}\tilde{\sigma}^{I CD}= -4
\delta_{[A}{}^{C}\delta_{B]}{}^{D}\; , \qquad {\sigma}^{\check{I}}_{AB}\, {\sigma}^{\check{I}}_{CD} = -2\epsilon_{ABCD}\; .
\qquad
\end{eqnarray}

Using (\ref{Ug8=bww}) and (\ref{bww=1}), it is not difficult to check that Eqs. (\ref{SO6proj=ww}) are satisfied
\footnote{\label{gI=Uww=foot} In this calculation and below the following identity is  useful
\begin{eqnarray}\label{gI=Uww}
\gamma^I_{q \dot{p}}= \bar{U}_I  \bar{w}_{{q}A} w_{\dot{p}}^A  + U_I  w_q^{A} \bar{w}_{\dot{p}A}  +
 U_I^{\check{J}} (i w_q^{A}\sigma^{\check{J}}_{AB}  w_{\dot{p}}^{B} + i \bar{w}_{qA}\tilde{\sigma}{}^{\check{J}AB } \bar{w}_{\dot{p}B} )
\nonumber \; . \qquad
\end{eqnarray}
}.

The above bridge coordinates or {\it harmonic variables} \cite{Galperin:1984av,Galperin:1984bu,Galperin:2001uw} can be used to define the SO(8) invariant version  of complex  covariant derivatives (\ref{D+A:=}),
and of complex linear combinations of $8$ bosonic superfields $W^I$
\begin{eqnarray}\label{Phi=WU}
 \Phi= W^I U_I \; , \qquad \bar{\Phi} = W^I \bar{U}_I \; , \qquad
\end{eqnarray}
which are analytic and anti-analytic, (\ref{bDPhi=0}).

The expression for fermionic superfield $\Psi_{\dot{q}}$ can be written in the form of  (\ref{Psi=DPhi}),
but now with $w$ and $\bar{w}$ factorizing the covariant projectors (\ref{SO6proj=ww}).
It is also not difficult to write the covariant counterpart of the expression (\ref{W6=DDPhi}) for other
$6$ projections $W^{\check{I}} = W^J U_J^{\check{I}}$  of the $8$-vector
superfield $W^I$.
However, a more straightforward expression for  $W^{\check{I}} = W^J U_J^{\check{I}}$ in terms of $\Phi$ reads \begin{eqnarray}\label{WchI=DchIP}
 W^{\check{I}} = - \overline{{\bb D}}{}^{\check{I}}\Phi \; , \qquad
\end{eqnarray}
where
\begin{eqnarray}
\label{bDchJ}
\overline{{\bb D}}{}^{\check{J}}&=&  \frac 1 2 \bar{U}_I \frac {\partial} {\partial U_I^{\check{J}}}-  U_I^{\check{J}} \frac \partial {\partial {U}_I} + \frac i 2 \sigma{}^{\check{J}}_{AB} w_q^B \frac \partial {\partial \bar{w}_{qA}} - \frac i 2 \tilde{\sigma}{}^{\check{J}AB} \bar{w}_{\dot{q}B} \frac \partial {\partial w_{\dot{q}}^A}\; , \qquad
\end{eqnarray}
is one of the covariant harmonic derivatives (first introduced in \cite{Galperin:1984av} and \cite{Galperin:1984bu} for $SU(2)/U(1)$ and $SU(3)/[U(1)\times U(1)]$ harmonic variables). In our case the other covariant derivatives
are
\begin{eqnarray}
\label{DchJ}
{\bb D}^{\check{J}} &=& \frac 1 2 U_I\frac {\partial} {\partial U_I^{\check{J}}}-  U_I^{\check{J}} \frac  {\partial} {\partial \bar{U}_I}  + \frac i 2 \tilde{\sigma}{}^{\check{J}AB} \bar{w}_{qB} \frac \partial {\partial w_q^A}- \frac i 2 \sigma{}^{\check{J}}_{AB} w_{\dot{q}}^B \frac \partial {\partial \bar{w}_{\dot{q}A}} \; , \qquad
\end{eqnarray}
conjugate to (\ref{bDchJ}), and
\begin{eqnarray}
\label{D0:=}
{\bb D}^{(0)} &=& U_I\frac {\partial} {\partial U_I}-  \bar{U}_I\frac  {\partial} {\partial \bar{U}_I}  + \frac 1 2\left(\bar{w}_{qA} \frac \partial {\partial \bar{w}_{qA}}- w_q^A \frac \partial {\partial w_q^A}\right) + \frac 1 2\left(w_{\dot{q}}^A \frac \partial {\partial w_{\dot{q}A} ^A}-\bar{w}_{\dot{q}A} \frac \partial {\partial \bar{w}_{\dot{q}A} }\right) \; , \qquad
\end{eqnarray}
\begin{eqnarray}
\label{DchIJ}
{\bb D}^{\check{I}\check{J}} &=& \frac 1 2 \left(  U_K^{\check{I}}\frac {\partial} {\partial U_K^{\check{J}}}-  U_K^{\check{J}} \frac  {\partial} {\partial U_K^{\check{I}} }\right)  + \frac i 2 {\sigma}{}^{\check{I}\check{J}}{}_B{}^A \left(  w_q^B \frac \partial {\partial w_q^A}-  \bar{w}_{qA }\frac \partial {\partial \bar{w}_{qB}} \right) + \qquad \nonumber \\
&&+ \frac i 2 {\sigma}{}^{\check{I}\check{J}}{}_B{}^A \left(  w_{\dot{q}}^B \frac \partial {\partial w_{\dot{q}}^A}-  \bar{w}_{\dot{q}A }\frac \partial {\partial \bar{w}_{\dot{q}B}} \right)
\; , \qquad
\end{eqnarray}
providing the differential operator representation of the $U(1)$ and $Spin(6)=SU(4)$ generators on the space of internal harmonics.
These covariant derivatives preserve all the constraints on harmonic variables, Eqs.  (\ref{wbw+cc=1}), (\ref{bww=1}), (\ref{Ug86=wg6w+cc}) and  (\ref{Ug8=bww}), and form the $so(8)$ algebra.

One can easily check that, by construction,  our analytic superfield (\ref{Phi=WU}) obeys
\begin{eqnarray}
\label{DchIP=0}
{\bb D}^{\check{J}} \Phi=0 \; , \qquad \\ \label{DchIJP=0}
{\bb D}^{\check{I}\check{J}} \Phi=0 \; , \qquad \\ \label{D0P=P}
{\bb D}^{(0)} \Phi=\Phi \; . \qquad
\end{eqnarray}
These equations are consistent with the analyticity conditions (\ref{bDPhi=0}) as
\begin{eqnarray}
\label{DJDf-=0}
{}[{\bb D}^{\check{J}}, \bar{D}{}^+_A]=0\; . \qquad
\end{eqnarray}

 \subsection{Analytic superfields and harmonic on-shell superspace}

Thus, we have solved the superfield equations for constrained on-shell superfields of $D=10$
SYM in term of one complex
analytic superfield $\Phi$ obeying the chirality-type equation (\ref{bDPhi=0})
with complex fermionic derivatives (\ref{D+A:=}) defined with the use of
$\frac {Spin(8)}{Spin(6)\otimes U(1)}= \frac {SO(8)}{SU(4)\otimes U(1)}$ coset coordinates
(\ref{wbw+cc=1}), (\ref{bww=1}), which we, following
\cite{Galperin:1984av,Galperin:1984bu,Galperin:2001uw}, call harmonic variables or internal harmonics.

These analytic superfields are actually defined on a 'harmonic on-shell superspace' which can be
understood as direct product of the on-shell superspace (\ref{On-shellSSP}) and the
$\frac {Spin(D-2)}{Spin(D-4)\otimes U(1)}$ coset
\begin{eqnarray}\label{HOn-shellSSP}
\Sigma^{(3(D-3)|2{\cal N})} \; = &&  \{ (x^= ,  v_{\alpha q}^{\; -}; \bar{w}_{qA}, w_q^A ; \theta^-_q
)\} \; ,  \qquad \\ \nonumber && \{ x^=\} = {\bb R}^1\; , \qquad \{ v_{\alpha q}^{\; -}\} = {\bb S}^{D-2}\; , \qquad \{ (\bar{w}_{qA}, w_q^A)\}= \frac {Spin(D-2)}{Spin(D-4)\otimes U(1)}\; .
\qquad
\end{eqnarray}
Here and below in (\ref{anHOn-shellSSP}), to exclude the literal repetition of the same equations, we write them in the form applicable both for
$D=10$ and $D=11$ cases, for which
\begin{eqnarray}\label{ind=10D11D}
q=1,..., 2{\cal N}\; , \qquad \alpha =1,..., 4{\cal N}\; , \qquad
{\cal N}= \begin{cases} \; 4\qquad for\; D=10 \cr  8\qquad  for\; D=11 \; . \end{cases}
\qquad
\end{eqnarray}

Supersymmetry acts on the coordinates of the harmonic on-shell  superspace by ({\it cf}. (\ref{susy=S}))
\begin{eqnarray}\label{susy=R}
\delta_\epsilon x^== 2i \theta^-_q \; \epsilon^\alpha v_{\alpha q}^{\; -} \; , \qquad
\delta_\epsilon \theta^-_q =  \epsilon^\alpha v_{\alpha q}^{\; -}
\; , \qquad \delta_\epsilon v_{\alpha q}^{\; -}=0\; , \qquad \delta_\epsilon \bar{w}_{qA}=0=\delta_\epsilon {w}_{q}^{\;A}\; , \qquad
\qquad
\end{eqnarray}
and leaves invariant the covariant derivatives (\ref{D+dq:=})
\begin{eqnarray}\label{D+q=C}
D^+_{q} = \partial^+_q +2i\theta^-_q \partial_=\; , \qquad
D_= =\partial_= \; , \qquad
\qquad
\end{eqnarray}
as well as $\bar{D}{}^+_A= \bar{w}_{qA} D^+_q$ used to define analytic superfields $\Phi$ by
$\bar{D}{}^+_A\Phi=0$,
(\ref{bDPhi=0}).

To see that the  analytic superfields are actually functions on a sub-superspace of (\ref{HOn-shellSSP}), we have to pass to the analytic coordinate basis.

 \subsection{Analytical basis and analytic subsuperspace of the harmonic on-shell superspace}

The presence of additional harmonic variables allows to change the coordinate basis of the
harmonic on-shell superspace $\Sigma^{(3(D-3)|2{\cal N})}$ to the following {\it analytical basis}
\begin{eqnarray}\label{anHOn-shellSSP}
\Sigma^{(3(D-3)|2{\cal N})} &=&   \{ (x_L^= ,  v_{\alpha q}^{\; -}; \bar{w}_{qA}, w_q^A ; \eta^-_A, \bar{\eta}{}^{-A}
)\} \; ,  \qquad \\ \nonumber && {}\qquad  x^=_L:= x^= + 2i \eta^-_A\bar{\eta}{}^{-A}\; , \qquad \eta^-_A:= \theta^-_q \bar{w}_{qA}\, , \qquad \bar{\eta}{}^{-A}= \theta^-_q {w}_q^{A} \; . \qquad
\end{eqnarray}
The supersymmetry acts on the coordinates of this basis by
\begin{eqnarray}\label{susy=An}
\delta_\epsilon x^=_L= 4i \eta^-_A\bar{\epsilon}{}^{-A}\; , \qquad \delta_\epsilon  \eta^-_A = {\epsilon}{}^{-}_{A} \; , \qquad  \delta_\epsilon \bar{\eta}{}^{-A}= \bar{\epsilon}{}^{-A}\; , \qquad
\qquad
\end{eqnarray}
where
\begin{eqnarray}\label{ep-A=}
{\epsilon}{}^{-}_{A}= \epsilon^\alpha v_{\alpha q}^{\; -}  \bar{w}_{qA}\; , \qquad \bar{\epsilon}{}^{-A}=  \epsilon^\alpha v_{\alpha q}^{\; -}  {w}_q^{\; A}\; . \qquad
\end{eqnarray}
It is generated by the differential operators
  \begin{eqnarray}\label{bQ+A=}
\bar{Q}{}^+_A= \bar{\partial}^+_A + 4i {\eta}{}^{-}_{A}  \partial^L_=\; , \qquad
Q^{+A}= \partial^+_A \;  \qquad
\end{eqnarray}
and leaves invariant the covariant derivatives
\footnote{To be rigorous, one might want to write the $L$ symbol also on the fermionic derivatives in
(\ref{D+q=An}), $ \bar{\partial}^+_A \mapsto  \bar{\partial}^{+}_{A\, L}$,
$\partial^{+A} \mapsto \partial^{+A}_{\, L}$. We, however, prefer to make the formulae lighter
and write this symbol on the bosonic derivative $\partial_=^L$ only. }
\begin{eqnarray}\label{D+q=An}
\bar{D}{}^+_A= \bar{\partial}^+_A \equiv \frac {\partial}{\partial \bar{\eta}{}^{-A}} \; , \qquad
D^{+A}= \partial^+_A +4i\bar{\eta}{}^{-A} \partial^L_=\; , \qquad
D_= =\partial^L_= \equiv \frac {\partial}{\partial x^{=}_L} \; . \qquad
\qquad
\end{eqnarray}
The harmonic covariant derivatives in the analytical basis have the form
\begin{eqnarray}
\label{DchJ=An}
{ D}^{\check{J}} &=& {\bb D}^{\check{J}} - \frac i 2 \eta^-_A \tilde{\sigma}{}^{\check{J}AB} \frac \partial {\partial \bar{\eta}{}^{-B}} \; , \qquad
\\
\label{bDchJ=An}
\overline{{D}}{}^{\check{J}}&=&\overline{{\bb D}}{}^{\check{J}}-  \frac i 2 \bar{\eta}{}^{-A} \sigma{}^{\check{J}}_{AB} \frac \partial {\partial \eta^-_B } \; , \qquad
\\
\label{D0=An}
{D}^{(0)} &=& {\bb D}^{(0)}   + \frac 1 2  \eta^-_A  \frac \partial {\partial  \eta^-_A } - \frac 1 2  \bar{\eta}{}^{-A} \frac \partial {\partial \bar{\eta}{}^{-A} }\; , \qquad
\\
\label{DchIJ=An}
{D}^{\check{I}\check{J}} &=& {\bb D}^{\check{I}\check{J}} + \frac i 2  \bar{\eta}{}^{-B} {\sigma}{}^{\check{I}\check{J}}{}_B{}^A  \frac \partial {\partial\bar{\eta}{}^{-A}}-  \frac i 2  {\sigma}{}^{\check{I}\check{J}}{}_B{}^A \eta^-_A \frac \partial {\partial \eta^-_B}
\; , \qquad
\end{eqnarray}
where ${\bb D}^{\check{J}}$,  $\overline{{\bb D}}{}^{\check{J}}$, ${\bb D}^{(0)}$ and ${\bb D}^{\check{I}\check{J}}$ formally coincides with
(\ref{DchJ}), (\ref{bDchJ}), (\ref{D0:=}) and  (\ref{DchIJ}).

It is not difficult to see that supersymmetry (\ref{susy=An})  leaves invariant
{\it analytical on-shell superspace} $\Sigma_L^{(3(D-3)|{\cal N})}$, a sub-superspace
of $\Sigma^{(3(D-3)|2{\cal N})}$ with coordinates
\begin{eqnarray}\label{an=subSSP}
\Sigma_L^{(3(D-3)|{\cal N})} &=&    \{ (x_L^= ,  v_{\alpha q}^{\; -}; \bar{w}_{qA}, w_q^A ; \eta^-_A
)\} \; .
\end{eqnarray}
The above defined analytic superfields are superfields on this analytic sub-superspace,
\begin{eqnarray}\label{an=SF}
 && \Phi=\Phi (x_L^= , \eta^-_A;  v_{\alpha q}^{\; -}; \bar{w}_{qA}, w_q^A ) \qquad \Leftrightarrow \qquad \bar{D}{}^+_A\Phi=0\; .
\end{eqnarray}
The supersymmetry transformation of the analytical superfields are defined by $\Phi^\prime  (x_L^{= \prime} , \eta^{-\prime}_A; ...  ) =
 \Phi (x_L^= , \eta^-_A; ... )$, or equivalently,
\begin{eqnarray}\label{anSF=susy}
\Phi^\prime  (x^=_L , \eta^{-}_A; ...  ) &=& e^{^{-(\bar{\epsilon}{}^{-A} \bar{Q}{}_A^++ {\epsilon}{}^{-}_{A} {Q}{}^{+A})}} \Phi (x^=_L , \eta^{-}_A; ...  ) =
\qquad \\
&=& e^{^{-4i\bar{\epsilon}{}^{-A} \eta_A^-\partial_{=}^L - {\epsilon}{}^{-}_{A} \partial_L^{+A}}} \Phi (x^=_L , \eta^{-}_A; ...  )\; . \qquad\nonumber
\end{eqnarray}

For our discussion of the amplitudes, it will be useful to consider a Fourier image of an analytic superfield with respect to $x_L^= $,
\begin{eqnarray}\label{anSF=FT}
 && \Phi (x_L^= , \eta^-_A;  v_{\alpha q}^{\; -}; \bar{w}_{qA}, w_q^A ) = \int d\rho^\# e^{^{-i\rho^\# x_L^=}}\Phi (\rho^\# , \eta^-_A;  v_{\alpha q}^{\; -}; \bar{w}_{qA}, w_q^A )
 \; . \qquad
\end{eqnarray}
The supersymmetry acts on this Fourier image as
\begin{eqnarray}\label{FanSF=susy}
\Phi^\prime  (\rho^\#  , \eta^{-}_A; ...  )
&=& e^{^{-4\rho^\#  \bar{\epsilon}{}^{-A} \eta_A^-- {\epsilon}{}^{-}_{A} \partial^{+A}}} \Phi (\rho^\# , \eta^{-}_A; ...  )\;  \qquad \\
&=& exp\{{-4\rho^\#  \bar{\epsilon}{}^{-A} \eta_A^-}\}\; \Phi (\rho^\# , \eta^{-}_A-{\epsilon}{}^{-}_{A}; ...  ) \; . \qquad
\end{eqnarray}

The analytic  on-shell superfield  can be decomposed in series on complex fermionic variable,
 \begin{eqnarray}\label{Phi=SYM}  \Phi (\rho^\# , v_{\alpha q}^{\; -}; w, \bar{w}; \eta_A) = \phi^{(+)} +
    \eta_A \psi^{+1/2\,  A} + \frac 1 2 \eta_B \eta_A \phi^{ AB }+ (\eta)^{\wedge 3\; A} \psi^{-1/2 }_{A}
  + (\eta^+)^{\wedge 4}  \phi^{(-)}\; . \qquad
\end{eqnarray}
In this description 8 fermions $\psi_{\dot{q}}$ and 8 bosons  $w^I$ of the SO(8) covariant constrained superfield formalism are split into  {\bf 4+4} and {\bf 1+6+1} representations of $SO(6)\approx SU(4)$
 \begin{eqnarray} \label{fermi=8}
 \psi^\alpha \; & \leftrightarrow & \qquad (\; \Psi_{\dot{q} }\; ) \; = \;
 (\psi^{+1/2\, A} , \psi^{-1/2}_A)\; , \qquad \\
 \label{bose=8}
  A_{\mu}\; &\leftrightarrow& \qquad (\;w^I\;) \; = \;
 (\phi^{(+)} , \phi^{AB}, \phi^{(-)} )\; . \qquad
\end{eqnarray}
The sign and numerical superscripts of the fields describe their charge with respect to $U(1)$ group acting on $\eta_A=\eta^-_A$. The origin of the analytic superfield in components of SO(8) vectors suggests that its charge is equal to +1, $  \Phi = \Phi ^{(+)}$.

This can be expressed by the differential equation
\begin{eqnarray}\label{hh-1=10D}
\hat{h}^{(10D)} \Phi (\rho^\# , \, v_{\alpha q}^{\; -}; w, \bar{w}; \eta_A) = \Phi (\rho^\# , v_{\alpha q}^{\; -}; w, \bar{w}; \eta_A)\; ,\qquad
\end{eqnarray}
where
\begin{eqnarray}
 \label{hh:=10D}
\hat{h}^{(10D)} := \frac{1}{2}  \bar{w}_{qA} \frac \partial {\partial\bar{w}_{qA}} - \frac{1}{2}w_q^{A} \frac \partial {\partial w_q^{A}} +\frac{1}{2}
\eta_A  \frac \partial {\partial \eta_A }\;    \qquad \end{eqnarray}
is the 10D counterpart of the superhelicity operator.
It is easy to see that, when acting on an analytic superfield, $\hat{h}^{(10D)}$ coincides with covariant harmonic derivative $D^{(0)}$ (\ref{D0=An}),
\begin{eqnarray}
 \label{hh:=10DAn}
\hat{h}^{(10D)} = D^{(0)} \vert_{on\; analytic\; superfields }\; ,   \qquad \end{eqnarray}
so that Eq. (\ref{hh-1=10D}) for analytical $\Phi$ coincides with
${D}^{(0)} \Phi=\Phi$.
Indeed, while the central basis form of this  latter equation is given by (\ref{D0P=P}) with ${\bb D}^{(0)}$ from (\ref{D0:=}),  in the analytical basis we have to use the covariant derivative  (\ref{D0=An}) so that our  analytic superfield, being independent of $\bar{\eta}{}^{-A}$, obeys
\begin{eqnarray}
\label{D0Phi-Phi=An}
({D}^{(0)} -1) \Phi (\rho^\# , \, v_{\alpha q}^{\; -}; w, \bar{w}; \eta_A)= \left( {\bb D}^{(0)}   + \frac 1 2  \left(\eta^-_A  \frac \partial {\partial  \eta^-_A } \right)-1\right) \Phi (\rho^\# , \, v_{\alpha q}^{\; -}; w, \bar{w}; \eta_A)=0\; . \qquad
\end{eqnarray}
This is identical to (\ref{hh-1=10D}) with $\hat{h}^{(10D)}$ given in (\ref{hh:=10D}).

Similarly, the analytic basis counterparts of Eqs. (\ref{DchIP=0}) and (\ref{DchIJP=0}), ${D}^{\check{J}} \Phi=0$ and ${D}^{\check{I}\check{J}} \Phi=0$, include the derivatives (\ref{DchJ=An}) and  (\ref{DchIJ=An}) and read
\begin{eqnarray}
\label{DchIP=0An}
& {D}^{\check{J}} \Phi (\rho^\# , \, v_{\alpha q}^{\; -}; w, \bar{w}; \eta_A)= {\bb  D}^{\check{J}}  \Phi (\rho^\# , \, v_{\alpha q}^{\; -}; w, \bar{w}; \eta_A) =0 \; , \qquad \\ \label{DchIJP=0An}
& {\bb D}^{\check{I}\check{J}}   \Phi (\rho^\# , \, v_{\alpha q}^{\; -}; w, \bar{w}; \eta_A) =  \frac i 2  {\sigma}{}^{\check{I}\check{J}}{}_B{}^A \eta^-_A \frac \partial {\partial \eta^-_B} \;  \Phi (\rho^\# , \, v_{\alpha q}^{\; -}; w, \bar{w}; \eta_A)\; .  \qquad
\end{eqnarray}
In both cases the analyticity (\ref{bDPhi=0}) of the superfield $\Phi$,
\begin{eqnarray}
\label{bDAP=0An}
\bar{D}{}^+_A \Phi  =  \frac {\partial}{\partial \bar{\eta}{}^{-A}} \Phi (\rho^\# , \, v_{\alpha q}^{\; -}; w, \bar{w}; \eta_A) =0 \;  ,  \qquad
\end{eqnarray}
has been used \footnote{A reader with experience in harmonic superspace formulation of
${\cal N}=2$ $D=4$ supersymmetric  matter and gauge theories might notice the similarity of Eqs. (\ref{bDPhi=0}) and
(\ref{DchIP=0}) with basic equations of the hypermultiplet superfield $q^{+}$, which read $\; \bar{D}{}^+_{\dot{\alpha}} q^{+}  =0={D}{}^+_{\alpha} q^{+}  $ and
$D^{++}q^{+}=0$. In central basis of   ${\cal N}=2$  harmonic superspace $\bar{D}{}^+_{\dot{\alpha}}= u^{+i} \bar{D}{}_{\dot{\alpha} i}$ , ${D}{}^+_{\alpha} = u^{+}_{i} {D}{}^i_{\alpha}$ and $D^{++}= u^+_i\frac {\partial} {\partial u^-_i}$ where ${D}{}^i_{\alpha}=(\bar{D}{}_{\dot{\alpha} i})^*$ are standard  ${\cal N}=2$ fermionic covariant derivatives,
$i,j=1,2$ and  $
\epsilon^{ij} u^+_iu^-_j=1$. It is well--known  \cite{Galperin:1984av,Galperin:2001uw} that the first of these equations (Grassmann analyticity conditions) are  dynamical and the last is purely algebraic in this basis. However, after passing to an analytical basis the role of the equations interchange: the Grassmann analyticity conditions define a subclass of superfields, analytic superfields,   while $D^{++}q^{+}=0$ becomes dynamical equation for the analytic superfield.
 One might wonder whether similar interchange effect occurs in our formalism.
The answer is negative as far as  our D=10 on-shell superspace description is oriented on collecting inside  an analytic  superfield the on-shell degrees of freedom of the SYM: we cannot distinguish algebraic and dynamical equations in this framework.
Furthermore, as we will stress and discuss below, our internal harmonics are actually pure gauge variables.
}.

As we have already noticed, the spectrum of the  component fields described by the analytical superfield (\ref{Phi=SYM}) formally coincides with the fields of  ${\cal N}=4$ D=4 SYM.
However, these fields  depend on different set of bosonic variables: on 1+8+12=21
\begin{eqnarray}\label{vin-w-in10}
\{ \rho^\# \}= {\bb R}_+^1 \; , \qquad \{ v_{\alpha q}^{\; -} \} \;  = {\bb S}^{8}\; , \qquad \{ (w_q^{\; A}, \bar{w}_{Aq}) \} \; = \;  \frac {SO(8)}{SU(4)\otimes U(1)}\;  \qquad
\end{eqnarray}
instead of 3=4-1 non-pure gauge components of  $\; (\lambda,\bar{\lambda})={\bb C} {\bb P}{}^2$ in 4D case.

The indices $A,B=1,...,4$ of the fermionic coordinates and of some of the on-shell component fields of
10D SYM are transformed by $SU(4)$. However, in distinction to the rigid $SU(4)$ R-symmetry group of
${\cal N}=4$ D=4 SYM, in ten dimensional theory  $SU(4)$ is a gauge symmetry:
it is used as identification relation on the set of harmonic  variables  $w_q^{\; A}, \bar{w}_{Aq}$ (\ref{bww=1})
 making them generalized homogeneous  coordinates of the $\frac {SO(8)}{SU(4)\otimes U(1)}$
coset.

\subsection{On $(w, \bar{w})$- dependence of the analytic superfields. Complex spinor harmonics}
The meaning of the  dependence of analytic superfields (\ref{an=SF}) on the additional set of internal harmonic variables $(w, \bar{w})$ requires some clarification. As we have discussed in sec. 3, the 10D counterpart of the 4D helicity spinors $(\lambda,\bar{\lambda})$ is provided by  the  spinor harmonic variables $(v^+_{\dot{q}}, v^-_q)$. However, these are real while to define an analytic or chiral superfield, similar to the ones used in the on-shell superfield description of ${\cal N}=4$ SYM in D=4, we need to have a complex structure. The role of internal harmonics $(w, \bar{w})$ is to introduce such a complex structure without breaking explicitly  the  $Spin(8)$ gauge symmetry. This `little group' symmetry,  acting on the spinor harmonics $(v^+_{\dot{q}}, v^-_q)$ and  used to identify them with the homogeneous coordinates of the celestial sphere
${\bb S}^8= \frac{Spin(1,9)}{[Spin(6)\otimes Spin(2)\otimes SO(1,1)]\subset\!\!\!\! \times K_{8}}$ (\ref{v-v+=G-H}),
  acts also on $(w, \bar{w})$. Moreover, $(w, \bar{w})$ are pure gauge with respect to this $Spin(8)$ symmetry,  so that the only invariant content encoded in them is the above mentioned complex structure.

One can formally fix the gauge with respect to $SO(8)$ symmetry by setting $U_I=\delta _I^7+i\delta_I^8$. The residual symmetry of this gauge,  in which
$(w, \bar{w})$ are determined by (\ref{p+=wbw}), is $SU(4)\otimes U(1)$, and the $Spin(8)$ symmetry acting on the spinor harmonic reduces to this smaller subgroup. In this language, the analytic superfields depend on the spinor harmonics only, but these parametrize the coset
\begin{eqnarray}\label{G-H=U4}
\frac{Spin(1,9)}{[Spin(6)\otimes Spin(2)\otimes SO(1,1)]\subset\!\!\!\!\!\! \times K_{8}}=\frac{Spin(1,9)}{[SU(4)\otimes U(1)\otimes SO(1,1)]\subset\!\!\!\!\!\! \times K_{8}}
\end{eqnarray}  instead of (\ref{v-v+=G-H}).

This is tantamount to saying  that the analytic superfields  (\ref{an=SF}) depend on the set of {\it complex spinor harmonics}
composed of $(v^+_{\dot{q}}, v^-_q)$ and $(w, \bar{w})$ according to
\begin{eqnarray}\label{v-A:==}
 v_{\alpha A}^{-}:= v_{\alpha q}^{-} \bar{w}_{qA}\; , \qquad \bar{v}{}_{\alpha}^{-A}:= v_{\alpha {p}}^{-} {w}_{{p}}^{\; A} \; , \qquad v_{\alpha A}^{+}:= v_{\alpha \dot{p}}^{+} \bar{w}_{\dot{p}A}\; , \qquad  \bar{v}{}_{\alpha}^{+A}:= v_{\alpha \dot{p}}^{+} {w}_{\dot{p}}^{\; A} \; ,  \qquad \\
\label{v-A:==-1}   v_{ A}^{-\alpha}:= v_{\dot{q}}^{-\alpha} \bar{w}_{\dot{q}A} , \qquad \bar{v}{}^{-A\alpha }:= v_{\dot{q}}^{-\alpha} {w}_{\dot{q}}^{\; A} , \quad v_{A}^{+\alpha }:= v_{q}^{+\alpha } \bar{w}_{qA}, \qquad \bar{v}{}^{+A\alpha}:= v_{{q}}^{+\alpha} {w}_{{q}}^{\; A} . \qquad
\end{eqnarray}
After taking into account the constraints and identification relations, one concludes that these parametrize
the coset (\ref{G-H=U4}).
Resuming, (\ref{an=SF}) can be equivalently written in the form
\begin{eqnarray}\label{an=SFc}
 && \Phi=\Phi (x_L^= , \eta^-_A;  v_{\alpha A}^{\; -}, \bar{v}{}_{\alpha}^{-A} ) \qquad \Leftrightarrow \qquad \bar{D}{}^+_A\Phi=0\; .
\end{eqnarray}
We, however, find more convenient at this stage to think about dependence of analytic superfields on real spinor harmonics, parametrizing the celestial sphere ${\bb S}^8$ ($=\frac{Spin(1,9)}{[Spin(8)\otimes SO(1,1)]\subset\!\!\!\!\times {K}_8}$), and on the set of internal harmonic variables parametrizing the coset $\frac{Spin(8)}{SU(4)\otimes U(1)}$, in spite of  these latter are pure gauge in our case.

\bigskip

\subsubsection{Origin of internal harmonics}

To clarify the passage from (\ref{an=SF}) to (\ref{an=SFc}) and the auxiliary nature of internal harmonics, it is instructive to discuss how the analytic superfield $\Phi$ can be obtained in quantization of massless 10D superparticle. The reader not interested in this issue can pass directly to sec. 5.

More details on quantization in present notation can be found in \cite{Bandos:2017eof} (see also \cite{Bandos:2007mi} for 11D case).
Here we begin from stating that in a Lorentz-analytic coordinate basis\footnote{The ten bosonic and 16 fermionic coordinates of the Lorentz-analytical coordinate  basis of Lorentz harmonic 10D superspace are constructed from the standard superspace coordinates $x^a, \theta^\alpha$ and Lorentz harmonics as
$ x^=:=  x^au_a^=$, $x^\#:= x^au_a^\#$, $x^I:= x^au_a^I+{i} \theta^-_{{q}}\gamma^I_{q\dot{q}}\theta^+_{\dot{q}}$,    $\theta^-_q= \theta^\alpha  v_{\alpha q}^{\; -}
$ and $\theta^+_{\dot{q}} = \theta^\alpha  v_{\alpha \dot q}^{\; +}$ (see \cite{Bandos:2017eof} and refs. therein).}  the 10D (and 11D) massless superparticle has no fermionic first class constraints but only 8 (16) second class fermionic constraints $d^+_q$. They have the form $d^+_q = \pi^+_q +2i \theta^-_q \rho^\# \approx 0$, where $\pi^+_q$ is the momentum variable conjugate to fermionic coordinate function $\theta^-_q=\theta^-_q(\tau)$ and $\rho^\#=\rho^\#(\tau)$ is momentum conjugate to the bosonic coordinate function $x^==x^=(\tau)$ depending  on particle proper time variable $\tau$. These $d^+_q$ are  the classical mechanics counterparts of the fermionic covariant derivatives in (\ref{D+dq:=}). Their second class nature is reflected by the Poisson bracket (P.B.) relations
\begin{eqnarray}\label{d+d+=PB}
\{ d^+_q, d^+_p\}_{_{P.B.}} =-4i\rho^\# \delta_{qp}\; . \end{eqnarray}
The second class constraints  can be resolved by passing to  Dirac brackets (D.B.). Then the dynamical system has no fermionic constraints but the fermionic coordinate variable
 obey
\begin{eqnarray}\label{t-t-=DB} {} \{ \theta^-_q , \theta^-_p\}_{_{D.B.}} = - \frac {i}{4\rho^{\#}}\delta_{qp}\;.  \end{eqnarray}
After quantization (in the momentum representation with respect to $x^=$), the algebra of fermionic operators $\hat{\theta}^-_q$ is
\begin{eqnarray}\label{t-t-=Cl}{} \{ \hat{\theta}^-_q , \hat{\theta}^-_p\}  = \frac {1}{4\rho^{\#}}\delta_{qp}\; , \end{eqnarray}
and we have to find a representation of this Clifford-like algebra on the superparticle state vectors ('wavefunctions').

The appearance on this way of the constrained superfields  (\ref{D+Phi=gPsi}) and of the Clifford superfield formalism by Caron-Huot and O'Connel is discussed in \cite{Bandos:2017eof} (see also concluding sec. 9 for a brief discussion). To arrive at the analytic superfields formalism, we need to split 8 (16) Clifford-like variables
$\hat{\theta}^-_q $ on the set of 4 complex fermionic coordinates and 4  momenta conjugate to these. Such an 'oscillator' (creation and annihilation operator) representation of Clifford algebra is well known, but it requires to introduce a complex structure, which   breaks the $SO(8)$  symmetry of the 8-dimensional Clifford algebra down to
$U(4)$.

A generic complex structure can be described by (complex linear combinations of the) columns of an $SO(8)$ valued matrix, $w_{q}^A$ and $\bar{w}_{qA}$ (\ref{wqA=})  obeying
(\ref{bww=1}). They can be used to split $\hat{\theta}^-_q $ on the counterparts of creation and annihilation operators,
\begin{eqnarray}\label{heta=}
\hat{\eta}{}^-_A= \hat{\theta}^-_q \bar{w}_{qA}\; , \qquad
\hat{\bar{\eta}}{}^{-A}= \hat{\theta}^-_q w_{q}^A\; , \qquad \{\hat{\eta}{}^-_A\, , \,
\hat{\bar{\eta}}{}^{-B} \}= \frac {1}{4\rho^{\#}} \delta_A{}^B\; .
\end{eqnarray}
Then, we can quantize superparticle in $\eta$-representation, in which $\hat{\eta}{}^-_A={\eta}{}^-_A$ and
$\hat{\bar{\eta}}{}^{-B}  \propto \frac \partial {\partial {\eta}{}^-B}$, and  the wavefunction depends on ${\eta}{}^-_A$.
Such a wavefunction do not depend on the complex  conjugate fermionic coordinate ${\bar{\eta}}{}^{-B}$ and, hence, is a counterpart of chiral superfield,  an analytical superfield.

Actually in such a way,  we arrive at the analytical superfield  (\ref{Phi=SYM}), which depend, besides the above mentioned
${\eta}{}^-_A$ and bosonic $\rho^\#$ (or its conjugate coordinate $x^=$) also on the Lorentz harmonic variables $v_{\alpha q}^{\; -}$, parametrizing the celestial sphere of the 10D observer (\ref{v-qi=G-H}) and the above variables $w_{q}^A, \bar{w}_{qA}$.

Such a description corresponds to introducing
$w_{q}^A(\tau), \bar{w}_{qA}(\tau)$ as additional variables of particle mechanics, so that the splitting of
real $\theta^-_q(\tau)$ on ${\eta}{}^-_A(\tau)$ and
${\bar{\eta}}{}^{-A}(\tau)$, as in (\ref{heta=}), can be performed already at the level of superparticle action. This is also in correspondence with the general ideology of harmonic superfield approach \cite{Galperin:1984av,Galperin:1984bu,Galperin:2001uw} (see \cite{Akulov:1988tm,Buchbinder:2008pn,Buchbinder:2008ub} for quantization of superparticle in the standard ${\cal N}\geq 2$ D=4 harmonic superspace). However, in our case there is a peculiarity related to the fact that such internal harmonic coordinate functions in the superparticle action will be pure gauge with respect to $SO(8)$ gauge symmetry. Furthermore, this is the same $SO(8)$ gauge symmetry  which was used as an identification relation on the set of Lorentz harmonic variables and allowed to treat them as  homogeneous coordinates of the celestial sphere (\ref{v-qi=G-H}). Hence, on one hand, we can fix the SO(8) gauge by setting $w_{q}^A, \bar{w}_{qA}$ to some constant values thus breaking  $SO(8)$ gauge symmetry down to $SU(4)\times U(1)$. But on the other hand, after that we cannot use $SO(8)$ gauge symmetry as an identification relation on the Lorentz harmonic variables. Then these latter cannot be considered as
parametrizing the celestial sphere, but rather are homogeneous coordinates of a bigger coset (\ref{G-H=U4}). The number of additional (with respect to celestial sphere) dimensions of this coset coincides with the dimension of the
 coset $SO(8)/[SU(4)\times U(1)]$.

Such a gauge fixing leads us to the wavefunction (\ref{an=SFc}) seemingly dependent on smaller number of harmonic variables. However, as we have just explained, the number of degrees of freedom in the Lorentz harmonic variables serving as an argument of the wavefunction (\ref{an=SFc}) is the same as the sum of the number of degrees of freedom in Lorentz harmonic variables (\ref{v-qi=G-H}) and internal harmonics  (\ref{vin-w-in10}) the superfield (\ref{an=SF}) depend on.

As we have already said, we prefer the second description, in which $SO(8)$ gauge symmetry is used as an identification relation on the set of Lorentz harmonics,  which then parametrize the celestial sphere ${\bb S}^8$, and the internal harmonics describe the degrees of freedom of the coset
$SO(8)/[SU(4)\times U(1)]$.

\subsubsection{Comment on harmonic integration}

If we were constructing the harmonic superspace actions for field theories in terms of our analytic superfields, then at some stage we would need to define and to use the  integration over the internal harmonic variables  $w_{q}^A, \bar{w}_{qA}$. In particular, the Lagrangians of such actions should be defined as an integral over $SO(8)/[SU(4)\times U(1)]$ coset.
Such a problem, although interesting, goes beyond the scope of this paper where we use only on-shell superfields and their  multiparticle generalizations, tree superamplitudes.

When working with the on-shell superfield description of free supermultiplets and tree amplitudes, we can always treat the analytic superfields/superamplitudes as encoding the constrained superfields/superemplitudes  and their components, particle amplitudes, which are independent of internal harmonics.
The internal harmonic variables  enter such encoding relations in a linear manner: see (\ref{Phi=WU}) and its superamplitude generalization (\ref{cA=etcAU}).
However, the further development of the formalism, and particularly its generalization to loop amplitudes, might require to introduce and to use the integration over $SO(8)/[SU(4)\times U(1)]$ coset parametrized by internal harmonics.

\section{Spinor helicity formalism and on-shell superfield descriptions of the linearized 11D SUGRA}

\subsection{Spinor helicity formalism in D=11}

In this section we develop  $D=11$ spinor helicity formalism \cite{Bandos:2016tsm} on the basis of the spinor moving frame approach to 11D superparticle  \cite{Bandos:2007mi,Bandos:2007wm}. This uses the Lorentz harmonics which can be considered as square roots of the vector frame variables, the 11D version of  vector harmonics introduced in \cite{Sokatchev:1985tc,Sokatchev:1987nk}. The description of this latter basically coincide with that given in sec.  \ref{vectorFrame} for their 10D cousin, but with setting D=11, $a,b,c=0,1...,10$, and $I,J,K=1,...,9$  in the appropriate places.

This is to say the vector frame is described by Eqs. (\ref{uaib=}) or (\ref{harm=++--I}) with $D=11$, and is ''attached'' to a light--like 11--momentum $k_{ai}$ by Eq. (\ref{kia=ru=}). The Lorentz harmonic variables forming the vector frame matrix are constrained by   (\ref{uu=0})--(\ref{I=UU}) and defined up to the transformations (\ref{SO1-1=}), (\ref{SOD-2=}), (\ref{KD-2=}) which allow to identify them as homogeneous coordinates of the coset
 (\ref{++--I=SD-2}) with $D=11$. This last equation is equivalent 11D version of (\ref{u--=SD-2}) where in the {\it l.h.s.} the light-like vector
 $u_a^=$ is defined modulo its scaling transformations (as resulting from acting on it by $SO(1,1)$ symmetry of the set of  vector harmonics  (\ref{++--I=SD-2})).

\subsection{Spinor frame and spinor helicity formalism  in  D=11 }

\label{sec=11Dharm}

11D spinor harmonics are defined as rectangular blocks of $Spin(1,10)$-valued spinor frame matrix
\begin{eqnarray}\label{harmV=11D}
V_{\alpha}^{(\beta)}= \left(\begin{matrix} v_{\alpha {q}}^{\; +} , & v_{\alpha q}^{\; -}
  \end{matrix}\right) \in Spin(1,10)\; . \qquad
\end{eqnarray}
This is defined as a kind of square root of the  vector frame matrix (\ref{++--I=SD-2}) by  constraints
\begin{eqnarray}\label{VGVt=G11} V\Gamma_b V^T =  u_b^{(a)} {\Gamma}_{(a)}\, , \qquad V^T \tilde{\Gamma}^{(a)}  V = \tilde{\Gamma}^{b} u_b^{(a)}\;
 \, , \qquad \\
 \label{VCVt=C11}
 VCV^T=C \; . \qquad
\end{eqnarray}
Here $C$ is the 11D charge conjugation matrix, which is imaginary and antisymmetric
$C_{\gamma\beta}= - C_{\beta\gamma}=- (C_{\gamma\beta})^*$. ${\Gamma}_{b}$ and $\tilde{\Gamma}^{b}$ in (\ref{VGVt=G11}) are  real symmetric  $32\times 32$ matrices $\Gamma^a_{\alpha\beta}=\Gamma^a_{\beta\alpha}= \Gamma^a_{\alpha}{}^{\gamma}C_{\gamma\beta}$  and $\tilde{\Gamma}^{a \; \alpha\beta}= \tilde{\Gamma}^{a \; \alpha\beta}= C^{\alpha\gamma} {\Gamma}^{a \;\beta}_\gamma$, obeying (\ref{GtG=I}). They  are constructed as  products of 11D Dirac matrices $\Gamma^a_{\alpha}{}^{\gamma}=- (\Gamma^a_{\alpha}{}^{\gamma})^*$ obeying the Clifford algebra, $\Gamma^a\Gamma^b+\Gamma^b\Gamma^a = 2\eta^{ab}{\bb I}_{32\times 32}$, and of the above described charge conjugation matrix.

In (\ref{harmV=11D}) $\alpha, \beta,\gamma$ are indices of 32 dimensional Majorana spinor representation of $SO(1,10)$ and $q,p$ are spinor indices of $SO(9)$,
\begin{eqnarray}\label{q=D11}
 D=11: \qquad \alpha,\beta,\gamma =1,...,32\;  \quad and \qquad q,p =1,...,16\;  \quad  \; .  \quad
\end{eqnarray}
Notice that, in distinction with D=10, both spinor harmonics in 11D spinor frame matrix (\ref{harmV=11D}) carry $Spin(9)$ indices of the same type. Furthermore, the existence of charge conjugation matrix allows to construct the elements of inverse spinor frame matrix obeying
\begin{eqnarray}\label{v-qv+p=11}
&
v^{-\alpha}_{{q}}   v_{\alpha {p}}^{\; +}=\delta_{{q}{p}}
 \; ,  \qquad & v^{-\alpha}_{{q}}   v_{\alpha q}^{\; -}=0 \;  , \qquad
 \nonumber  \\
 & v^{+\alpha}_{{q}}  v_{\alpha {p}}^{\; +}=0
 \;  , \qquad & v^{+\alpha}_{{q}} v_{\alpha {p}}^{\; -} = \delta_{qp} \;   \qquad
\end{eqnarray}
 in terms of the same spinor harmonics:  \begin{eqnarray}
\label{V-1=CV-A} D=11\; : \qquad  v_{q}^{+ \alpha}  =  i C^{\alpha\beta}v_{\beta q}^{\; +  }   \, ,
\qquad  v_{q}^{- \alpha}  =  -i C^{\alpha\beta}v_{\beta q}^{\; -  }   \, . \qquad
 \end{eqnarray}

The constraints (\ref{VGVt=G11}) can be split on the following
set of $SO(1,1)\otimes SO(9)$ covariant relations
\begin{eqnarray}\label{u==v-v-=11D}
 & u_a^= \Gamma^a_{\alpha\beta}= 2v_{\alpha q}{}^- v_{\beta q}{}^-  \; , & \qquad
 u_a^= \delta_{{q}{p}} = v^-_{{q}} \tilde{\Gamma}_{a}v^-_{{p}}  \qquad  \\
\label{v+v+=u++}
& v_{{q}}^+ \tilde{\Gamma}_{ {a}} v_{{p}}^+ = \; u_{ {a}}^{\# } \delta_{{q}{p}}\; , & \qquad 2 v_{{\alpha}{q}}{}^{+}v_{{\beta}{q}}{}^{+}= {\Gamma}^{ {a}}_{ {\alpha} {\beta}} u_{ {a}}^{\# }\; , \qquad \\
 \label{uIs=v+v-=11D}
& v_{{q}}^- \tilde {\Gamma}_{ {a}} v_{{p}}^+=u_{ {a}}^{I} \gamma^I_{q{p}}\; , &\qquad
 2 v_{( {\alpha}|{q} }{}^- \gamma^I_{q{q}}v_{|{\beta}){q}}{}^{+}= {\Gamma}^{a}_{\alpha\beta} u_{ {a}}^{I}\; , \quad  \end{eqnarray}
where  $\gamma^I_{qp}=\gamma^I_{pq}$ are nine dimensional 16$\times$16 Dirac matrices, $I=1,...,9$. In the Majorana spinor representation of SO(9) the charge conjugation matrix is symmetric and we identify it with $\delta_{qp}$.

The $K_{D-2}$ transformations of Lorentz harmonics in D=11 are given by   \begin{eqnarray}\label{K9=v}
K_{9}\; :\qquad v_{\alpha {q} }^{\; +} \mapsto v_{\alpha {q} }^{\; +} + \frac 1 2 K^{\# I}v_{\alpha p }^{\; -}\gamma^I_{p{q}}\; ,  \qquad v_{\alpha {q} }^{\; -} \mapsto v_{\alpha {q} }^{\; -}\; .
\end{eqnarray}
When $[SO(1,1)\otimes Spin(9)]\subset\!\!\!\!\!\!\times  K_{9}\;$ can be used as an identification relation, {\it i.e.} in the model which possesses gauge symmetry under these transformations, the spinor harmonics can be considered as the coordinates of the coset of the Lorentz group isomorphic to ${\bb S}^{9}$ sphere,
\begin{eqnarray}\label{v-v+=G-H11}
\{ v_{\alpha \dot{q} }^{\; +}, v_{\alpha q }^{\; -} \}  \; \in  \; \frac {Spin(1,10)} {[SO(1,1)\otimes Spin(9)]\subset\!\!\!\!\!\!\times  K_{9}} = \; {\bb S}^{9}\; .   \qquad
\end{eqnarray}

\bigskip

The fact that the  vector frame is adapted to the light-like 11--momentum $k_a$ by Eq. (\ref{kia=ru=})
imply  \begin{eqnarray}\label{k=pv-v-=11}
k_a
\Gamma^a_{\alpha\beta}= 2\rho^{\#} v_{\alpha q}^{\; - } v_{\beta q}^{\; - } \; , \qquad
 \rho^{\#} v^-_{{q}} \tilde{\Gamma}_{a}v^-_{{p}}= k_{a} \delta_{{q}{p}}\;  , \qquad  \end{eqnarray}
 which can be equivalently written as
 \begin{eqnarray}\label{k=pv-v-1=11}  && k_{a} \tilde{\Gamma}{}^{a\, \alpha\beta}= 2\rho^{\#}v^{-\alpha}_{ q}   v^{-\beta}_{ q} \; , \qquad \rho^{\#}  v^-_{{q}} {\Gamma}_{a}v^-_{{p}}=  k_{a}  \delta_{{q}{p}}
 \; . \qquad
\end{eqnarray}
These relations imply that the spinor harmonics  $ v_{\alpha q}^{\; - } $ obey the Dirac equation
\begin{eqnarray}\label{Dirac=11}
 k_{a} \tilde{\Gamma}{}^{a\, \alpha\beta}v_{\beta  q}^{\; -} =0
\;  \qquad \Leftrightarrow \qquad k_a\Gamma^a_{\alpha\beta}  v^{-\beta}_{ q}=0\;  \qquad
\end{eqnarray}
and hence  define the helicity spinor
\begin{eqnarray}\label{shel=l=11D}
\lambda_{\alpha q}= \sqrt{\rho^{\#}} v_{\alpha  q}^{\; -}  \; . \qquad
\end{eqnarray}
The polarization spinor in D=11 can be obtained from helicity spinor with the use  of charge conjugation matrix,
\begin{eqnarray}\label{spol=l=11D} D=11\; : \qquad \lambda^{\alpha}_{{q}}= \sqrt{\rho^{\#}} v^{-\alpha}_{{q}}= i C^{\alpha\beta} \lambda_{\beta q}  \; . \qquad
\end{eqnarray}

\subsection{Linearized D=11 SUGRA  in the Lorentz  harmonic spinor helicity formalism}

The linearized on-shell field strength of 3-form gauge field of 11D SUGRA (called 'formon' in \cite{Deser:2000xz}) can be  expressed by
\begin{eqnarray}\label{FabcdIJK=}
 F_{abcd}=k_{[a}u_{b}{}^I u_{c}{}^J u_{d]}{}^K\, A_{IJK} \;  \qquad
\end{eqnarray}
in terms of light-like momentum (\ref{k=pu--}), spacelike vectors $u_{b}{}^I $ of the frame
adapted to the momentum by  (\ref{k=pu--}), and an
antisymmetric SO(9) tensor $A_{IJK}=A_{[IJK]}$ (in {\bf 84} of SO(9)). The linearized on-shell expression for the Riemann tensor reads
\begin{eqnarray}\label{Rabcd=}
 R_{ab}{}^{cd}= k_{[a}u_{b]}{}^I k^{[c}u^{d]J} h_{IJ} \; , \qquad
\end{eqnarray}
where the second rank SO(9) tensor  $ h_{IJ}$ is symmetric and traceless (in {\bf 44} of SO(9))
\begin{eqnarray}\label{hIJ=hJI}
 h_{IJ}=h_{JI} \; , \qquad h_{II}=0  \; . \qquad
\end{eqnarray}
We express these properties by writing $h_{IJ}=h_{((IJ))}$.
Finally, the gravitino field strength solving the Rarita-Schwinger equation is  expressed
in terms of  $\gamma$--traceless SO(9) vector-spinor  $\Psi_{I{q}}$ ({\bf 128} of SO(9)) by
\begin{eqnarray}\label{Psiaal=}
{\cal T}_{ab}{}^\alpha = k_{[a} u_{b]}^I  v^{-\alpha}_q \Psi_{Iq}= \rho^\# u^=_{[a} u_{b]}^I  v^{-\alpha}_q \Psi_{Iq}
\; , \qquad \gamma^I_{qp}  \Psi_{Ip}=0\; . \qquad
\end{eqnarray}

The set of on-shell fields $h_{IJ}$, $A_{IJK}$, $ \Psi_{Ip}$ can be used to describe the supergravity multiplet in light-cone gauge \cite{Green:1999by}. In our spinor helicity/spinor frame description, which can be deduced  from the on-shell superfield formalism of \cite{Galperin:1992pz}, these fields depend on the density $\rho^{\#}$ and spinor harmonics $v_{\alpha q}^{\; -}$ (homogeneous coordinates of ${\bb S}^9$ (\ref{v-qi=G-H})) related to the momentum by  (\ref{k=pv-v-}),
\begin{eqnarray}\label{A=A-rv}
A_{IJK}=A_{[IJK]}(\rho^{\#}, v_{q}^{ -})\; , \qquad  h_{IJ}=h_{((IJ))}(\rho^{\#}, v_{q}^{ -})\; , \qquad
\Psi_{Iq}=\Psi_{Iq}(\rho^{\#}, v_{q}^{ -})\; .  \qquad
\end{eqnarray}
In the next section we will use the (superfield generalization) of the Fourier images of the above fields defined on  ${\bb R}\otimes  {\bb S}^9$ space,
\begin{eqnarray}\label{A=A-xv}
A_{IJK}=A_{[IJK]}(x^{=}, v_{q}^{ -})\; , \qquad  h_{IJ}=h_{((IJ))}(x^{=}, v_{q}^{ -})\; , \qquad
\Psi_{Iq}=\Psi_{Iq}(x^{=}, v_{q}^{ -})\; .  \qquad
\end{eqnarray}

\subsection{Constrained on-shell superfield description of 11D SUGRA}

A constrained on-shell superfield formalism for linearized 11D SUGRA was proposed in
\cite{Galperin:1992pz} and was generalized for the case of  superamplitudes in
\cite{Bandos:2016tsm} (see \cite{Bandos:2017eof} for details).

The constrained on-shell superfields are functions  on the real on-shell superspace
\begin{eqnarray}\label{On-shellSSP=11D}
\Sigma^{(10|16)} : \qquad &&  \{ (x^= , \theta^-_q, v_{\alpha q}^{\; -})\} \; ,  \qquad \{ v_{\alpha q}^{\; -}\} = {\bb S}^{9}\; , \qquad \\ \nonumber &&
q=1,..., 16\; , \qquad \alpha =1,..., 32\; , \qquad
\qquad \nonumber
\end{eqnarray}
where the 11D supersymmetry acts  as follows ({\it cf.} (\ref{susy=S}))
\begin{eqnarray}\label{susy=S=11D}
\delta_\epsilon x^== 2i \theta^-_q \; \epsilon^\alpha v_{\alpha q}^{\; -} \; , \qquad
\delta_\epsilon \theta^-_q =  \epsilon^\alpha v_{\alpha q}^{\; -}
\; , \qquad \delta_\epsilon v_{\alpha q}^{\; -}=0\; .  \qquad
\qquad
\end{eqnarray}
The fermionic covariant derivatives have the form
\begin{eqnarray}\label{D+dq:=11D}
D^+_{{q}}={\partial}^+_{{q}} + 2i  \theta^-_{{q}}\partial_{=} \; , \qquad \partial_{=}
:={\partial\over \partial x^=}\; , \qquad \partial^+_{{q}}
:={\partial\over \partial \theta^-_{{q}}}\; , \qquad q=1,..., 16\; .
\end{eqnarray}
They obey  $d=1$  $\frak{N}=16$ supersymmetry algebra ({\it cf.} (\ref{D+qD+p=I}))
\begin{eqnarray}\label{D+qD+p=I=11}
\{ D^+_{{q}},  D^+_{{p}}\} = 4i \delta_{qp}\partial_=\,  .
\end{eqnarray}

\label{sec=OnSh11D}
The linearized 11D supergravity was described in \cite{Galperin:1992pz}  by a bosonic  antisymmetric tensor superfield
\begin{eqnarray}\label{AIJK:=}
A^{IJK}=
A^{[IJK]}(x^=, \theta^-_{{q}}, v_{\alpha {q}}^{\; -})
 \end{eqnarray}
 which obeys the superfield equation
\begin{eqnarray}\label{D+Phi=gPsi}
 D^+_{{q}}A^{IJK}  = 3i\gamma^{[IJ}_{qp}  \Psi^{K]}_{p}\, , \qquad \gamma^I_{qp}\Psi^I_{p}=0\; , \qquad
q,p= 1,...,16\; , \quad  I =1,..,8,9\, . \qquad
\end{eqnarray}
The consistency of Eq. (\ref{D+Phi=gPsi}) requires that gamma-traceless fermionic $\Psi_{Iq}$ obeys
\begin{eqnarray}\label{D+Psi=11D}
 D^+_{{q}}\Psi^{I}_{p}  = \frac 1 {18} \left(\gamma^{IJKL}_{qp}+ 6 \delta ^{I[J}\gamma^{KL]}_{qp}  \right) \partial_{=}A^{JKL}   + 2  \partial_{=}H_{IJ}\gamma^{J}_{qp}  \;  \qquad \end{eqnarray}
with symmetric traceless tensor superfield $H_{IJ}$ satisfying
\begin{eqnarray} \label{D+h=11D} D^+_{{q}}H_{IJ} = i \gamma^{(I}_{qp} \Psi^{J)}_{p} \; , \qquad H_{IJ}= H_{JI}\; , \qquad H_{II}=0 \; . \qquad
\end{eqnarray}
Actually any of the three equations, (\ref{D+Phi=gPsi}), (\ref{D+Psi=11D}) or (\ref{D+h=11D}), can be chosen as the fundamental; then other two will be reproduced as its consistency conditions. All the on-shell degrees of freedom of the 11D SUGRA can be extracted from any of the three constrained superfields $H_{IJ}$, $A_{IJK}$ or $\Psi_{Iq}$.

\subsection{Analytic  on-shell superfields of  11D SUGRA}

Similar to the case of 10D SYM, the 11D SUGRA can be also described by one complex superfield in  ${\cal N}=8$ extended analytic superspace. This is to say, the system of superfield equations (\ref{D+Phi=gPsi}), (\ref{D+Psi=11D}), (\ref{D+h=11D}) can be solved in terms of one analytic
(chiral-like) superfield carrying charge $2$ under the $U(1)$ subgroup of ${SO(9)} \subset  Spin(1,10)$ acting naturally on the  $\frac {Spin(9)}{Spin(7)\otimes U(1)}$ coset.

\subsubsection{$\frac{SO(9)}{SO(7)\times SO(2)}$ harmonic variables}

Following the line described in sec. \ref{sec=U-harm} for the case of D=10 SYM, let us  introduce internal vector harmonics providing a set of constrained homogeneous coordinates for the $\frac{SO(9)}{SO(7)\times SO(2)}$ coset ({\it cf.} (\ref{UinSO8}))
\begin{eqnarray}\label{UinSO9}
&  U_I^{(J)}= \left(U_I{}^{\check{J}}, U_I{}^{(8)}, U_I{}^{(9)}\right)= \left(U_I{}^{\check{J}}, \frac 1 2 \left( U_I+ \bar{U}_I\right), \frac 1 {2i} \left( U_I- \bar{U}_I \right)\right) \; \in \; SO(9)
 \; . \qquad
\end{eqnarray}
The condition (\ref{UinSO9}) is equivalent to the set of relations which are described by (\ref{U2=0}) and (\ref{UU6=0}), but now with $I,J=1,...,9$ and $\check{I}, \check{J}=1,...,7$. Eqs. (\ref{U2=0}), in their turn, imply that the symmetric 16$\times$16 matrices
\begin{eqnarray}\label{U/=Ug}
U\!\!\!\!/{}_{qp} := U_I \gamma^I_{qp}\; , \qquad \bar{U}\!\!\!\!/{}_{qp} := \bar{U}{}_I\gamma^I_{qp}\;  \qquad
\end{eqnarray}
are nilpotent
\begin{eqnarray}\label{U/U/=0}
U\!\!\!\!/{}U\!\!\!\!/{}=0 \; , \qquad  \bar{U}\!\!\!\!/{}\bar{U}\!\!\!\!/{}=0 \; , \qquad
\end{eqnarray}
and their anticommutator is proportional to unity matrix
\begin{eqnarray}\label{bU/bU/=0}
U\!\!\!\!/{}\bar{U}\!\!\!\!/{}+\bar{U}\!\!\!\!/{}{U}\!\!\!\!/{}=4\; .
\end{eqnarray}
Hence, $U\!\!\!\!/{}\bar{U}\!\!\!\!/{}/4$ and its complex conjugate $\bar{U}\!\!\!\!/{}{U}\!\!\!\!/{}/4$ are  orthogonal projectors and thus can be factorized
\begin{eqnarray}\label{P+=11}
{\cal P}^+_{qp} = \frac 1 4 \bar{U}\!\!\!\!/{}{U}\!\!\!\!/{}= w_q{}^A\bar{w}_{pA}\; , \qquad
{\cal P}^-_{qp} = \frac 1 4 {U}\!\!\!\!/{}\bar{U}\!\!\!\!/{}= \bar{w}_{pA}w_q{}^A\; \; \qquad
\end{eqnarray}
in terms of complex 16$\times$8 matrices $w_q{}^A=(\bar{w}_{pA})^*$, which obey Eqs. (\ref{wbw+cc=1})  and (\ref{bww=1}).

Let us introduce the Spin(9) valued matrix $w_q^{(p)}$ providing a bridge between
spinor representations of SO(9) and SO(7), and also a kind of  'square root' of  $U_I^{(J)}$ of (\ref{UinSO9}).
\begin{eqnarray}\label{w=inSO9}
w_q^{(p)} \; \in\; Spin(9)\; , \qquad
 U_I^{(J)} \gamma^I_{qp} = w_q^{(q')} \gamma^{(J)}_{(q')(p')} w_p^{(p')} \; . \qquad
\end{eqnarray}
Then we can calculate the projectors (\ref{P+=11}) in its term and find
\begin{eqnarray}\label{wbw=wggw}
2w_q{}^A\bar{w}_{pA}=  w_q^{(q')} (I + i \gamma^{8}\gamma^{9})_{(q')(p')} w_p^{(p')} \; ,  \qquad
2\bar{w}_{qA}w_p{}^A=  w_q^{(q')} (I - i \gamma^{8}\gamma^{9})_{(q')(p')} w_p^{(p')} \; . \qquad
\end{eqnarray}
These equations  make manifest that complex 16$\times$ 8 matrices  $w_q{}^A$ and  $\bar{w}_{qA}$ are combinations of the columns of a real 16$\times$16 $Spin(9)$ valued matrix $ w_q^{(p)}$.
Thus, the space parametrized by $w_q{}^A$ and  $\bar{w}_{pA}$ is $Spin(9)$ group manifold. Now,  if we assume the $Spin(7)\otimes Spin(2)$ gauge symmetry and use it as an identification relation in this space, we can treat $w_q{}^A$ and  $\bar{w}_{pA}$  as homogeneous coordinates of the coset $\frac {Spin(9)}{Spin(7)\otimes Spin(2)}$,
 \begin{eqnarray}\label{wbw=G-H}
\{ w_q{}^A, \bar{w}_{qA}\} \; = \frac {Spin(9)}{Spin(7)\otimes U(1)}  \; , \qquad
\end{eqnarray}
and call them  $\frac {Spin(9)}{Spin(7)\otimes Spin(2)}$  harmonic variables  \cite{Galperin:1984av,Galperin:1984bu,Galperin:2001uw}.

Let us stress that the conditions (\ref{w=inSO9}) are stronger than the ones imposed by (\ref{bww=1}) with $q,p=1,...,16$. These latter would imply $w_q^{(p)}\; \in \; SO(16)$ only, while
 (\ref{w=inSO9}) results in  $w_q^{(p)}\; \in \; Spin(9)\; \subset SO(16)$.

Another useful observation is that  the second equation in (\ref{w=inSO9}) with $J=8,9$ can be written in the form
\begin{eqnarray}\label{U/=wcUw}
U\!\!\!\!/{}_{qp} = 2\bar{w}_{qA}{{\cal U}}{}^{AB} \bar{w}_{pB}\; , \qquad
{}
\bar{U}\!\!\!\!/{}_{qp}= 2w_q^{\;A}\bar{{\cal U}}{}_{AB} w_p^{\; B}\; ,
\end{eqnarray}
where the complex symmetric matrices $ {\cal U}_{AB}$ and $ \bar{{\cal U}}{}^{AB} =({\cal U}_{AB})^*$ obey
\begin{eqnarray}\label{cbUcU=I}
\bar{{\cal U}}{}_{AC} {{\cal U}}{}^{CB}=\delta_A{}^B  \; . \qquad
\end{eqnarray}
These matrices can be identified with the charge conjugation matrix of SO(7), which is symmetric and can be chosen to be the unity matrix. Then $ {\cal U}_{AB}=  \delta_{AB}= {\cal U}^{AB}$ and (\ref{U/=wcUw}) simplifies to
\begin{eqnarray}\label{U/=ww}
U\!\!\!\!/{}_{qp} = 2\bar{w}_{qA}\bar{w}_{pA}\; , \qquad
{}
\bar{U}\!\!\!\!/{}_{qp}= 2w_q^{\;A} w_p^{\; A}\; ,
\end{eqnarray}
which gives more reasons to state that   $\bar{w}_{qA}$ is a square root of the complex nilpotent matrix $U\!\!\!\!/{}_{qp}$.

The covariant harmonic derivatives preserving all the constraints on the
$\frac{Spin(9)}{Spin(7)\otimes U(1)}$ harmonic variables, Eqs. (\ref{wbw+cc=1}), (\ref{bww=1}), (\ref{w=inSO9}) and  (\ref{U/=ww}), have the form
\begin{eqnarray}
\label{DchJ=11D}
{\bb D}^{\check{J}} &=& \frac 1 2 U_I\frac {\partial} {\partial U_I^{\check{J}}}-  U_I^{\check{J}} \frac  {\partial} {\partial \bar{U}_I}  + \frac i 2 \tilde{\sigma}{}^{\check{J}AB} \bar{w}_{qB} \frac \partial {\partial w_q^A}\; , \qquad
\\
\label{bDchJ=11D}
\overline{{\bb D}}{}^{\check{J}}&=&  \frac 1 2 \bar{U}_I \frac {\partial} {\partial U_I^{\check{J}}}-  U_I^{\check{J}} \frac \partial {\partial {U}_I} + \frac i 2 \sigma{}^{\check{J}}_{AB} w_q^B \frac \partial {\partial \bar{w}_{qA}}  \; , \qquad
\\
\label{D0:=11D}
{\bb D}^{(0)} &=& U_I\frac {\partial} {\partial U_I}-  \bar{U}_I\frac  {\partial} {\partial \bar{U}_I}  + \frac 1 2\left(\bar{w}_{qA} \frac \partial {\partial \bar{w}_{qA}}- w_q^A \frac \partial {\partial w_q^A}\right) \; , \qquad
\\
\label{DchIJ=11D}
{\bb D}^{\check{I}\check{J}} &=& \frac 1 2 \left(  U_K^{\check{I}}\frac {\partial} {\partial U_K^{\check{J}}}-  U_K^{\check{J}} \frac  {\partial} {\partial U_K^{\check{I}} }\right)  + \frac i 2 {\sigma}{}^{\check{I}\check{J}}{}_B{}^A \left(  w_q^B \frac \partial {\partial w_q^A}-  \bar{w}_{qA }\frac \partial {\partial \bar{w}_{qB}} \right)
\; , \qquad
\end{eqnarray}
where $ \sigma{}^{\check{J}}_{AB}=\tilde{\sigma}{}^{\check{J}AB}$ are $SO(7)$ Clebsch-Gordan coefficients and
${\sigma}{}^{\check{I}\check{J}}= \sigma{}^{[\check{I}}\tilde{\sigma}{}^{\check{J}]}$.

As far as we have chosen 7d charge conjugation matrix to be equal to unity matrix, the contraction of two spinor subindices is allowed (see (\ref{U/=ww})), and we can write all the SO(7) spinor indices as subindices. However we find convenient to keep  some part of these as superindices as an indication of the origin of variables which carry them as well as with the aim to keep as manifest as possible the similarity of our 11D SUGRA formalism to the 10D SYM case.

 \subsubsection{Analytic on-shell superfields from constrained on-shell superfields}

Using the above described harmonic variables it is not difficult to  find a projection of the symmetric traceless tensor superfield $H_{IJ}$ which, as a result of (\ref{D+h=11D}), obeys an analyticity equation. Indeed, let us define a complex superfield
\begin{eqnarray}\label{Phi=HUU}
\Phi= H^{IJ}U_IU_J\; .
\end{eqnarray}
Multiplying (\ref{D+h=11D}) on $U_IU_J$ we find that this superfield satisfies
\begin{eqnarray}\label{D+qP=UUPsi}
D^+_q \Phi = i U\!\!\!\!/{}_{qp} \, \Psi^J_pU^J\; , \qquad
\end{eqnarray}
which, in the light of (\ref{U/U/=0}), implies  $(\bar{U}\!\!\!\!/{}{U}\!\!\!\!/)_{pq}D^+_q \Phi =0$. Using
the factorization of the projector (\ref{P+=11}) and Eqs.  (\ref{bww=1}) we find that this is equivalent to the  analyticity  condition
\begin{eqnarray}\label{bDAP=0}
\bar{D}_A^+  \Phi=0 \; , \qquad \bar{D}_A^+ = \bar{w}_{qA}D_q^+\; . \qquad
\end{eqnarray}
Besides (\ref{bDAP=0}), the analytic superfield (\ref{Phi=HUU}) obeys
\begin{eqnarray}
\label{DchIP=0=11D}
{\bb D}^{\check{J}} \Phi=0 \; , \qquad \\ \label{DchIJP=11D}
{\bb D}^{\check{I}\check{J}} \Phi=0 \; , \qquad \\ \label{D0P=P=11D}
{\bb D}^{(0)} \Phi=2\Phi \; . \qquad
\end{eqnarray}

As in the case of 10D SYM, using the harmonic derivative (\ref{bDchJ=11D}) and the  remaining parts of (\ref{D+qP=UUPsi}) and (\ref{D+h=11D}) we can obtain the expression for all other components of
the constrained superfield $H_{IJ}$
 in terms of $\Phi$ and its complex conjugate $\bar{\Phi}$.

Passing to the analytic coordinate basis of the on-shell superspace (\ref{On-shellSSP=11D}), which is described by  $D=11$ ${\cal N}=8$ version of (\ref{anHOn-shellSSP}), one can check that a superfield $\Phi$ obeying
Eq. (\ref{bDAP=0})  depends on $\eta_A$ but not on its c.c. $\bar{\eta}{}^A=
(\eta_A)^*$. This is reflected by the name  analytic  superfield which we have attributed to such $\Phi$.

Notice that the complex fermionic variable  $\eta_A $ is almost identical with the one used in the description of ${\cal N}=8$ 4D supergravity; the word 'almost' here refers to the fact that
only $SO(7)$ subgroup of $SU(8)$ acts on its index $_A$
\footnote{
Probably, to observe the SU(8) symmetry, one has  to consider $(w,\bar{w})$ as parametrizing the coset
$SO(16)/[SU(8)\otimes H]$ with some $H\subset SO(16)$. (This in its turn would require to consider  ${\cal U}^{AB}$ in (\ref{U/=wcUw})  to be an  independent spin-tensor coordinate).
Thus a hidden $SO(16)$
symmetry of 11D SUGRA might be relevant in this problem. It is tempting to speculate that $E_{8}$
hidden symmetry might also happen to be useful in this context. }. The decomposition of our analytic superfield in $\eta_A \; $
\footnote{To streamline the presentation at this stage we prefer to pass to the Fourier image of the
superfields with respect to $x^=$ (actually  $x^=_L=x^=+2i\eta_A\bar{\eta}{}^A$) coordinate ((\ref{A=A-rv}) {\it vs} (\ref{A=A-xv})).
}
 \begin{eqnarray}\label{Phi=SUGRA} \Phi (\rho^\# , v_{\alpha q}^{\; -}; w, \bar{w}; \eta_A)= \phi^{(+2)} +
  \eta_A \psi^{(+3/2)A}  + \ldots + (\eta)^{\wedge 7\; A} \psi^{(-3/2)}_{A}
  + (\eta)^{\wedge 8}  \phi^{(-2)}\;
 \end{eqnarray}
 looks very much the same as chiral  superfield (\ref{Phi=3,4}) describing the linearized ${\cal  N}=8$ supergravity. However, all the fields in its decomposition depend on a different set of variables:
 \begin{eqnarray}\label{charges}\phi^{(+2)} =\phi^{(+2)}  (\rho^\# , v_{\alpha q}^{\; -}; w, \bar{w})\; , \qquad \psi^{(+3/2)A}=\psi^{(+3/2)A} (\rho^\# , v_{\alpha q}^{\; -}; w, \bar{w})\, , \qquad etc. \;
 \end{eqnarray}  {\it versus}  $\phi^{+2}=\phi^{+2} ( \lambda, \bar{\lambda })$ with
complex  two component $\lambda=( \bar{\lambda })^*$ in D=4.

The set of 24=1+9+14 bosonic variables our on-shell fields depend on  includes 'energy' $\rho^\#$,
spinor harmonic variables $v_{\alpha q}^{\; -}$, which are considered as  homogeneous coordinates
of the celestial sphere $ {\bb S}^{9}$ realized as a coset of Lorentz group
$\frac {Spin(1,10)}{[SO(1,1)\times Spin(9)]\subset\!\!\!\!\times K_{9}} $
(\ref{v-qi=G-H}),
and a set of   $ \frac {Spin(9)}{Spin(7)\otimes Spin(2)}$ internal harmonic variables
$w_q^{\; A}, \bar{w}_{Aq}$  (\ref{wbw=G-H}).

The signs and numerical superscribes of the component fields in (\ref{Phi=SUGRA}) and (\ref{charges}) indicate their charges under $U(1)$ symmetry transformations acting on $\eta_A$, $\bar{w}$ and   $w$. These can be easily calculated  in the assumption that the overall charge of the superfield is equal to $+2$
\footnote{\label{P=HUU} This assumption is suggested by the origin of the analytic superfield in $SO(D-2)$ tensor,
$\Phi = H_{IJ}U_IU_J$. }.

The above statements about charges of variables and superfields under
$U(1)\subset SO(9)\subset SO(1,10)$ can be formulated as a differential equation
({\it cf.} (\ref{hhiA=hiA}))
\begin{eqnarray}\label{hi=11D}
\hat{h}^{(11D)} \Phi (\rho^\# , v_{\alpha q}^{\; -}; w, \bar{w}; \eta_A) = 2 \Phi (\rho^\# , v_{\alpha q}^{\; -}; w, \bar{w}; \eta_A)\; ,\qquad
\end{eqnarray}
where
\begin{eqnarray}
 \label{hh:=11D}
\hat{h}^{(11D)} :=  \frac 1 2 \bar{w}_{qA} \frac \partial {\partial\bar{w}_{qA}} -
 \frac 1 2 w_q^{A} \frac \partial {\partial w_q^{A}}+
\frac 1 2 \eta_A  \frac \partial {\partial \eta_A } \;    \qquad \end{eqnarray}
is the 11D counterpart of the helicity operator (\ref{Ui=}). On the analytic superfields it coincides  with
the harmonic covariant derivative $D^{(0)}$ (\ref{D0:=11D}),
\begin{eqnarray}
 \label{hh:=11DAn}
\hat{h}^{(11D)} = D^{(0)} \vert_{on\; analytic\; superfields }\; ,    \qquad \end{eqnarray}
so that Eq. (\ref{hh:=11D}) actually coincide with (\ref{D0P=P=11D}).

In the above presented  description by analytic superfield the 44+84=128 bosonic fields of the on-shell
11D supergravity ($h_{IJ}=h_{((IJ))}$ and $A_{IJK}=A_{[IJK]}$ in the $SO(9)=SO(D-2)$
covariant notation) are split onto
{\bf 1+28+70+28+1} representations of $SO(7)=SO(D-4)$,
 \begin{eqnarray}\label{bose=128}
 e_\mu^a, A_{\mu\nu\rho}\; \leftrightarrow \qquad (h_{((IJ))}\, , \, A_{[IJK]}) \; = \;
 (\phi^{(-4)} , \phi^{(-2)AB}, \phi^{ABCD} , \phi^{(+2)}_{AB}, \phi^{(+4)} )\; , \qquad
\end{eqnarray}
and 128 fermionic fields splits on {\bf 1+28+70+28+1}
\begin{eqnarray}\label{fermi=128}
 \psi_\mu^\alpha \quad  \leftrightarrow \qquad (\Psi_{Iq }\; |\;  \gamma^I_{qp}\Psi_{Ip}=0) \; = \;
 (\psi^{(-3)}_A , \psi^{(-1)}_{ABC}, \psi^{(+1)ABC} , \psi^{(+3)A} )\; . \qquad
\end{eqnarray}

\subsection{Supersymmetry transformation of the analytic superfields}

As in the case of 10D SYM, we can find that  rigid 11D supersymmetry  acts on the analytic on-shell superfield of 11D supergravity by
 \begin{eqnarray}\label{Psi=susy}
   \Phi^\prime (\rho^\# , v_{q}^{\; -}; w, \bar{w}; \eta_A)&=&
   \exp\{-4{\rho^{\#}} \bar{\epsilon}{}^{\, -A}\eta_A\} \Phi (\rho^\# , v_{q}^{\; -}; w, \bar{w}; \eta_A - \epsilon^{-}_{A})\qquad \nonumber \\
   &=&
   \exp\left\{-4{\rho^{\#}}\bar{\epsilon}{}^{\, -A}\eta_A- \epsilon^{-}_{A} \frac { \partial}{ \partial \eta_A} \right\} \;  \Phi (\rho^\# , v_{q}^{\; -}; w, \bar{w}; \eta_A )\; ,
   \end{eqnarray}
where
\begin{eqnarray}\label{bep-A=}
   \bar{\epsilon}{}^{\, -A} =  \epsilon_q^{-} w^{\; A}_q =  \epsilon^\alpha v_{\alpha q}^{\; -} w^{\; A}_q \; , \qquad  \epsilon^{-}_{ A} =  \epsilon_q^{-} \bar{w}{}_{qA}= \epsilon^\alpha v_{\alpha q}^{\; -} \bar{w}{}_{qA} \end{eqnarray}
 and $ \epsilon^\alpha $ is constant fermionic parameter.

The  supersymmetry generator defined by $\Phi^\prime= e^{-\epsilon^\alpha Q_\alpha}\Phi$,
\begin{eqnarray}\label{susyQ=}
Q_\alpha =  \tilde{q}_\alpha+ \hat{q}_\alpha = 4 {\rho^{\#}}v_{\alpha q}^{\; -} \eta^-_A w^{\; A}_q  + v_{\alpha q}^{\; -} \bar{w}{}_{qA}  \frac { \partial}{ \partial \eta^-_A}\;   \end{eqnarray}
is given by the sum of the algebraic part $\tilde{q}_\alpha$ and of the  differentail operator $ \hat{q}_\alpha$,
\begin{eqnarray}\label{tq11D=-4i}
\tilde{q}_\alpha &=& 4 {\rho^{\#}}v_{\alpha q}^{\; -} w^{\; A}_q \eta_A  \; , \qquad \\
\label{hq=i} \hat{q}_\alpha &=&  v_{\alpha q}^{\; -} \bar{w}{}_{qA}  \frac { \partial}{ \partial \eta_A}\; .
\end{eqnarray}

However, in distinction to the D=4 case, to split the parameter of rigid supersymmetry $\epsilon^\alpha$ on the parts corresponding to $\tilde{q}_\alpha$ and  $ \hat{q}_\alpha$ we need to use the composite complex spinor harmonic variables
$v^{+\alpha}_q  w^{\; A}_q$ and  $v^{+\alpha}_q   \bar{w}{}_{qA}$ (while in D=4 the splitting appears automatically because
 $\tilde{q}^{(D=4)}_\alpha$ and $ \hat{q}^{(D=4)}_{\dot{\alpha}}$ carry different type of Weyl spinor indices).

To show that the algebra of  supersymmetry generators (\ref{susyQ=}) is closed on the momentum,
\begin{eqnarray}\label{QQ=p}
\{ Q_\alpha , Q_\beta \} = 4{\rho^{\#}}v_{\alpha q}^{\; -} v_{\beta q}^{\; -} \, =
2p_a\Gamma^a_{\alpha \beta}\; ,  \end{eqnarray}
we have to use (\ref{wbw+cc=1}), $ 2w^{\; A}_{(q }\bar{w}{}_{p)A} =\delta_{qp}$.

\bigskip

\section{Analytic superamplitudes in D=10 and D=11 }

In this section and below, to avoid doubling of the formulae, we will tend to write the universal equations describing simultaneously D=10 and D=11 case whenever it is possible.    (See sec. 1.1 for the universal description of our index notation).

\subsection{Properties of analytic superamplitudes}

The simplest superamplitudes  are multiparticle counterparts of the analytic  superfields (\ref{Phi=SYM}) and (\ref{Phi=SUGRA})
 \begin{eqnarray}\label{cA=alt}
 {\cal A}_n\; \delta^D \left(\sum\limits_i^n k_{ai} \right) = {\cal A}_n (\{ \rho^\#_{i}, v_{\alpha qi}^{\; -}; w_i, \bar{w}_i; \eta_{A i}\}) \; \delta^D \left(\sum_i\limits^n \rho^{\#}_{i} u_{a i}^= \right)\; . \qquad \end{eqnarray}
They do not carry indices,
are Lorentz invariant,  invariant under
$\prod\limits_{i=1}^n [ SO(1,1)_i\otimes SO(D-2)_i\otimes SO(D-4)_i]$
and covariant under $\prod\limits_i SO(2)_i=U(1)_i$ symmetry transformations.

The Lorentz group $SO(1,D-1)$ acts nontrivially on spinor harmonic variables $v^{\; -}_{\alpha qi}$  only,  $SO(D-2)_i$ act on
$v^{\; -}_{\alpha qi}$ and on the internal harmonic variables  $(w^{A}_{qi} ,\bar{w}{}_{Aqi})$, $SO(1,1)_i$ act on $v^{\; -}_{\alpha qi}$
and on the fermionic $\eta_{Ai}=\eta_{Ai}^-$, and  $SO(D-4)_i$ transform
$w^{A}_{qi}$ and $\bar{w}{}_{Aqi}, \eta_{Ai}$ in {\bf 4s} and ${\bf\overline{ 4s}}$, respectively.
Finally, $ SO(2)_i=U(1)_i$ symmetries act nontrivially on $w^{A}_{qi}, \bar{w}{}_{Aqi}, \eta_{Ai}$
with the same value of $i$,  and on the amplitude which carries the charge $s={\cal N}/4$
($+2$ for 11D SUGRA and $+1$ for 10D SYM) with respect to each  of the $U(1)_i$ group.

The gauge symmetry  $\prod\limits_i SO(1,1)_i\otimes SO(D-2)_i\otimes SO(D-4)_i$ make possible to identify each set of harmonic variables, $ (v^-_{\alpha qi}, v^+_{\alpha \dot{q}i})$ and $(w^{A}_{qi}, \bar{w}{}_{Aqi}$), with generalized homogeneous coordinates of the cosets:
 \begin{eqnarray}\label{v-i=G-H}
 \{ (v^-_{\alpha qi}, v^+_{\alpha \dot{q}i})\}    = \left(\frac {Spin(1,D-1)}{[SO(1,1)\otimes Spin(D-2)]\subset\!\!\!\!\!\!\times K_{(D-2)}}\right)_i={\bb S}^{(D-2)}_i \qquad  \end{eqnarray}
 and\footnote{$K_{(D-2)i}$ symmetry  (\ref{KD-2=v}) implies an   independence of the amplitude on the complementary $ v^+_{\alpha \dot{q}i}$ harmonics. This is reflected by the list of arguments of the amplitude in (\ref{cA=alt}). Let us recall that for $D=11$ case $ \dot{q}=q=1,..,16$, while for D=10 $\dot{q}=1,...,8$ and $q=1,...,8$ are indices of different spinor representations of SO(8). }
  \begin{eqnarray}\label{wi=G-H=alt}
 \{ w^{A}_{qi}, \bar{w}{}_{Aqi}\} = \left(\frac {SO(D-2)}{SO(D-4)\otimes U(1)}\right)_i \; . \qquad \end{eqnarray}

Notice that here, in distinction to sec. 2 devoted to D=4 case, we prefer to write explicitly
the momentum preserving delta function
$ \delta^{D}(\sum_i\limits^n k_{a i} )=  \delta^{D}( \sum_i\limits^n \rho^{\#}_{i} u_{a i}^= )$
and denote by ${\cal A}$ the amplitudes with arguments obeying the overall momentum conservation
 \begin{eqnarray}\label{P=cons}
 \sum_i\limits^n k_{a i} = \sum_i\limits^n \rho^{\#}_{i} u_{a i}^==0 \; , \qquad
 u_{a i}^= \Gamma^a_{\alpha\beta}= 2 v_{\alpha q\; i}^{\; -} v_{\beta q\; i}^{\; -} \; . \end{eqnarray}

The representations of the variables  and superamplitude with respect to the  symmetry groups are summarized in Table 1, where the parameter $s={\cal N}/4$ distinguish the cases of D=10 SYM (s=1) and D=11 SUGRA (s=2).
\begin{center}
    \begin{tabular}{ | l | l | l | l | l |l |}
    \hline
    variables/representations  & SO(1,1)$_i$ & SO(D-2)$_i$  & SO(D-4)$_i$   & SO(2)$_i$=U(1)$_i$ &  Spin(1,D-1) \\
    &  weight &  repr. & repr.   &  charge &  \\ \hline  & & & & &
    \\   ${\cal A}_n(\{ \rho^\#_{i}, v_{i}^{\; -}; w_i, \bar{w}_i; \eta_{i}\}) $ &  &  &  & $+s$ &
    \\  $\rho^{\#}_{(i)}$ & +2 &  &  & &
    \\ $v_{\alpha q(i)}^{\; -}$ & $-1$ & {\bf 8s} &  &  & {\bf 16s}
    \\ $w_{q(i)}^{\; A}$ &  & {\bf 8s} &  {\bf 4s} & $-1/2$ &
    \\ $\bar{w}_{qA(i)}$ &  & {\bf 8s} &  $\overline{{\bf 4s}}$ & $+1/2$ &
    \\ $\eta_{A(i)}\equiv \eta_{A(i)}^-$ & -1 &  &  $\overline{{\bf 4s}}$ & $+1/2$  & \\
     \hline
     $U_I$ &  &  {\bf 7+s} && +1 & \\  $\bar{U}_I$ &  &  {\bf 7+s} && $-$1 & \\
     \hline
    \end{tabular}

\medskip

Table 1. {\small\it  SO(D-4) and SO(D-2) representations, SO(1,1) weights and U(1) charges of the analytic  superamplitude and its arguments; $s=1$ for D=10 SYM and $s=2$ for 11D SUGRA.  $U_I$ and $\bar{U}_I$   are bilinears of $w$ and $\bar{w}$ as defined in (\ref{Ug8=bww}) for D=10 and  (\ref{U/=ww}) for D=11. }
\end{center}

This table also indicates that the simplest superamplitudes (\ref{cA=alt}) are Lorentz scalars, have charges $+s$ with respect to all the $U(1)_i$ symmetry groups and are inert under all other bosonic symmetry transformations $SO(1,1)_i\otimes SO(D-2)_i\otimes SO(D-4)_i$. As we will discuss below, the analytic superamplitudes also obey a set of equations with harmonic covariant derivatives  which provide us with   counterparts of the 4D helicity constraints (\ref{hhiA=hiA}).

More complicated superamplitudes, which do carry the nontrivial representations of $SO(D-4)_i$ and different charges under $SO(2)_i=U(1)_i$ can be obtained by acting on the analytic superamplitude  (\ref{cA=alt}) by fermionic  covariant derivatives
$D_{A(i)}^+$ and by harmonic covariant derivatives.

\subsection{From constrained to analytic superamplitudes. 10D SYM}

Let us discuss the relation of the above described analytic superamplitude (\ref{cA=alt}) with the constrained superamplitude formalism \cite{Bandos:2016tsm,Bandos:2017eof}.

The basic constrained superamplitude of 10D SYM theory
\begin{eqnarray}\label{cAI1In=10D}
 {\cal A}_{I_1... I_n}(k_1,\theta^-_1; ...;  k_n, \theta^-_n)\; \delta^D \Big(\sum\limits_i^n k_{ai} \Big)  = {\cal A}_{I_1... I_n} (\{ \rho^\#_{i}, v_{\alpha qi}^{\; -}; \theta_{ qi}^{-}\}) \; \delta^D \Big(\sum_i\limits^n \rho^{\#}_{i} u_{a i}^= \Big)\; , \qquad \end{eqnarray}
carry $n$ vector indices of $SO(8)_i$ groups. It  obeys the equations (see  \cite{Bandos:2017eof} for details)
\begin{eqnarray}\label{D+qAI=GA}
D^{+j}_{{q}} {\cal A}^{(n)}_{I_1... I_j...I_n}
=   2 \rho^{\#}_{j}\, \gamma^{I_j}_{q\dot{q}_j} {\cal A}^{(n)}_{I_1... I_{j-1}\dot{q}_j I_{j+1}... I_n}\; ,
\end{eqnarray}
where
\begin{eqnarray}\label{D+j=d+rho-th}
D^{+j}_{{q}} =\partial^{+j}_{{q}} + 2 \rho^{\#}_{j}\theta^-_{{q}j}  \qquad     \; ,
\qquad \partial^{+j}_{{q}}:= \frac \partial{ \partial\theta^-_{{q}j}}\; .
\end{eqnarray}
To express the analytic superamplitude  (\ref{cA=alt}) through the constrained superamplitude (\ref{cAI1In=10D}), let us first contract  the $SO(8)_i$ vector indices of this latter with the complex null-vectors $U_{I_i\, i}$ of the corresponding internal frames
(\ref{wi=G-H=alt}),
\begin{eqnarray}\label{cAn=cAIUI}
\tilde{{\cal A}}_n (\{ \rho^\#_{i}, v_{\alpha qi}^{\; -}; w_i, \bar{w}_i; \theta^-_{q i}\})=
 U_{I_1 1}\ldots U_{I_n n}\; {\cal A}_{I_1... I_n} (\{ \rho^\#_{i}, v_{\alpha qi}^{\; -}; \theta^-_{q i}\}) \; , \qquad
 \\ \label{Uig8=wbw}
U\!\!\!\!/{}_{q \dot{p}\, i }:=  \gamma^I_{q \dot{p}} U_{I\, i} = 2  \bar{w}_{{q}A i} w_{\dot{p}i}^A  \; . \qquad
\end{eqnarray}
Using (\ref{Uig8=wbw}) and the properties (\ref{bww=1}) of the internal harmonics
(\ref{wi=G-H=alt}), one can easily check that
\begin{eqnarray}\label{D+AcAcn=0}
\bar{D}{}^{+j}_{{A}} \tilde{{\cal A}}_n (\{ \rho^\#_{i}, v_{\alpha qi}^{\; -}; w_i, \bar{w}_i; \theta^-_{q i}\})= 0 \qquad \forall j=1,...,n\; . \qquad
\end{eqnarray}
In these equations
\begin{eqnarray} \label{bD+jA:=}
\bar{D}{}^{+j}_{{A}}= \bar{w}_{q{A}j} {D}{}^{+j}_{{q}} =
\frac \partial {\partial \bar{\eta}^{-A}_j} \; + 2\rho^{\#}_j \eta^-_{Aj}
 \; , \qquad \eta^-_{Aj} = \theta^-_{qj} \bar{w}_{q{A}j} \; ,   \qquad  \bar{\eta}^{-A}_j= \theta^-_{qj} {w}^{\; A}_{qj} \; . \qquad
\end{eqnarray}
Our analytic 10D SYM superamplitude is related to
 (\ref{cAn=cAIUI}) by
\begin{eqnarray}\label{cA=etcA}
{\cal A}_n (\{ \rho^\#_{i}, v_{\alpha qi}^{\; -}; w_i, \bar{w}_i; \eta_{A i}\}) &=& e^{^{-2\sum_j  \rho^\#_{j} \eta^-_{Bj}  \bar{\eta}^{-B}_j }}
 \tilde{{\cal A}}_n (\{ \rho^\#_{i}, v_{\alpha qi}^{\; -}; w_i, \bar{w}_i;\; \eta^-_{Ai}  {w}^{\; A}_{qi} +   \bar{\eta}^{-A}_i \bar{w}_{q{A}i}\,   \})\; .   \qquad  \end{eqnarray}
Indeed, one can easily check that  $$\bar{D}{}^{+(j)}_{{A}}e^{^{2\sum_j  \rho^\#_{j} \eta^-_{Bj}  \bar{\eta}^{-B}_j }} = e^{^{2\sum_j  \rho^\#_{j} \eta^-_{Bj}  \bar{\eta}^{-B}_j }} \frac \partial {\partial \bar{\eta}^{-A}_j}$$ so that ${\cal A}_n $ of (\ref{cA=etcA})  is
 $\bar{\eta}^{-A}_i$--independent due to (\ref{D+AcAcn=0}).

Resuming, the analytic superamplitude (\ref{cA=alt}) is expressed in terms of constrained superamplitude (\ref{cAI1In=10D}) by contracting its SO(8) vector indices $_{I_i}$ with appropriate null-vectors $U_{I_i i}$ constructed from internal harmonics as in (\ref{Ug8=bww}):
\begin{eqnarray} \label{cA=etcAU}
{\cal A}_n (\{ \rho^\#_{i}, v_{\alpha qi}^{\; -}; w_i, \bar{w}_i; \eta_{A i}\}) &=& e^{^{-2\sum_j  \rho^\#_{j} \eta^-_{Bj}  \bar{\eta}^{-B}_j }}
 U_{I_1 1}\ldots U_{I_n n}\; {\cal A}_{I_1... I_n} (\{ \rho^\#_{i}, v_{\alpha qi}^{\; -}; \; \eta^-_{Ai}  {w}^{\; A}_{qi} +   \bar{\eta}^{-A}_i \bar{w}_{q{A}i}\}) . \nonumber \\ {}
\end{eqnarray}

It is not difficult to check that this amplitude also obeys
\begin{eqnarray} \label{DchJcA=0}
D_j^{\check{J}} {\cal A}_n (\{ \rho^\#_{i}, v_{\alpha qi}^{\; -}; w_i, \bar{w}_i; \eta_{A i}\})& =& 0 \; , \qquad j=1,...,n\; , \\
\label{DchJchK=0}
D_j^{\check{J}\check{K}} {\cal A}_n (\{ \rho^\#_{i}, v_{\alpha qi}^{\; -}; w_i, \bar{w}_i; \eta_{A i}\}) &=& 0 \; , \qquad j=1,...,n\; , \\
\label{D0cA=0}
D_j^{(0)} {\cal A}_n (\{ \rho^\#_{i}, v_{\alpha qi}^{\; -}; w_i, \bar{w}_i; \eta_{A i}\}) &=& {\cal A}_n (\{ \rho^\#_{i}, v_{\alpha qi}^{\; -}; w_i, \bar{w}_i; \eta_{A i}\}) \; , \qquad j=1,...,n\; ,
\end{eqnarray}
with the derivative defined as in (\ref{DchJ=An})--(\ref{DchIJ=An}), (\ref{DchJ})--(\ref{DchIJ}), but for $j$-th internal harmonic variables.
Eqs.  (\ref{DchJcA=0}) --(\ref{D0cA=0}) can be considered as counterparts of the D=4 super-helicity constraints (\ref{UPhi=1}).

\subsection{Analytic superamplitudes of 11D SUGRA from constrained superamplitudes }

Eq. (\ref{cA=etcA}) with $q=1,.., 2{\cal N}$ and $A=1,..., {\cal N}$ describes also  the relation of the constrained and analytic superamplitudes of 11D SUGRA
if we set ${\cal N}=8$ and
\begin{eqnarray}\label{cAn=cAIUI=11D}
\tilde{{\cal A}}_n (\{ \rho^\#_{(i)}, v_{\alpha q(i)}^{\; -}; w_i, \bar{w}_i; \theta^-_{q i}\})=
 U_{I_1 1}U_{J_1 1}\ldots U_{I_n n} U_{J_n n}\; {\cal A}_{((I_1J_1))... ((I_nJ_n))} (\{ \rho^\#_{i}, v_{\alpha qi}^{\; -}\}) \; . \qquad
\end{eqnarray}
Here the basic superamplitude of the constrained superfield formalism,  ${\cal A}_{((I_1J_1))... ((I_nJ_n))} $ symmetric and traceless on each pair of $SO(9)_i$ vector indices enclosed in doubled brackets, obeys the equation \cite{Bandos:2016tsm,Bandos:2017eof}
\begin{eqnarray}\label{Df-AIJ}
&&  D^{+(j)}_{{q}_j} {\cal A}^{(n)}_{((I_1J_1))...((I_jJ_j))...((I_nJ_n)) } =
 \, i \gamma_{(I_j| q_jp_j} \,  {\cal A}^{(n)}_{((I_1J_1))...  |J_j) p_j\, ...((I_nJ_n))  } \, .  \qquad
\end{eqnarray}
The r.h.s. of this equation contains  $\gamma$-- traceless
$ {\cal A}^{(n)}_{((I_1J_1))...((I_{j-1}J_{j-1}))\; J_j p_j\,  ((I_{j+1}J_{j+1}))...((I_nJ_n))  }$,
\begin{eqnarray}\label{gJcAJp=0}
&&  \gamma^{J_j}_{qp_j}\;  {\cal A}^{(n)}_{((I_1J_1))...\; J_j p_j\,  ...((I_nJ_n))  } =0\,  . \qquad
\end{eqnarray}
Finally, $U_{I_1 i}$ in the {\it r.h.s.} of (\ref{cAn=cAIUI=11D}) is expressed through bilinear of the internal harmonics
by  \begin{eqnarray}\label{Uig9=wbw}
U\!\!\!\!/{}^{(i)}_{q {p}}:=  \gamma^I_{q {p}} U_{I\, i} = 2  \bar{w}_{{q}A i}\bar{w}_{{p}A i} \qquad
\end{eqnarray}
(see (\ref{U/=ww})).

Due to (\ref{Df-AIJ}) and (\ref{gJcAJp=0}),
$\tilde{{\cal A}}_n$ of (\ref{cAn=cAIUI=11D}) obeys (\ref{D+AcAcn=0})
and the 11D superamplitude (\ref{cA=etcA}) is analytic, {\it i.e.} it depends on
$\eta^-_{Ai}$ but is independent of its complex conjugate $\bar{\eta}^{-A}_i$.

The analytic amplitude (\ref{cAn=cAIUI=11D}) also obeys
 the equations
 \begin{eqnarray} \label{DchJcA=011}
D_j^{\check{J}} {\cal A}_n (\{ \rho^\#_{i}, v_{\alpha qi}^{\; -}; w_i, \bar{w}_i; \eta_{A i}\})& =& 0 \; , \qquad j=1,...,n\; , \\
\label{DchJchK=011}
D_j^{\check{J}\check{K}} {\cal A}_n (\{ \rho^\#_{i}, v_{\alpha qi}^{\; -}; w_i, \bar{w}_i; \eta_{A i}\}) &=& 0 \; , \qquad j=1,...,n\; , \\
\label{D0cA=011}
D_j^{(0)} {\cal A}_n (\{ \rho^\#_{i}, v_{\alpha qi}^{\; -}; w_i, \bar{w}_i; \eta_{A i}\}) &=& 2{\cal A}_n (\{ \rho^\#_{i}, v_{\alpha qi}^{\; -}; w_i, \bar{w}_i; \eta_{A i}\}) \; , \qquad j=1,...,n\; ,
\end{eqnarray}
 with $\check{J},\check{K}=1,...,7$.

\subsection{Supersymmetry transformations of the analytic superamplitudes}
\label{susy-sA}
 The supersymmetry acts on our analytical superamplitudes as
 \begin{eqnarray}\label{cAn=susy}
 {\cal A}^\prime_n (\{ \rho^\#_{i}, v_{\alpha qi}^{\; -}; w_{qi}^{\; A}, \bar{w}_{qAi}; \eta_{A i}\})
   & =& e^{-\epsilon^\alpha(\tilde{q}_\alpha + \hat{q}_\alpha) }\,
   {\cal A}_n (\{ \rho^\#_{i}, v_{i}^{ -}; w_i, \bar{w}_i; \eta_{A i}\}) \qquad  \nonumber \\ & =&
   e^{-\epsilon^\alpha\tilde{q}_\alpha }\,
   {\cal A}_n (\{ \rho^\#_{i}, v_{i}^{ -}; w_i, \bar{w}_i; \eta_{A i}- \epsilon^{-}_{Ai}\}) \; , \qquad \end{eqnarray}
   where (see (\ref{tq11D=-4i}) and (\ref{hq=i}))
    \begin{eqnarray}\label{exponent}
   \epsilon^\alpha\tilde{q}_\alpha = 4\sum\limits_{i=1}^n {\rho_i^{\#}}
   \bar{\epsilon}{}_i^{A-} \eta_{Ai}  \; &,& \qquad
   \epsilon^\alpha\hat{q}_\alpha =  \sum\limits_{i=1}^n {\epsilon}{}^-_{Ai} \frac {\partial} {\partial \eta_{Ai}}  \;  \qquad
   \end{eqnarray}
 and  (see (\ref{bep-A=}))
  \begin{eqnarray}\label{e-+iA=}
{\epsilon}^{-}_{ Ai} = \epsilon^\alpha \bar{v}{}_{\alpha Ai}^{\; -}  =
\epsilon^\alpha v_{\alpha qi}^{\; -} \bar{w}{}_{qAi} \; , \qquad
\bar{\epsilon}^{A-}_i = \epsilon^\alpha v_{\alpha i}^{\; -A} = \epsilon^\alpha v_{\alpha qi}^{\; -} w^{A}_{qi}\; . \qquad  \end{eqnarray}
As in the case of the on-shell superfields, the supersymmetry generator acting on superamplitude splits
\begin{eqnarray}\label{susyQ=tq+hq}
Q_\alpha =  \tilde{q}_\alpha+ \hat{q}_\alpha
\end{eqnarray}
onto the purely algebraic part and the differential operator
 \begin{eqnarray}\label{tqal=}
\tilde{q}_\alpha = 4 \sum\limits_{i=1}^n {\rho_i^{\#}} v_{\alpha (i)}^{\; -A}  \eta^-_{Ai} = 4 \sum\limits_{i=1}^n {\rho_i^{\#}} v_{\alpha q(i)}^{\; -} w^{A}_{q(i)} \eta_{Ai} \; , \\
\label{hqal=}
\hat{q}_\alpha =  \sum\limits_{i=1}^n  v_{\alpha Ai}^{\; -} \frac {\partial} {\partial \eta_{Ai}}
=  \sum\limits_{i=1}^n v_{\alpha qi}^{\; -}\bar{ w}_{qAi} \frac {\partial} {\partial \eta_{Ai}}    \; .
 \end{eqnarray}

It is easy to check (using $2w_{(q|i}{}^A\bar{w}_{|p)Ai}=\delta_{qp}$ (\ref{wbw+cc=1})) that  the
generators (\ref{susyQ=tq+hq}) obey  the supersymmetry algebra, and actually anti-commute as far as
the momentum is conserved:
\begin{eqnarray}\label{QQ=Sp=0}
\{ Q_\alpha , Q_\beta \} = 4 \sum\limits_{i=1}^n{\rho^{\#}_i}v_{\alpha qi}^{\; -} v_{\beta qi}^{\; -} \, =
2\Gamma^a_{\alpha \beta} \sum\limits_{i=1}^n p_{ai}=0 \; . \qquad \end{eqnarray}

\bigskip

\subsection{Supermomentum in D=10 and D=11}

Although we have succeeded in writing the supersymmetry generator in terms of complex $\eta^-_{Ai}$ and its derivative, the simplest way to write a supersymmetric invariant linear combination  of fermionic variables uses the real fermionic $\theta^-_{qi}$ of the constrained superfield formalism (see (\ref{anHOn-shellSSP})).  Indeed, the real fermionic spinor
\begin{eqnarray}\label{qf:=}
q_{\alpha} &:=& \sum\limits_{i=1}^n \rho^{\#}_{i}
v_{\alpha qi}^{\; -} \theta^-_{qi}  \; , \qquad
\end{eqnarray}
which can be called {\it supermomentum}, is transformed into the  momentum by supersymmetry
\begin{eqnarray}\label{susy-qf:=}
\delta_\epsilon q_{\alpha} &=& \epsilon^\beta \sum\limits_{i=1}^n \rho^{\#}_{i}
v_{\alpha qi}^{\; -}v_{\beta qi}^{\; -} =  \frac 1 2  \Gamma^{a}_{\alpha\beta}\epsilon^\beta \sum\limits_{i=1}^n p_{ai}
  \qquad
\end{eqnarray}
and, hence is supersymmetric invariant  when momentum is conserved,
\begin{eqnarray}\label{susy-qf=0}
\delta_\epsilon q_{\alpha} &=& 0\qquad when \qquad \sum\limits_{i=1}^n p_{ai} =0 \; .
  \qquad
\end{eqnarray}

\section{Convenient parametrization of spinor harmonics
(convenient  gauge fixing of the auxiliary gauge symmetries)}

\subsection{Reference spinor frame  and minimal parametrization of spinor harmonics }
\label{ref-frame}

It looks convenient to fix the gauge with respect to the defining gauge symmetries of the spinor frame variables
${[SO(1,1)_i\otimes SO(D-2)_i]\subset\!\!\!\!\!\!\times K_{(D-2)\, i}}$ by setting
\begin{eqnarray}\label{v-=v+Kv=G}
v_{\alpha qi}^{\; -} &=&
v_{\alpha  q }^{\; -}+ {1\over 2} K^{=I}_{i}  \gamma^I_{q\dot{p}} v_{\alpha  \dot{p}}^{\; +}
\; , \qquad
v_{\alpha \dot{q}i}^{\;+}= v_{\alpha \dot{q}}^{\; +}
 \; . \qquad
\end{eqnarray}
Here $(v_{\alpha  q }^{\; -}, v_{\alpha  \dot{q}}^{\; +})$ is an auxiliary reference spinor frame the
components  of which can be identified with homogeneous coordinates of an auxiliary coset (or reference coset)
$ \frac {SO(1,D-1)}{[SO(1,1)\otimes SO(D-2)]\subset\!\!\!\!\times K_{D-2}}={\bb S}^{D-2}$.
Clearly, the reference spinor frame can be chosen arbitrary, in a way convenient for the problem under consideration.

Eq. (\ref{v-=v+Kv=G}) provides the explicit parametrization of the spinor harmonics  $(v_{\alpha  q i}^{\; -}, v_{\alpha  \dot{q}i}^{\; +})$ describing a celestial sphere of $i$-th $D$-dimensional observer by single $SO(D-2)$ vector $ K^{=I}_{i}$. This is manifestly invariant
under one set of  ${[SO(1,1)\otimes SO(D-2)]\subset\!\!\!\!\!\!\times K_{D-2}}$ gauge symmetries acting
on the reference spinor frame variables.

Eqs.  (\ref{v-=v+Kv=G}) lead to  the following expressions for vector harmonics in terms of reference vector frame
 \begin{eqnarray}\label{u--=KuI=11G}
u^=_{ai}= u^=_{a}   + K^{=I}_{i}u^I_{a} + \frac 1 4 (\vec{K}^{=}_{i} )^2 u^\#_{a}
 \; , \qquad \\
 \label{uI=K--uiG11}
u^I_{ai} = u^I_{a} + \frac 1 2  K^{= I }_{i}\, u^\#_{a}\; , \qquad \\ \label{u++=KuiG11}
u^\#_{ai} =u^\#_{a} \; . \qquad
\end{eqnarray}
The momentum of $i$-th particle is expressed through  $K^{=I}_{i}$ and  density $\rho^\#_{i}$ by
\begin{eqnarray}\label{kia=K--Igauge}
& k_{ai}= \rho^\#_{i}u^=_{ai}=
\rho^\#_{i}\left( u^=_{a}   + K^{=I}_{i}u^I_{a} + \frac 1 4 (\vec{K}^{=}_{i} )^2 u^\#_{a}\right)
  \; . \qquad
\end{eqnarray}
Thus, in the gauge (\ref{v-=v+Kv=G}), the $n$-point  amplitude (\ref{cA=alt}) is a function of
energies $ \rho^\#_{i}$, of $SO(D-2)$ vectors $K_i^I$, of the fermionic $\eta_{Ai}$ and also of the constrained complex bosonic $w_i, \bar{w}_i$ variables,
 \begin{eqnarray}\label{cA=altG}
 {\cal A}_n = {\cal A}_n (\{ \rho^\#_{i}, K_{i}^{=I}; w_i, \bar{w}_i; \eta_{A i}\}) \; .
 \end{eqnarray}
This latter dependence will be specified below.

\subsection{Generic parametrization of spinor harmonic variables and $K^{\# I}=0$ gauge}

A generic parametrization  of the spinor harmonics   (\ref{harmV=D}) is
\begin{eqnarray}\label{v-j=+Kv-}
v_{\alpha qi}^{\; -} &=& e^{-\alpha_i} {\cal O}_{i{q}{p}}  \left(
v_{\alpha  p}^{\; -}+ {1\over 2} K^{=I}_{i}  \gamma^I_{p\dot{q}} v_{\alpha  \dot{q}}^{\; +}
\right) \; ,  \qquad \\
\label{v+j=+Kv-}
v_{\alpha  \dot{q}i}^{\; +}&=&  {\cal O}_{i\dot{q}\dot{p}}
e^{\; \alpha_{i}} v_{\alpha {\dot{p}}}^{\;+}+  {1\over 2} K^{\# I}_{i}  v_{\alpha p i}^{\; -} \gamma^I_{p\dot{q}}
 \; , \qquad
\end{eqnarray}
where the `physical' degrees of freedom are carried by  $SO(D-2)$ vector $ K^{=I}_{i}$
parametrizing the celestial sphere ${\bb S}^{(D-2)}$ (through a kind of stereographic projection). Besides this, the r.h.s. of Eqs. (\ref{v-j=+Kv-}), (\ref{v+j=+Kv-}) contain
 $\alpha_i$, which is the parameter of $SO(1,1)$,  ${\cal O}_{i{q}{p}}$ and ${\cal O}_{i\dot{q}\dot{p}}$,
which  are the $Spin(D-2)$ matrices\footnote{Notice that for both  D=10 and  D=11  the $Spin(D-2)$ valued matrices ${\cal O}_{{q}{p}i}$ obey also
 ${\cal O}_{{q}{p}_1i}{\cal O}_{{p}{p}_1i}= \delta_{{q}{p}}$; this is to say $Spin(D-2)\subset SO(2{\cal N})$ for these cases. }, and $K^{\# I}_{i}$, which parameterizes the
${K}_{(D-2)}$ symmetry transformations.  All these transformations  are used as identification relations
on the set of  spinor harmonic variables. This is tantamount to saying that they are the gauge symmetry
of the spinor frame construction. We can fix their values arbitrarily
thus providing an  explicit parametrization of the coset (\ref{v-i=G-H}).
A particular  choice $\alpha_i=0=K^{\# I}_{i}$, ${\cal O}_{i{q}{p}}= \delta_{{q}{p}}$ gives us
the simple expressions (\ref{v-=v+Kv=G}) and (\ref{u--=KuI=11G}).

In $D=10$ case we have to complete (\ref{v-j=+Kv-}), (\ref{v+j=+Kv-}) with
\begin{eqnarray}\label{v-j=-Kv-}
v_{\dot{q}i}^{-\alpha} &=& e^{-\alpha_{i}} {\cal O}_{i\dot{q}\dot{p}}  \left(
v_{\dot{p}}^{-\alpha} -  {1\over 2} K^{=I}_{i}  v_{p}^{+\alpha}\gamma^I_{p\dot{p}}
\right) \; ,  \qquad \\
\label{v+j=-Kv-}
v_{q i}^{+\alpha } &=&  e^{\; \alpha_{i}}  {\cal O}_{i{q}{p}}
v_{{{p}}}^{+\alpha }- {1\over 2} K^{\# I}_{i}   \gamma^I_{q\dot{p}}v_{\dot{p} i}^{-\alpha}
 \; , \qquad
\end{eqnarray}
while in $D=11$, where $\dot{q}=q$, these equations are equivalent to (\ref{v-j=+Kv-}) and (\ref{v+j=+Kv-}).

Eq. (\ref{v-j=+Kv-}) and (\ref{v+j=+Kv-}) imply
\begin{eqnarray}\label{v-iv-j=}
v_{\dot{p}i}^{-\alpha }
v_{\alpha qj}^{\; -} &=& \frac 1 2 e^{-\alpha_{i}-\alpha_{j}}  K^{= I}_{ji} ({\cal O}_{j}\gamma^I{\cal O}_{i}^T)_{{q}\dot{p}}
 \; , \qquad
\end{eqnarray}
 where
 \begin{eqnarray}
 \label{Kij=Ki-Kj}
 && {K}{}^{=I}_{ji}={K}{}^{=I}_{j}-{K}{}^{=I}_{i}\; .
\end{eqnarray}
In the gauge  (\ref{v-=v+Kv=G}) this simplifies to
\begin{eqnarray}\label{v-iv-j=Kij}
v_{\dot{p}i}^{-\alpha }
v_{\alpha qj}^{\; -} &=& \frac 1 2  K^{= I}_{ji} \gamma^I_{{q}\dot{p}}
 \;  \qquad
\end{eqnarray}
and becomes antisymmetric in $i,j$. This latter fact suggests to search the 10D (and 11D) counterparts of the 4D expression $<ij>$ (\ref{ij=lili}) on the basis of  (\ref{v-iv-j=Kij}).

The complete parametrization of the vector frame variables corresponding to (\ref{v-j=+Kv-}), (\ref{v+j=+Kv-}) is given by
\begin{eqnarray}\label{u--=KuI=11}
u^=_{ai} &=& e^{-2\alpha_{i}}\left(u^=_{a} + \frac 1 4 (\vec{K}^{=}_{i} )^2 u^\#_{a}  + K^{=I}_{i}u^I_{a} \right)
 \;  \qquad \end{eqnarray}
and quite complicated expressions for $u^\#_{ai}$ and $u^I_{ai}$. The light-like momentum of $i$-th particle
has the form of (\ref{kia=K--Igauge}), but with a redefined $\rho_i^\#$,
\begin{eqnarray}\label{kia=K--I}
& k_{ai}=
\tilde{\rho}{}^\#_{i} \left( u^=_{a}   + K^{=I}_{i}u^I_{a} + \frac 1 4 (\vec{K}^{=}_{i} )^2 u^\#_{a}\right)\; , \qquad  \\ \label{trho:=} & \tilde{\rho}{}^\#_{i}=e^{-2\alpha_i}
\rho^\#_{i}
  \; . \qquad
\end{eqnarray}

The  expressions for $u^\#_{ai}$ and $u^I_{ai}$ simplify essentially if we use the ${K}_{(D-2)i}$ symmetry to fix the gauge
\begin{eqnarray} \label{K++I=0}
K^{\# I}_{i}=0\;
  \end{eqnarray}
  in which case
 \begin{eqnarray} \label{u++=Kui}
u^\#_{ai} &=& e^{+2\alpha_{(i)}} u^\#_{a} \; , \qquad
 \label{uI=K--ui}
u^I_{ai} = \, {\cal O}^{IJ}_{i}\, \left(u^J_{a}+ \frac 1 2 u^\#_{a}  K^{= J }_{i}
  \right) \; , \qquad \\ \label{gIO=OgO} && \gamma^I_{q\dot{p}}{\cal O}^{IJ}_{i}  =\gamma^J_{p\dot{q}} {\cal O}_{pqi} {\cal O}_{\dot{q}\dot{p}i} \; . \qquad
  \end{eqnarray}
The spinor frame parametrization  with    $K^{\# I}_{i}=0$ (\ref{K++I=0})
  is given by the same Eqs.
(\ref{v-j=+Kv-}) and (\ref{v-j=-Kv-}), while Eqs. (\ref{v+j=+Kv-}) and  (\ref{v+j=-Kv-}) simplify essentially:
\begin{eqnarray}
\label{v+j=Ov+}
v_{\alpha  \dot{q}i}^{\; +}&=& e^{\; \alpha_{i}}  {\cal O}_{i\dot{q}\dot{p}}
 v_{\alpha {\dot{p}}}^{\;+}
 \; , \qquad
v_{q i}^{+\alpha } =  {\cal O}_{i{q}{p}}
e^{\; \alpha_{i}} v_{{{p}}}^{+\alpha } \; . \qquad
\end{eqnarray}

\subsection{Internal harmonics and reference internal frame}

As $SO(D-2)_i$ auxiliary gauge symmetry acts not only on $i$-th set of spinor harmonics but also on  i-th set of internal harmonics
$(\bar{w}_{{q}A\, i}, {w}_{q\, i}^{\; A})$, the introduction of the reference spinor frame in (\ref{v-j=+Kv-}), (\ref{v+j=+Kv-}) should be accompanied by
the introduction  of the reference 'internal frame'. This is described by  the set of harmonic variables  $(\bar{w}_{{q}A}, {w}_{q}^{\; A})$ parametrizing the coset
$\left(\frac {SO(D-2)} {SO(D-4)\otimes SO(2)}\right)$ (reference coset).  The i-th internal harmonic  variables can be decomposed on this reference frame,
\begin{eqnarray}\label{bwi=bwOU}
\bar{w}_{{q}A\, i}= {\cal O}_{{q}{p}\, i} \bar{w}_{{p}B} \; e^{-i\beta _i} \, {\cal U}_{A\, i}^{\dagger\, B}\; , \qquad
   {w}_{q\, i}^{\; A} = {\cal O}_{{q}{p}\, i} {w}_{p}^{\; B} \;  e^{+i\beta _i}  {\cal U}_{B\, i}^{\;  A}\; , \qquad\\ \label{cUinOinU}
    {\cal U}_{B\, i}^{\;  A} \in SO(D-4) \subset SU({\cal N})\; . \qquad
\end{eqnarray}
In the case of D=10, we must also introduce the internal reference frame with c-spinor $SO(8)$ indices
$(\bar{w}_{\dot{q}A}, {w}_{\dot q}^{\; A})$ and relate it to $i$-th internal harmonics by
\begin{eqnarray}\label{bwid=bwOU}
\bar{w}_{\dot{q}A\, i}= {\cal O}_{\dot{q}\dot{p}\, i} \bar{w}_{\dot{p}B} \; e^{i\beta _i} \, {\cal U}_{A\, i}^{\dagger\, B}\; , \qquad
   {w}_{\dot{q}\, i}^{\; A} = {\cal O}_{\dot{q}\dot{p}\, i} {w}_{\dot{p}}^{\; B} \;  e^{-i\beta_i}  {\cal U}_{B\, i}^{\;  A}\; . \qquad
\end{eqnarray}
The $Spin(D-2)$  valued matrices ${\cal O}_{{q}{p}\, i}$ in (\ref{bwi=bwOU}) and ${\cal O}_{\dot{q}\dot{p}\, i} $
in (\ref{bwid=bwOU}) are bridges between $SO(D-2)_i$ acting on
$i$-th spinor frame and  $SO(D-2)$ acting on the reference spinor frame. In other words, the first index of
${\cal O}_{{q}{p}\, i}$ (${\cal O}_{\dot{q}\dot{p}\, i}$) matrix is transformed by   $SO(D-2)_i$ and
the second- by $SO(D-2)$  group. One can also consider them as compensators for  $SO(D-2)_i$ auxiliary
gauge symmetry.

Similarly, the unitary $Spin(D-4)$ valued matrices ${\cal U}_{B\, i}^{\;  A}$ (\ref{cUinOinU}) are bridges  between  $Spin(D-4)_i$ and $Spin(D-4)\subset SU({\cal N})$  groups, and the phase factor $ e^{i\beta _i} $
serves as a bridge between $U(1)_i$ and $U(1)$ acting on the reference 'internal frame'. Notice the opposite phases $e^{-i\beta _i}$ and $e^{+i\beta _i}$ in the expressions for  $\bar{w}_{{q}A\, i}$ and $\bar{w}_{\dot{q}A\, i}$ of 10D case. These are needed to make charged the complex null-vectors $U_{I i}$,
$\bar{U}_{I i}$ which are related with reference internal vector  frame by
\begin{eqnarray}\label{UIi=Ui}
{U}_{I i}=e^{-2i\beta_i}{U}_{J }{\cal O}^{JI}_{i}\; , \qquad \bar{U}_{I i}=e^{+2i\beta_i}\bar{U}_{J }{\cal O}^{JI}_{i}\; ,\qquad
\end{eqnarray}
where ${\cal O}^{JI}_{i}$ is $SO(D-2)$ matrix related to ${\cal O}_{{q}{p} i}$ and  ${\cal O}_{\dot{p}\dot{q}i}$ from (\ref{bwi=bwOU}) and (\ref{bwid=bwOU})  by (\ref{gIO=OgO}).

Generically ${\cal O}_{{q}{p}\, i}$ and ${\cal O}_{\dot{q}\dot{p}\, i} $
in  (\ref{bwi=bwOU}) and  (\ref{bwid=bwOU}) can be different from the matrices denoted by the same  symbols in
(\ref{v-j=+Kv-}) and  (\ref{v+j=+Kv-}). We however impose the condition that they are the same.

Actually this implies that we do not have $n$ independent sets of internal harmonics,
but only one reference internal frame, and that the derivatives with respect to $j$-th internal harmonics
does not live inert its $i$-th cousin, for example
  \begin{eqnarray}
 \label{DchJbwd=}
D_j^{\check{J}}\bar{w}_{\dot{q} A\, i}= - \frac i 2 e^{\beta_{ij}}{\cal U}^\dagger{}_{A ij}^{\; B}\sigma^{\check{J}}_{BC}
{w}_{\dot{q}j}^{\; C}\equiv  e^{\beta_{ij}}{\cal U}^\dagger{}_{A ij}^{\; B}D_j^{\check{J}}\bar{w}_{\dot{q} B\, j}
\; , \qquad \beta_{ij}:= \beta_{i}-\beta_{j}\; , \qquad
\end{eqnarray}
but
\begin{eqnarray}
 \label{DchJbw=0}
D_j^{\check{J}}\bar{w}_{{q} A\, i}= 0\; , \qquad D_j^{\check{J}}{w}_{\dot{q}\, i}^A= 0\; , \qquad
\end{eqnarray}
and
\begin{eqnarray}
 \label{DchJbw=0}
D_j^{\check{J}}U_{I\, i}= 0\; . \qquad
\end{eqnarray}
This implies that the analytic superamplitude of D=10 SYM obeying (\ref{DchJcA=0})
 have to be constructed with the use of
$\bar{w}_{{q} A\, i}, {w}_{\dot{q}\, i}^A$ and $U_{I\, i}$ variables only. Similarly, the  analytic  11D superamplitudes
(\ref{DchJcA=011}) are  constructed with the use of
$\bar{w}_{{q} A\, i}$ and $U_{I\, i}$ variables.

To motivate this identification, let  us recall that the only role of the internal harmonics is to split the
real fermionic variables $\theta^-_{qi}$ on a pair of complex conjugate
$\eta^-_{Ai}$ and $\bar{\eta}{}^{-A}_i$ thus introducing a complex structure
(see discussion in sec. 4.6). Our choice implies that we induce all the  complex structures, for all $i=1,...,n$, from a single complex structure. This latter  is introduced with the reference internal frame $(\bar{w}_{qA}, w_q^A)$ which serves  as a compensator for $Spin(D-2)$ gauge symmetry of the reference spinor frame.

\subsection{Complex spinor frames  and reference complex spinor frame}

The identification of all the sets of internal harmonics through (\ref{bwi=bwOU}) and (\ref{bwid=bwOU}) automatically implies that the $SO(D-2)$ symmetry transformations of the reference spinor frame acts also on the reference internal frame. This allows to introduce a
complex reference spinor frame ({\it cf.} (\ref{v-A:==}))
\begin{eqnarray}\label{v-A:=}
 v_{\alpha A}^{-}:= v_{\alpha q}^{-} \bar{w}_{qA}\; , \qquad \bar{v}{}_{\alpha}^{-A}:= v_{\alpha {p}}^{-} {w}_{{p}}^{\; A} \; , \qquad v_{\alpha A}^{+}:= v_{\alpha \dot{p}}^{+} \bar{w}_{\dot{p}A}\; , \qquad  \bar{v}{}_{\alpha}^{+A}:= v_{\alpha \dot{p}}^{+} {w}_{\dot{p}}^{\; A} \; ,  \qquad \\
\label{v-A:=-1}   v_{ A}^{-\alpha}:= v_{\dot{q}}^{-\alpha} \bar{w}_{\dot{q}A} , \qquad \bar{v}{}^{-A\alpha }:= v_{\dot{q}}^{-\alpha} {w}_{\dot{q}}^{\; A} , \quad v_{A}^{+\alpha }:= v_{q}^{+\alpha } \bar{w}_{qA}, \qquad \bar{v}{}^{+A\alpha}:= v_{{q}}^{+\alpha} {w}_{{q}}^{\; A} \qquad
\end{eqnarray}
and to express the complex spinor harmonics
\begin{eqnarray}\label{v-Ai:=}
v_{\alpha A\, i}^{-}:= v_{\alpha q\, i}^{-} \bar{w}_{{q}A\, i}\; , \qquad
\bar{v}_{\alpha  i}^{-A}:= v_{\alpha q i}^{-}  {w}_{{q} i}^{\; A} \; , \qquad {v}_{\alpha Ai}^{\; +}:= v_{\alpha \dot{q} i}^{\; +}\bar{w}_{\dot{q}A i} \; ,\qquad \bar{v}_{\alpha i}^{\; +A}:= v_{\alpha \dot{q} i}^{\; +}{w}_{\dot{q} i}^{\; A} \; ,\qquad
 \\ \label{v-Ai-1:=}
v_{ A i}^{-\alpha}:= v_{ \dot{q} i}^{-\alpha} \bar{w}_{\dot{q}  A i} \; , \qquad  \bar{v}{}_{ i}^{-A\alpha\,}:= v_{\dot{q}\, i}^{-\alpha } {w}_{\dot{q}i}^{\; A}\; , \qquad  v_{ A i}^{+\alpha } := v_{{q} i}^{+\alpha} \bar{w}_{{q} i A }\; , \qquad  \bar{v}_{i }^{+\alpha A} := v_{ {q} i}^{ +\alpha} {w}_{{q} i}^{ A }\;  \qquad
\end{eqnarray}
in terms of that.

In particular, one finds
\begin{eqnarray}\label{v-Ai=}
v_{\alpha A i}^{-}&=&
e^{-\alpha_i-i\beta_i} {\cal U}_{Ai}^{\dagger\, B}\left(v_{\alpha B}^{-} + \frac 1 2 K^{=I}_i U_I v_{\alpha B }^{+} +\frac i 2 K^{=I}_i U_I^{\check{J}} \sigma^{\check{J}}_{BC} \bar{v}{}_{\alpha }^{+C}\right)\; , \qquad
\\ \label{bv-Ai-1=}
\bar{v}_{i}^{-\alpha A}&=&
e^{-\alpha_i-i\beta_i}\left(\bar{v}^{-\alpha B}-  \frac 1 2 K^{=I}_i {U}_I\bar{v}^{+\alpha B} - \frac i 2 K^{=I}_i U_I^{\check{J}} v_{ C }^{+\alpha}\tilde{\sigma}{}^{\check{J} CB} \right)\,  {\cal U}_{Bi}^{\; A}\; , \qquad
\end{eqnarray}
and
\begin{eqnarray}\label{bv-Ai=} \bar{v}{}_{\alpha  i}^{-A}=
e^{-\alpha_i+i\beta_i} \left(v_{\alpha}^{-B} + \frac 1 2 K^{=I}_i \bar{U}_I v_{\alpha }^{+B} +\frac i 2 K^{=I}_i U_I^{\check{J}} \tilde{\sigma}{}^{\check{J} BC}\bar{v}{}_{\alpha C }^{+}\right)\; {\cal U}_{Bi}^{\; A}\; , \qquad \\
 \label{v-Ai-1=}
v_{ A i}^{-\alpha}=
e^{-\alpha_i+i\beta_i} {\cal U}_{Ai}^{\dagger\, B}\left(v_{ B}^{-\alpha }-  \frac 1 2 K^{=I}_i \bar{U}_I v_{ B}^{+\alpha } +\frac i 2 K^{=I}_i U_I^{\check{J}} \sigma^{\check{J}}_{BC} \bar{v}{}^{ +C\alpha}\right)\; . \qquad
\end{eqnarray}

The complex spinor harmonics
(\ref{v-A:=}) and (\ref{v-A:=-1})
obey
\begin{eqnarray}\label{v-Av-1B=}
 v_{ A}^{-\alpha} v_{\alpha  }^{-B}= 0\; , \qquad   v_{ A}^{+\alpha} v_{\alpha  }^{+B}= 0 , \qquad v^{\pm A\alpha} v_{\alpha  }^{\pm B}= 0\; , \qquad  v_{A}^{\pm\alpha } v^{\; \pm}_{\alpha  B}= 0 \; , \qquad \nonumber \\
 v_{A}^{+\alpha } v_{\alpha  }^{-B}= \delta_A{}^B \; , \qquad  v_{A}^{-\alpha } v_{\alpha  }^{+B}= \delta_A{}^B\; . \qquad
\end{eqnarray}
The  product of harmonics from different frames, say $i$-th and $j$-th,  can be calculated using  (\ref{v-Ai=}), (\ref{bv-Ai=}) and (\ref{v-Av-1B=}).
Clearly for  $j=i$ the relation of the form of (\ref{v-Av-1B=}) are reproduced for $i$-th set of complex spinor  harmonics.

In particular, Eqs. (\ref{v-Ai=}) and (\ref{bv-Ai-1=}) imply
\begin{eqnarray}\label{bv-Biv-Aj=}
<i^{-B} j^-_A>:= \bar{v}_{i}^{-\alpha B}v_{\alpha A j}^{-}&=&  {\cal U}_{E i}^{\;\, B} e^{-\alpha_i-i\beta_i} \frac 1 2 K^{=I}_{ji}U_I  \;     {\cal U}_{A j}^{\dagger\, E} e^{-\alpha_j-i\beta_j}\;  \qquad   \nonumber \\ &=&   \frac 1 2 K^{=I}_{ji}U_I  \;     {\cal U}_{A ji}^{\dagger\, B} e^{-\alpha_i-\alpha_j-i\beta_i-i\beta_j}\; . \qquad \end{eqnarray}
The expression in the first line of (\ref{bv-Biv-Aj=}) is convenient to calculate products of bracket matrices, while the second is more compact due to the use of notation
 \begin{eqnarray}\label{cUij:=} && {}\qquad {\cal U}_{Aji}^{\dagger\, B}:=  \, {\cal U}_{Aj}^{\dagger\, C}\,   {\cal U}_{Ci}^{\; B}\, =   {\cal U}_{Aij}^{\; B}\,
\; .  \qquad
\end{eqnarray}
When deriving Eqs. (\ref{v-Ai=})-- (\ref{bv-Biv-Aj=}) the following consequences   of (\ref{wbw+cc=1}), (\ref{Ug86=wg6w+cc}) and
 (\ref{Ug8=bww}) are useful
\begin{eqnarray}\label{wgI=}
\bar{w}_{{q}A}\gamma^I_{q\dot{p}}= \bar{w}_{\dot{p}A}  U_I+i \sigma^{\check{J}}_{AB}  w_{\dot{p}}^{B} U_I^{\check{J}}\; , \qquad
w_q^{A} \gamma^I_{q\dot{p}} =  w_{\dot{p}}^A \bar{U}_I+  i \tilde{\sigma}{}^{\check{J}AB } \bar{w}_{\dot{p}B} U_I^{\check{J}}\, , \;
\\ \label{gammaIbw=wU}
\gamma^I_{q\dot{p}} {w}_{\dot{p}}^{\; A} = U_I w_q^{\; A}  + i \bar{w}_{qB}\tilde{\sigma}{}^{\check{J}BA } U_I^{\check{J}} \; , \qquad
 \gamma^I_{q\dot{p}} \bar{w}_{\dot{p}A}= \bar{w}_{{q}A}  \bar{U}_I+  i{w}_{q}^{\; B}{\sigma}{}^{\check{J}}_{BA } U_I^{\check{J}} \; .
\end{eqnarray}

One can also calculate the expressions for $<i_B^{-} j^{-A}>:= {v}_{ Bi}^{- \alpha }\bar{v}_{\alpha j}^{-A}$. However, in our perspective the contraction (\ref{bv-Biv-Aj=}) is much more interesting as far as it obeys
\begin{eqnarray}\label{DchJbv-Biv-Aj=0}
D_{l}^{\check{J}}  <i^{-B} j^-_A>= 0 \; , \qquad \forall l=1,...,n
\; .  \qquad
\end{eqnarray}

The expressions for complimentary harmonics in terms of complex reference  spinor frame simplify essentially in the  gauge
(\ref{K++I=0}),  $K^{\# I}_i=0$,  where (\ref{v+j=Ov+}) and (\ref{bwi=bwOU}), (\ref{bwid=bwOU}) result in
\begin{eqnarray}\label{v+Ai=U+v}
v_{\alpha A i}^{\; +}= {\cal U}_{Ai}^{\dagger\, B} v_{\alpha  B}^{\; +} e^{\alpha_{i}+i\beta_{i}} \; ,  \qquad
\bar{v}_{i}^{+A \alpha }  = \bar{v}^{+B \alpha }{\cal U}_{Bi}^{\; A} e^{\alpha_{i}+i\beta_{i}} \; ,  \qquad
\end{eqnarray}
and
\begin{eqnarray}\label{v+Ai=v+U}
\bar{v}_{\alpha i}^{\; +A} = \bar{v}_{\alpha  }^{\, +B} {\cal U}_{Bi}^{\; A}  e^{\alpha_{i}-i\beta_{i}} \; ,  \qquad
{v}_{i A}^{+\alpha }  = {\cal U}_{Ai}^{\dagger\, B}   {v}_B^{+\alpha } e^{\alpha_{i}-i\beta_{i}} \; .  \qquad
\end{eqnarray}
This allows to find
\begin{eqnarray}\label{i+Bj-A=}
&& <i^{+B}j^-_A>:= \bar{v}_{i}^{+\alpha B}v_{\alpha A j}^{-}= \; {\cal U}_{Aji}^{\dagger\, B} e^{-\alpha_{ji}-i\beta_{ji}}\; ,  \qquad \\
\label{i-Bj+A=} && <i^{-B}j^+_A>:= \bar{v}_{i}^{-\alpha B}v_{\alpha A j}^{+}= \; {\cal U}_{Aji}^{\dagger\, B} e^{+\alpha_{ji}+i\beta_{ji}}=\; {\cal U}_{Aij}^{\;\, B} e^{-\alpha_{ij}-i\beta_{ij}}\; .  \qquad
\end{eqnarray}
Let us stress that these are gauge fixed expressions: when $K^{\# I}\not=0$ the {\it r.h.s}-s will acquire the contributions proportional to
(\ref{v-iv-j=}).

Using the bridges $e^{\alpha_i}$, $e^{i\beta_i}$ and $ {\cal U}_{A\, i}^{\; B}$ we can transform  the complex fermionic variable $\eta^-_{Ai}$  carrying $SU(4)_i$ index,
$U(1)_i$ charge and $SO(1,1)_i$ weight to
\begin{eqnarray}\label{teta:=Ueta}
\tilde{\eta}{}^-_{A i}:=
 e^{\alpha_i+i\beta_i} {\cal U}_{A\, i}^{\; B}\;
{\eta}{}^-_{B i}
\; , \qquad
\end{eqnarray}
which is inert under $SU(4)_i \otimes U(1)_i\otimes SO(1,1)_i$ but transforms nontrivially under the gauge symmetry
$SU(4) \otimes U(1)\otimes SO(1,1)$ of the reference complex spinor frame.  The advantage of such variables is that
$\tilde{\eta}{}^-_{A i} + \tilde{\eta}{}^-_{A j}$ is covariant for any values of $i$ and $j$.
The expressions in the {\it r.h.s}-s of Eqs. (\ref{i+Bj-A=}) and (\ref{i-Bj+A=}), as well as of their counterparts  with $j=0$, corresponding to the reference complex spinor frame,
\begin{eqnarray}\label{i+B0-A=}
&& <i^{+B}\; .^-_A> := \bar{v}_{i}^{+\alpha B}v_{\alpha A }^{-}= \; {\cal U}_{Ai}^{\; B} e^{\alpha_{i}+i\beta_{i}}\; ,  \qquad
 <i^{-B}\; .^+_A> := \bar{v}_{i}^{-\alpha B}v_{\alpha A}^{+}= \; {\cal U}_{Ai}^{\;\, B} e^{-\alpha_{i}-i\beta_{i}}\; ,  \qquad
\\
\label{0+Bj-A=}
&& <.^{+B}j^-_A>:= \bar{v}^{+\alpha B}v_{\alpha A j}^{-}= \; {\cal U}_{Aj}^{\dagger\, B} e^{-\alpha_{j}-i\beta_{j}}\; ,  \qquad  <.^{-B}j^+_A>:= \bar{v}^{-\alpha B}v_{\alpha A j}^{+}= \; {\cal U}_{Aj}^{\dagger\, B} e^{+\alpha_{j}+i\beta_{j}}\; .  \qquad
\end{eqnarray}
 can be used as covariant counterparts of the above $SU(4)_i \otimes U(1)_i\otimes SO(1,1)_i$ bridges. They will be useful  below in the discussion on 3-point superamplitude of 10D SYM.

In particular, it will be important that
\begin{eqnarray}\label{i+0-0-j+=}
&& <i^{+B}\; .^-_C>  <.^{-C}i^+_A>= \; \delta_{A}^{\; B} e^{2\alpha_{i}+2i\beta_{i}}
\; ,  \qquad  < .^{-B}i^+_C><i^{+C}\; .^-_A> = \; \delta_{A}^{\; B} e^{+2\alpha_{i}+2i\beta_{i}}\; . \qquad
\\
\label{i-0+0+j-=}
&& <i^{-B}\; .^+_C>  <.^{+C}i^-_A>= \; \delta_{A}^{\; B} e^{-2\alpha_{i}-2i\beta_{i}}
\; ,  \qquad  < .^{+B}i^-_C><i^{-C}\; .^+_A> = \; \delta_{A}^{\; B} e^{-2\alpha_{i}-2i\beta_{i}}\; ,\qquad \end{eqnarray}
and
\begin{eqnarray}\label{deti-0+0=}
&& \det <i^{-B}\; .^+_A>= e^{-4\alpha_{i}-4i\beta_{i}}\; , \qquad
\det <i^{+B}\; .^-_A>= e^{4\alpha_{i}+4i\beta_{i}}\; , \qquad
\end{eqnarray}
represent the scale and phase factors corresponding to i-th particle.

\section{3-point analytic superamplitudes in 10D and 11D }

\subsection{Three particle kinematics and supermomentum}
\label{3p=K}

Let us study 3-particle kinematics in the vector  frame formalism.
With (\ref{kia=ru=}) we can write the  momentum conservation as
\begin{eqnarray}\label{P1-3=0}
 \rho^{\#}_1{u^{=a}_{1}}+ \rho^{\#}_2u^{=a}_{2}+ \rho^{\#}_3u^{=a}_{3}=0\; .
 \qquad
\end{eqnarray}
Then, using (\ref{u--=KuI=11G})--(\ref{u++=KuiG11}) we split (\ref{P1-3=0}) into
\begin{eqnarray}\label{P3r=0}
 && \rho^{\#}_1 +{\rho}_2^{\# } + {\rho}_3^{\# }=0 \; ,\qquad  \\
 \label{P3rK=0}
 && \rho^{\#}_1  {K}{}^{=I}_{1}  + \rho^{\#}_2 {K}{}^{=I}_{2} + \rho^{\#}_3 {K}{}^{=I}_{3} =0 \; ,\qquad \\
 \label{P3rK2=0}
 &&  \rho^{\#}_1  ({\overrightarrow{K}}{}_{1} )^2 + \rho^{\#}_2 (\overrightarrow{K}{}_{2})^2 +
 \rho^{\#}_3 (\overrightarrow{K}{}_{3})^2=0\; .  \qquad
\end{eqnarray}
Eq. (\ref{P3r=0}) makes  (\ref{P3rK=0}) equivalent to
\begin{eqnarray}
 \label{rK31=}
\frac { {K}{}^{=I}_{32}} {\rho^{\#}_1} =\frac { {K}{}^{=I}_{13}}  {\rho^{\#}_2} =
\frac {{K}{}^{=I}_{21}} {\rho^{\#}_3} \; , \qquad
\end{eqnarray}
 where ${K}{}^{=I}_{ji}\equiv {K}{}^{=I}_{[ji]}={K}{}^{=I}_{j}-{K}{}^{=I}_{i}$ (\ref{Kij=Ki-Kj}).
Using (\ref{rK31=}) and (\ref{P3r=0})   we find that (\ref{P3rK2=0}) implies
\begin{eqnarray}
 \label{K12K12=0}
 && ({\overrightarrow{K}}{}_{21} )^2  =0 \qquad \Rightarrow \qquad  ({\overrightarrow{K}}{}_{13} )^2  =0\; ,
 \qquad  ({\overrightarrow{K}}{}_{32} )^2  =0\; . \qquad
\end{eqnarray}
The solution of Eqs. (\ref{K12K12=0}) for real vectors ${K}{}^{=I}_{ji}$ are trivial. Thus a nontrivial  on-shell 3-particle amplitude can be defined only for complexified ${K}{}^{=I}_{ji}$ which implies that the light-like momenta $k_i^a$ of the scattered particles are complex.

The general solution of the momentum conservation conditions can be written in terms of say
 ${K}{}^{=I}_{1}$ and complex null vector ${\cal K}^I$ as
\begin{eqnarray}
 \label{K2=K1+q}
 && {K}{}^{=I}_{2}={K}{}^{=I}_{1}+ {\cal K}^I, \qquad K{}^{=I}_{3}= {K}{}^{=I}_{1} +
 {\cal K}^I \frac  {\rho^{\#}_2 } { \rho^{\#}_1 + \rho^{\#}_2 }, \qquad {\cal K}^I{\cal K}^I=0 . \qquad
\end{eqnarray}

Notice that, to make the above equations valid for arbitrary parametrization
(\ref{u--=KuI=11}), it is sufficient just to rescale the scalar densities as in (\ref{trho:=})
\begin{eqnarray}
 \label{r--tr}
 && \rho^{\#}_i \;\longrightarrow \; \tilde{\rho}{}^{\#}_i= e^{-2\alpha_{i}}\rho^{\#}_i\; . \qquad
\end{eqnarray}
In particular,
\begin{eqnarray}
 \label{Kij=q}
 &&  \frac {{K}{}^{=I}_{32}}{\tilde{\rho}^\#_1}= \frac {{K}{}^{=I}_{21}}{\tilde{\rho}^\#_3}= \frac {{K}{}^{=I}_{13}}{\tilde{\rho}^\#_2} =: {\cal K}^{==I}\; , \qquad  {\cal K}^{==I}{\cal K}^{==I}=0 \;  \qquad
\end{eqnarray}
are valid for a generic parametrization of the spinor harmonics.

\subsection{3-points analytical superamplitudes in 10D SYM and 11D SUGRA}
\label{3cA=sec}

A suggestion about the structure of 10D and 11D tree superamplitudes may be gained from the observation that, when the external momenta belong to a 4d subspace of the D-dimensional space, they should reproduce the known answer
for 4-dimensional tree superamplitudes of ${\cal N}=4$ SYM and ${\cal N}=8$ SUGRA, respectively. Due to the momentum conservation, this is always the case for a three point amplitude and superamplitude.

In this section we find the gauge fixed form  of the 10D 3-point superamplitude in the gauge (\ref{K++I=0}) and also present its 11D cousin.
We also describe the first stages in search for covariant form of the three point superamplitudes, which,  although have not allowed to succeed yet, might be suggestive for further study.

\subsubsection{ 3-points analytical superamplitude of 10D SYM. Gauge fixed form }

We chose as  D=4 reference point the anti-MHV superamplitude of ${\cal N}=4$ SYM (\ref{cA3g=bMHV}).
As we show in Appendix A, using an explicit parametrization of 4D helicity spinors in terms of reference spinor frame we can write it in the following form  (see (\ref{cA3g=K21}))
\begin{eqnarray}\label{cA3g=K21d4}
 && {\cal A}^{\overline{{\rm MHV}}}(1,2,3)= {\bb K}{}^{==}\; e^{-2i(\beta_1+\beta_2+\beta_3)}\, \delta^4 \left( \tilde{\rho}{}^{\#}_{1} \tilde{\eta}^-_{1A}   + \tilde{\rho}{}^{\#}_2 \tilde{\eta}^-_{2A}  + \tilde{\rho}{}^{\#}_3\tilde{\eta}^-_{3A}  \right)\; ,
 \qquad \nonumber \\  &&{} \qquad =
 -\frac { {\bb K}{}^=_{21}} { (\tilde{\rho}{}^{\#}_{1} +\tilde{\rho}{}^{\#}_{2} )} e^{-2i(\beta_1+\beta_2+\beta_3)}\; \delta^4 \left( \tilde{\rho}{}^{\#}_{1} \tilde{\eta}{}^-_{[13]A}   + \tilde{\rho}{}^{\#}_2 \tilde{\eta}^-_{[23]A}   \right)\; .
 \qquad
  \end{eqnarray}
Here $\tilde{\eta}{}^-_{A i}= {\eta}{}_{Ai} /\sqrt{\tilde{\rho}{}^{\#}_i}$ and $\tilde{\rho}{}^{\#}_i$
 are D=4 counterparts of the rescaled 10D variables (\ref{teta:=Ueta}) and (\ref{r--tr}) (see (\ref{tileta=})
 and (\ref{tilrho=}) in Appendix A),
${\bb K}^{==}= {\bb K}_{21}^{=}/\rho^\#_3$ (see (\ref{bbK==:=})) and  ${\bb K}_{21}^{=}$ is a complex number, which can be associated through ${\bb K}_{21}^{=}={ K}_{21}^{=1}+i{K}_{21}^{=2}$ with a real 2-component vector $({ K}_{21}^{=1}, {K}_{21}^{=2})$, the D=4 counterpart of the generic
  $(D-2)$-vector ${K}_{21}^{=I}$ in (\ref{Kij=Ki-Kj}).
 It is tempting to identify  ${\bb K}_{21}^{=}$ with ${K}_{21}^{=I} {U}_I$ of the previous section:
\begin{eqnarray}\label{bbK=KbU}
 ^{^{D=4}}\; {\bb K}_{21}^{=} \quad \longleftrightarrow \quad  {K}_{21}^{=I} {U}_I\; {}^{^{D=10 \; (and \; D=11)}} .\qquad
  \end{eqnarray}

The argument of the fermionic delta function in (\ref{cA3g=K21d4})  also has the straightforward 10D counterpart
$\tilde{\rho}{}^{\#}_1\tilde{\eta}{}^-_{A 1} + \tilde{\rho}{}^{\#}_2\tilde{\eta}{}^-_{A 2}+  \tilde{\rho}{}^{\#}_3\tilde{\eta}{}^-_{A 3}$ where $\tilde{\eta}{}^-_{A i}$ and $\tilde{\rho}{}^{\#}_i$ are defined in (\ref{teta:=Ueta}) and (\ref{r--tr})
in such a way that all of them carry indices, charges and weights with respect to the same  $SO(1,1) \otimes SU(4) \otimes U(1)$ acting on the reference complex frame,
\begin{eqnarray}\label{teta:=2}
\tilde{\eta}{}^-_{A i}:=
 e^{\alpha_i+i\beta_i} {\cal U}_{A\, i}^{\; B}\;
{\eta}{}^-_{B i}
\; , \qquad  \tilde{\rho}{}^{\#}_i= e^{-2\alpha_{i}}\rho^{\#}_i\; . \qquad
\end{eqnarray}


Thus, the straightforward generalization of (\ref{cA3g=K21d4}) to the case of $D=10$ SYM theory reads
\begin{eqnarray}\label{cA3g=10DK}
 {\cal A}_3^{D=10\; SYM} &=&
- \frac {{K}{}^{= I}_{21}U_I } {2 (\tilde{\rho}{}^{\#}_{1} +\tilde{\rho}{}^{\#}_{2} )} \; e^{-2i(\beta_1+\beta_2+\beta_3)}\,  \delta^4 \left( \tilde{\rho}{}^{\#}_{1} \tilde{\eta}{}^-_{[13]A}   + \tilde{\rho}{}^{\#}_2 \tilde{\eta}^-_{[23]A}   \right)\; \nonumber \\
&& = \frac 1 2  {{\cal K}{}^{==I}U_I} \; e^{-2i(\beta_1+\beta_2+\beta_3)}\, \delta^4 \left( \tilde{\rho}{}^{\#}_{1} \tilde{\eta}^-_{1A}   + \tilde{\rho}{}^{\#}_2 \tilde{\eta}^-_{2A}  + \tilde{\rho}{}^{\#}_3\tilde{\eta}^-_{3A}  \right)\; ,
 \qquad
  \end{eqnarray}
where the complex null-vector ${\cal K}{}^{==I}$ is defined in (\ref{Kij=q}).

The multiplier $e^{-2i(\beta_1+\beta_2+\beta_3)} $  makes the superamplitude invariant under
$U(1)$ symmetry acting on the reference internal frame variables and supplies it instead  with  charges
$+1$ with respect to all $U(1)_i$ groups, $i=1,2,3$,  related to scattered particles. All other variables  in (\ref{cA3g=10DK}) are redefined in such a way that they are inert under $\prod\limits_i^3 [SO(D-2)_i\otimes SO(1,1)_i\otimes U(1)_i\otimes SO(D-4)_i]$ and are transformed only by
$SO(D-2)\otimes SO(1,1)\otimes U(1)\otimes SO(D-4)$ acting on the reference spinor frame and reference internal frame.

\subsubsection{ Searching for a gauge covariant form of the  3-points superamplitude}

Let us try to search for a covariant expression for amplitude which, upon gauge fixing, reproduce  (\ref{cA3g=10DK}).
Again, a guideline can be found in 4D expression (\ref{cA3g=bMHV}). Counterparts of $<ij>$ blocks are given by
the matrices $\sqrt{\rho^\#_i\rho^\#_j }<i^{-B} j^-_A>$ with $<i^{-B} j^-_A>$ defined in
(\ref{bv-Biv-Aj=}) so that a possible 10D cousin of the denominator in  (\ref{cA3g=bMHV}) is given by  the trace of the product of three such matrices,
\begin{eqnarray}\label{1A2B-1A=}
& \rho^\#_1\rho^\#_2 \rho^\#_3 & <1^{-A} 2^-_B><2^{-B} 3^-_C><3^{-C} 1^-_A>= \qquad \nonumber \\
&=& \frac 1 {2^3} \, \tilde{\rho}{}^{\#}_1\tilde{\rho}{}^{\#}_2\tilde{\rho}{}^{\#}_3 \;
K^{=I}_{21}U_I\,K^{=J}_{32}U_J\, K^{=K}_{21}U_K\, e^{-2i(\beta_1+\beta_2+\beta_3)}
\; \nonumber \\
&& =  \frac 1 {2^3} \, (\tilde{\rho}{}^{\#}_1\tilde{\rho}{}^{\#}_2\tilde{\rho}{}^{\#}_3)^2 \;
({\cal K}^{==I} U_I)^3\, e^{-2i(\beta_1+\beta_2+\beta_3)}
\; .  \qquad
\end{eqnarray}

The next problem is to search for a counterpart of $\eta_{A1}<23>$ expression in the argument of fermionic delta function in
 (\ref{cA3g=bMHV}). Here the straightforward generalization $\propto \eta^-_{B1}<2^{-B}3^-_A>$ does not work: it is not covariant under
 $SU(4)_1$ and $SU(4)_2$. The covariance may be restored by using the matrices (\ref{i-Bj+A=}): the matrix
 \begin{eqnarray}\label{f122-3-=cov}
\eta^-_{C1} <1^{-C}2^+_B><2^{-B}3^-_A> = \tilde{\eta}^-_{B1} e^{-2\alpha_1 -2i\beta_1} {\cal U}_{A3}^{\dagger\,  B} e^{-\alpha_3 -i\beta_3}
\;   \qquad
\end{eqnarray}
is transformed in
$(1,1,4)$ of $SU(4)_1\otimes SU(4)_2 \otimes SU(4)_3$. However its nontrivial weights $(-2,0, -1)$ and charges $(+1,0,+1/2)$, indicated by
multipliers $e^{-2\alpha_1 -2i\beta_1}$ and $e^{-\alpha_3 -i\beta_3}$, do not allow  to sum it with its $\eta^-_2$ and $\eta^-_3$ counterparts without breaking the gauge symmetries.
To compensate the above multipliers one can use the matrices (\ref{i+0-0-j+=}).
   In such a way we arrive at the expression
   \begin{eqnarray}\label{eta1-23-=cov}
 \ll \eta^-_{1}2^{-}3^-\gg_A &:=&  \nonumber \\  &:=&  \eta^-_{B1} <1^{-B}2^+_C><2^{-C}3^-_D> <3^{+D}.^-_E><.^{-E}1^+_F><1^{+F}.^-_A> =  \nonumber \\ &&{} \qquad =  \frac 1 2 \tilde{\eta}{}^{-}_{i A} {K}_{32}^{=I}U_I  =
\frac 1 2 \tilde{\rho}{}^{\#}_1\tilde{\eta}{}^{-}_i {\cal K}^{==I}U_I
\; , \qquad \end{eqnarray}
which is invariant under $\prod\limits_{i=1}^3 SU(4)_i\otimes U(1)_i\otimes SO(1,1)_i$ and
carries the nontrivial representations of only $SU(4)\otimes U(1)\otimes SO(1,1)$ group acting
on the reference complex spinor frame. Then, as $\ll\eta^-_{i}j^{-}k^-\gg_A$ with $i,j,k$ given by
arbitrary permutation of $123$ has the same transformation properties, we can sum them and
write the 10D counterpart of the fermionic delta function in (\ref{cA3g=bMHV}),
\begin{eqnarray}\label{delta4=10D}
&& \delta^4\left((\rho^{\#}_1\rho^{\#}_2\rho^{\#}_3)^{1/2}\left( \ll \eta^-_{1}2^{-}3^-\gg_A +\ll \eta^-_{2}3^{-}1^-\gg_A +\ll \eta^-_{3}1^{-}2^-\gg_A \right)\right)=
\; \qquad  \nonumber \\ &&{} \qquad  = (\rho^{\#}_1\rho^{\#}_2\rho^{\#}_3)^{2}\delta^4\left(\ll \eta^-_{1}2^{-}3^-\gg_A +\ll \eta^-_{2}3^{-}1^-\gg_A +\ll \eta^-_{3}1^{-}2^-\gg_A \right)=
\;  \qquad \nonumber  \\
&&{} \qquad = \, \frac {({\cal K}{}^{==I}U_I)^4 } {2^4}
  ({\rho}{}^{\#}_{1} {\rho}{}^{\#}_{2} {\rho}{}^{\#}_{3})^2   \;   \delta^4 \left( \tilde{\rho}{}^{\#}_{1} \tilde{\eta}^-_{1A}   + \tilde{\rho}{}^{\#}_2 \tilde{\eta}^-_{2A}  + \tilde{\rho}{}^{\#}_3\tilde{\eta}^-_{3A}  \right)=\;  \qquad\nonumber  \\
 &&{} \qquad = \, \frac {({\cal K}{}^{==I}U_I)^4 } {2^4}
  (\tilde{\rho}{}^{\#}_{1} \tilde{\rho}{}^{\#}_{2} \tilde{\rho}{}^{\#}_{3})^2   \; \frac {e^{-4i(\beta_1+\beta_2+\beta_3)} } {\prod_{i=1}^{3} \det <i^{-B}.^+_A>}\;    \delta^4 \left( \tilde{\rho}{}^{\#}_{1} \tilde{\eta}^-_{1A}   + \tilde{\rho}{}^{\#}_2 \tilde{\eta}^-_{2A}  + \tilde{\rho}{}^{\#}_3\tilde{\eta}^-_{3A}  \right)\;  , \qquad \end{eqnarray}
where in the last lines we have used (\ref{eta1-23-=cov}) and (\ref{deti-0+0=}).

Then, the  covariant candidate  amplitude is given by  (\ref{delta4=10D}) divided by the product of (\ref{1A2B-1A=})
and multiplied by three determinants (\ref{deti-0+0=}),
\begin{eqnarray}\label{cA3g=10D}
&& {\cal A}_3^{D=10\; SYM} =^? \nonumber   \\  && {}\quad  \frac {\delta^4\left((\rho^{\#}_1\rho^{\#}_2\rho^{\#}_3)^{1/2}\left( \ll \eta^-_{1}2^{-}3^-\gg_A  +  \ll \eta^-_{2}3^{-}1^-\gg_A+ \ll \eta^-_{3}1^{-}2^-\gg_A
\right)\right)} {\,
 \rho^\#_1\rho^\#_2 \rho^\#_3  <1^{-A} 2^-_B><2^{-B} 3^-_C><3^{-C} 1^-_A>
 }  \prod_{i=1}^{3} \det <i^{-B}.^+_A> \nonumber   \\ {} \nonumber
 \\ \label{cA3g=10D=2} && {}{}\hspace{2cm} =
\, \rho^\#_1\rho^\#_2 \rho^\#_3 \; \delta^4\left(\ll \eta^-_{1}2^{-}3^-\gg_A +\ll \eta^-_{2}3^{-}1^-\gg_A +\ll \eta^-_{3}1^{-}2^-\gg_A \right)\;  \times \;
\nonumber \\ && {}\hspace{4cm} \times    \frac { \det <1^{-B}.^+_A>\,  \det <2^{-C}.^+_D>\,  \det <3^{-E}.^+_F>\,}
{ <1^{-A} 2^-_B><2^{-B} 3^-_C><3^{-C} 1^-_A>
  } \qquad
  \end{eqnarray}
One can easily check that in the gauge (\ref{K++I=0}) (see (\ref{v-j=-Kv-}),  (\ref{v+j=-Kv-}) with explicit
 parametrization (\ref{v-Ai=}), (\ref{bv-Ai-1=}), (\ref{v+Ai=U+v})) this expressions reduces to (\ref{cA3g=10DK}).

However, the main problem of the above covariant expression (besides that it depends explicitly on reference complex spinor
frame) is that apparently  it does not obey  (\ref{DchJcA=0}),
\begin{eqnarray}\label{DchJcA3g=10D-}
&& D_j^{\check{J}}{\cal A}_3^{D=10\; SYM \; of \;  Eq. (\ref{cA3g=10D})} (\{ \rho^\#_{i}, v_{\alpha qi}^{\; -}; w_i, \bar{w}_i; \eta_{A i}\})\not= 0  \; .
\qquad
  \end{eqnarray}
Indeed,  it is constructed with the use of blocks (\ref{i-Bj+A=}) and (\ref{i+B0-A=}) and, if we consider the complex spinor frames as composed from spinor and internal harmonics as in (\ref{v-Ai:=})
and use (\ref{DchJbwd=}), we find, for instance,
\begin{eqnarray}\label{DchJi-B0+A}
&& D^{\check{J}} <i^{-B}j^+_A> = - \frac i 2  e^{2i\beta_j} ({\cal U}^{\dagger}_j\sigma^{\check{J}}{\cal U}^{\dagger T}_j)_{AC}<i^{-B}j^+_C>
\not=0\; , \nonumber \\
\qquad && D^{\check{J}} <i^{-B}.^+_A>= - \frac i 2   \sigma^{\check{J}}_{AC}<i^{-B}.^+_C> \not=0 \; ,
\qquad
  \end{eqnarray}

Thus we should find either  a  different covariant representation for the gauge fixed amplitude  (\ref{cA3g=10DK}), or a way to relax/to modify  the condition (\ref{DchJcA=0}) for the analytic superamplitudes.

Alternatively, one can use the gauge fixed form of the 3-point superamplitude as a basis of a gauge fixed superamplitude formalism. In it all the
$K_{8 i}$ symmetries acting on $i$-th spinor frame variables are gauge fixed by the conditions (\ref{K++I=0}). This gauge fixing is performed with respect to a symmetry acting  on the auxiliary variables, complementary spinor harmonics. The gauge fixed expressions  (\ref{v+Ai=U+v}) and (\ref{v+Ai=v+U}),  as well as the expressions  for the physically relevant spinor harmonics (\ref{v-Ai:=})--
(\ref{v-A:=-1}), use the reference spinor frame. This makes our gauge fixed superamplitude formalism  manifestly Lorentz covariant. Such a role of reference spinor frame is in consonance with the original idea of introducing Lorentz harmonics to covariantize the light--cone gauge \cite{Sokatchev:1985tc}.

\subsubsection{Analytical 3-point superamplitude of  $D=11$  supergravity}

Similarly, the form of 3-point ${\cal N}=8$ 4D supergravity superamplitude, which is essentially the square of the  ${\cal N}=4$ 4D SYM one (see e.g. \cite{ArkaniHamed:2008gz,Heslop:2016plj}),  suggests the following gauge fixed expression  for the basic 3-point superamplitude of 11D supergravity,
 \begin{eqnarray}\label{cA3g=11DK}
 {\cal A}_3^{D=11\; SUGRA} &=&
\left(\frac {{K}{}^{= I}_{21}U_I } {2 (\tilde{\rho}{}^{\#}_{1} +\tilde{\rho}{}^{\#}_{2} )}\right)^2 \; e^{-4i(\beta_1+\beta_2+\beta_3)}\,
\delta^8 \left( \tilde{\rho}{}^{\#}_{1} \tilde{\eta}{}^-_{[13]A}   + \tilde{\rho}{}^{\#}_2 \tilde{\eta}^-_{[23]A}   \right)\; \nonumber \\
&& = \left(\frac 1 2  {{\cal K}{}^{==I}U_I} \right)^2\; e^{-2i(\beta_1+\beta_2+\beta_3)}\, \delta^8 \left( \tilde{\rho}{}^{\#}_{1} \tilde{\eta}^-_{1A}   + \tilde{\rho}{}^{\#}_2 \tilde{\eta}^-_{2A}  + \tilde{\rho}{}^{\#}_3\tilde{\eta}^-_{3A}  \right)\; ,
 \qquad
  \end{eqnarray}

 Eq. (\ref{cA3g=11DK}) can be obtained by gauge fixing from
  \begin{eqnarray}\label{cA3g=11D}
&& 
\, (\rho^\#_1\rho^\#_2 \rho^\#_3)^2 \; \delta^8\left(\ll \eta^-_{1}2^{-}3^-\gg_A +\ll \eta^-_{2}3^{-}1^-\gg_A +\ll \eta^-_{3}1^{-}2^-\gg_A \right)\;  \times \;
\nonumber \\ && {}\hspace{2cm} \times    \left(\frac { \det <1^{-B}.^+_A>\,  \det <2^{-C}.^+_D>\,  \det <3^{-E}.^+_F>\,}
{ <1^{-A} 2^-_B><2^{-B} 3^-_C><3^{-C} 1^-_A>
  } \right)^2\; . \qquad
  \end{eqnarray}
However, as (\ref{cA3g=10D}) in the case  of 10D SYM, this expression does not obey Eq.   (\ref{DchJcA=0}), so that
we should find either the reason to relax/to modify these equations, or to search for a different covariant expression reproducing (\ref{cA3g=11DK}) upon gauge fixing.

Another  interesting possibility is to use the gauge fixed spinor frame variables, obeying (\ref{K++I=0}) for all sets of spinor harmonics. As we have already stressed above, in distinction with
 light-cone gauge,  such a gauge fixed superamplitude formalism possesses manifest Lorentz invariance and supersymmetry.  This possibility is also under study now.

\section{Conclusion and discussion}

In this paper we have constructed the basis of  the analytic superfield formalism to calculate (super)amplitudes of 10D SYM and 11D SUGRA theories. This is alternative to the constrained superamplitude formalism of \cite{Bandos:2016tsm,Bandos:2017eof}  and also to the `Clifford superfield' approach of \cite{CaronHuot:2010rj}. The fact that it has more similarities with D=4 superamplitude calculus of ${\cal N}=4$ SYM and ${\cal N}=8$ SUGRA promises to allow us to use more efficiently the D=4 suggestions for its further development. In particular, such a suggestion was used  to find the gauge fixed form of the 3-point analytic superamplitude of 10D SYM and 11D SUGRA, Eqs. (\ref{cA3g=10DK}) and (\ref{cA3g=11DK}).

We have begun by solving the equations of the constrained  on-shell superfield formalism of 10D SYM and 11D SUGRA \cite{Galperin:1992pz,Bandos:2016tsm,Bandos:2017eof}
in terms of single analytic  superfield depending holomorphically on ${\cal N}=4$ and ${\cal N}=8$ complex coordinates, respectively. These ${\cal N}$ complex coordinates $\eta^-_{A}$ are related to $2{\cal N}$ real fermionic coordinates $\theta^-_q$ of the constrained superfield formalism  by complex rectangular matrix $\bar{w}_{qA}$ ($ =(w_q^A)^*$). This and its conjugate $w_q^A=(\bar{w}_{qA})^*$ obey some constraints which allow us to consider them as homogeneous coordinates  of the coset $\frac {Spin(D-2)}{Spin(D-4)\otimes U(1)}$ and to call them internal harmonic variables.

Similarly, the constrained $n$--point superamplitudes of the $\prod\limits_{i=1}^n SO(D-2)_i$ covariant constrained superfield formalism can be expressed in terms of analytic superamplitudes which depend, besides the $n$ sets of 10D or 11D spinor helicity  variables,
also on $n$ sets $(\bar{w}_{qA\,i}, w_{q\, i}^A)$ of
$\frac {Spin(D-2)_i}{Spin(D-4)_i\otimes Spin(2)_i}$ internal harmonic variables.
The sets of 10D and 11D spinor helicity variables include Lorentz harmonics or spinor frame variables  $v_{\alpha q\, i}^{\; -}$ which, after the constraints and gauge symmetries are taken into account,  parametrize the  celestial sphere ${\bb S}^{(D-2)}$. Together with scalar  densities $\rho^{\#}_i$,  they describe the light-like momenta and the ''polarizations'' ($SO(D-2)_i$ small group
 representations) of the scattered particles. The constrained superamplitudes, which depend on these spinor helicity variables and $(2{\cal N})$-component real fermionic variables $\theta^-_{q i}$, carry indices of the small groups $SO(D-2)_i$. In contrast, the analytic superamplitudes do not carry indices but only charges $s={\cal N}/4$ of $U(1)_i$
which act on the internal frame variables $(\bar{w}_{qA\,i}, w_{q\, i}^A)$ and on the complex fermionic
$\eta^-_{A i}=\theta^-_{qi} \bar{w}_{qA\,i}$. They may be constructed from the basic constrained superamplitudes by contracting their $ SO(D-2)_i$ vector indices with complex null vectors $U_{Ii}$ constructed from bilinear combinations of $(\bar{w}_{i}, w_{ i})$.

The dependence of the analytic superamplitudes on internal harmonics is restricted by the equations in terms of harmonic covariant derivatives which reflect the fact that the original constrained superamplitudes  are
independent of $(\bar{w}_{i}, w_{i})$. Moreover, the internal harmonics  $(\bar{w}_{i}, w_{i})$ are pure gauge with respect to the
$SO(D-2)_i$ symmetry which acts also on the spinor harmonics $(v_{\alpha q\, i}^{\; -},  v_{\alpha \dot{q}\, i}^{\; +})$. We have shown that internal harmonics can be defined in such a way that analytic superamplitudes actually depend only on complex spinor harmonics $(  v_{\alpha A\, i}^{\mp}, \bar{v}{}_{\alpha \, i}^{ A \mp})$ (\ref{v-Ai:=})
parametrizing the coset
$\frac{Spin(1,D-1)}{[SO(1,1) \otimes Spin(D-4)\otimes U(1)]\subset\!\!\!\! \times K_{D-2}}$,
\begin{eqnarray}\label{c-harmV=G-H}
{}\{ (  v_{\alpha A\, i}^{\mp}, \bar{v}{}_{\alpha \, i}^{ A \mp})\} \; = \; \frac{Spin(1,D-1)}{[SO(1,1) \otimes Spin(D-4)\otimes U(1)]\subset\!\!\!\!\!\! \times K_{D-2}} \; . \qquad
  \end{eqnarray}
However, we find convenient to consider these complex spinor harmonics to be composed from the real spinor harmonics, parametrizing  the coset isomorphic to the celestial sphere ${\bb S}^{(D-2)}=
\frac{Spin(1,D-1)}{[SO(1,1) \otimes Spin(D-2)]\subset\!\!\!\! \times K_{D-2}}$ (\ref{v-i=G-H}), and the above mentioned internal harmonics $(\bar{w}_{i}, w_{i})$, in spite of that these latter are pure gauge with respect to $Spin(D-2)_i$ symmetry
(see (\ref{bwi=bwOU}) and (\ref{v-j=+Kv-})).

We have found a  parametrization of the  spinor frame  variables and of the internal frame which is especially convenient for the analysis of the analytic superamplitudes. This has allowed us to establish the correspondence of higher dimensional quantities with  basic building blocks of 4D superamplitudes and to use it  to find  the expressions for
analytic 3-point superamplitudes of D=10 SYM and D=11 SUGRA theories. These are the necessary basic ingredients for calculation of the n-point superamplitudes with the use of on-shell recurrent relations, the problem which we intend to address in a forthcoming  paper.

The first stages in this direction should include a better understanding of the structure of the 3-point analytic superamplitudes, in particular the search for its more convenient, parametrization independent form, as well as the derivation of the BCFW-type recurrent relations for the  analytic superamplitudes. These should be more closely related to the relations for D=4 superamplitudes \cite{Britto:2005fq,ArkaniHamed:2008gz} than the BCFW-type  recurrent relations for real constrained 11D and 10D  superamplitudes  presented in \cite{Bandos:2016tsm,Bandos:2017eof}.

In particular, one expects the BCFW deformations used in such recurrent relations to have an intrinsic complex structure, similar to the one in  D=4 equations of  \cite{Britto:2005fq}. As we show in Appendix B, starting from
BCFW deformations of spinor frame and fermionic variables in \cite{Bandos:2016tsm}, which are essentially real,  this is indeed the case. The resulting BCFW-like deformations of the  complex spinor frame variables  (\ref{v-Ai:=})
and of the  complex fermionic variables (\ref{anHOn-shellSSP})
\begin{eqnarray}\label{BCWF=vnfM}
\widehat{v_{\alpha {A}(n)}^{\; -}}= v_{\alpha {A}(n)}^{\; -} + z \;
v_{\alpha {A}(1)}^{\; -} \; \sqrt{{\rho^{\#}_{1}}/{\rho^{\#}_{n}}}
\; , \qquad \widehat{\bar{v}_{\alpha (n)}^{\; A -} }= \bar{v}{}_{\alpha (n)}^{\; A -}
\; , \qquad \\ \label{BCWF=v1fM} \widehat{v_{\alpha {A}(1)}^{\; -}}= v_{\alpha {A}(1)}^{\; -}
\; , \qquad  \widehat{\bar{v}_{\alpha(1)}^{\; A -} }= \bar{v}{}_{\alpha(1)}^{\; A -}
- z \;  \bar{v}{}_{\alpha(n)}^{\; A -} \;
 \sqrt{{\rho^{\#}_{n}}/{\rho^{\#}_{1}}}
\; , \qquad \end{eqnarray}
\begin{eqnarray}\label{BCWF=eta}
\widehat{\eta^-_{A\,n}}= \eta^-_{A\,n}  + z \;
\eta^-_{A\,1}\; \sqrt{{\rho^{\#}_{1}}/{\rho^{\#}_{n}}}
\; , \qquad \widehat{\eta^-_{A\,1}}= \eta^-_{A\,1}
\;   \qquad \end{eqnarray}
have the structure quite similar to that of the 4D super-BCFW deformations from \cite{ArkaniHamed:2008gz} (see  (\ref{BCFWln=4D})--(\ref{BCWF=eta4D}) in Appendix B).

\bigskip

Thus presently their exist three alternative superamplitude formalisms for 10D SYM, two of which have been also generalized for the case of 11D supergravity. These are
Clifford superfield approach of \cite{CaronHuot:2010rj}, constrained superamplitude approach  of    \cite{Bandos:2016tsm,Bandos:2017eof} and the analytic superamplitude formalism of the present paper.  As  discussed in  \cite{Bandos:2017eof}, and also briefly commented in sec. 4.6.1, the one particle counterparts of all three types of superamplitudes can be obtained by different ways of covariant quantization of 10D and 11D massless superparticles. In short, the separation point   is how to deal with the Poisson brackets of the fermionic second class constraints, (\ref{d+d+=PB}).

The formalism of \cite{CaronHuot:2010rj} and the analytic superfield approach of the present paper imply 'solving' the constraints by passing to the
Dirac brackets (\ref{t-t-=DB}) and quantizing these. In such a way we obtain  the Clifford algebra like anticommutation relation (\ref{t-t-=Cl}) for 8 (16 in D=11 case) real fermionic variables $\hat{\theta}^-_q$. To arrive at the one-particle counterpart of the superamplitudes from \cite{CaronHuot:2010rj}, one should consider the superparticle 'wavefunction' to be dependent on the whole set of Clifford algebra valued variables $\hat{\theta}^-_q$, i.e. to be a 'Clifford superfield'. In contrast, to obtain an analytic superfield as superparticle wavefunction, we need to split 8 real $\hat{\theta}^-_q$ on 4 complex $\eta^-_A$ and its complex conjugate $\bar{\eta}^{-A}$, which obey the Heisenberg-like algebra. This implies that $\bar{\eta}^{-A}$ can be considered as creation operator or complex momentum conjugate to the annihilation operator $\eta^-_A$. Then in the $\eta^-_A$--coordinate (or holomorphic) representation the superparticle quantum state vector depends on $\eta^-_A$, but not on $\bar{\eta}^{-A}$. In other words, it will  be described by analytic superfield, the one-particle counterpart of our analytic superamplitudes.

From this perspective, one can arrive at doubts in consistency of the Clifford superamplitude approach of \cite{CaronHuot:2010rj}.
Indeed, in terms of complex fermionic variables the above described appearance of an unconstrained Clifford superfield in superparticle quantization requires to allow the wavefunction to depend  on both coordinate $\eta^-_A$ and momentum $\bar{\eta}^{-A}$ variables in an arbitrary manner.
Then such a  Clifford superfield wavefunction is not allowed in quantum mechanics in its generic form  and some conditions need to be imposed to restrict its dependence on   $\bar{\eta}^{-A}$ and/or $\eta^-_A$. The analytic superfields and superamplitudes can be obtained on this way: by imposing on Clifford superfields/superamplitudes just the conditions to be independent of $\bar{\eta}^{-A}$.

The constrained superfields, the one-particle counterparts of the constrained superamplitudes, appear as a result of superparticle quantization if, instead of passing to Dirac brackets (\ref{t-t-=DB}), we realize the fermionic second class constrains as differential operators $D^+_q=
\frac {\partial}{\partial \theta^-_q}+...$ obeying the quantum counterpart (\ref{D+qD+p=I}) of (\ref{d+d+=PB}). The 'imposing' of the quantum second class constraint is then achieved by considering a $\theta^-_q$-dependent multicomponent state vectors $\Psi_Q\;$ ($=(\Psi_{\dot{q}}, W^I)$ in D=10) and  requiring them  to obey a set of linear differential equations $D^+_q\Psi_Q=\Delta_{qQP}\Psi_P$ ((\ref{D+Psi=gV}) and (\ref{D+V=gdPsi}) in D=10;  see \cite{Bandos:2017eof} for details of this procedure). The advantages of this approach is the  use of Grassmann fermionic coordinates (rather than Clifford algebra valued ones) as well as  its manifest covariance under the 'small group' SO(8) (SO(9)) symmetry. The disadvantage is that superfields and superamplitudes are subject to the above mentioned set of quite complicated equations, which have no clear counterpart in D=4 case. This makes the calculations in the constrained superamplitude framework quite involving (in comparative terms) and creates  difficulties for the (straightforward) use of the experience gained in D=4.
Also the decomposition of constrained superfields on components looks quite non-minimal: in the 10D case, 9 components of constrained superfield, all nonvanishing, are constructed of two fields describing the on-shell degrees of freedom of SYM, bosonic $w^I$ and fermionic $\psi_{\dot{q}}$, appearing already in first two terms of the decomposition.

In contrast, the components of the analytic superfields include different components of $w^I=(\phi^{(+)}, \phi^{AB}, \phi^{(-)})$ and $\psi_{\dot{q}}
=(\psi^{+1/2\, A}, \psi^{-1/2}_A)$ only ones. Thus  the great advantage of the analytic superamplitude formalism is its minimality.
It is also much more similar to the on-shell superfield and superamplitude description used for maximal D=4 SYM and SUGRA theories.
In particular, this similarity helped us to find the gauge fixed expression for the 3-point analytic superamplitudes of 10D SYM and 11D SUGRA.
The price to be paid for these advantages  is the harmonic superspace type realization of the SO(8) (SO(9)) symmetry and, consequently,  dependence on additional set of harmonic variables $\bar{w}_{qA}, w^A_q$ parametrizing  $Spin(8)/[SU(4)\otimes U(1)]$ coset. Presently the analytic superemplitude formalism is under further development which, as we hope, will result in a significant progress in 10D and 11D amplitude calculations.

\bigskip

An alternative  direction we are also working out is  to use the structure of the analytic 3-point
superamplitude for deriving the expression for its cousin from  the real constrained superamplitude
formalism \cite{Bandos:2016tsm,Bandos:2017eof}, and
to use the interplay of the constrained and analytic superamplitude approaches
for their mutual development.

\bigskip

It will be also interesting to reproduce the analytic superamplitudes from an appropriate formulation of the ambitwistor string \cite{Mason:2013sva,Adamo:2013tsa,Adamo:2014wea}. Notice that, although original ambitwistor string model  \cite{Mason:2013sva} had been  of  NSR-type and had been formulated in D=10, quite  soon \cite{Bandos:2014lja} it was appreciated its relation with
null--superstring \cite{Bandos:1991my}
(see \cite{Casali:2016atr,Casali:2017zkz} for related results and  \cite{Bandos:2006af} for more references on null-string)
 and with twistor string \cite{Witten:2003nn,Berkovits:2004hg,Siegel:2004dj,Bandos:2006af}. This suggested its existence in spacetime of arbitrary dimension, including D=11 and D=4, and the last possibility was intensively elaborated in \cite{Geyer:2014fka,Lipstein:2015vxa,Bork:2017qyh,Farrow:2017eol}.
An approach  to derive the analytic superamplitudes from the Green-Schwarz type spinor moving frame formulation of D=10 and D=11 ambitwistor superstring \cite{Bandos:2014lja} looks promising and we plan to address it in the future publications.

\bigskip

\acknowledgments{
 This work has been supported in part by the
Spanish Ministry of Economy, Industry and Competitiveness  grants FPA 2015-66793-P, partially financed with FEDER/ERDF (European Regional Development Fund of the European
Union), by the Basque Government Grant IT-979-16, and the Basque Country University program UFI 11/55.

The author is thankful to Theoretical Department of CERN (Geneva, Switzerland),
to the Galileo Galilei Institute for Theoretical Physics and INFN (Florence, Italy), as well as to the Simons Center for Geometry and Physics, Stony Brook University (New York, US) for the hospitality and partial support of his visits at certain stages of this work. He is grateful to
Dima Sorokin for the interest to this work and reading the draft, to Emeri Sokatchev  for useful discussions and suggestions,  and to Luis Alvarez-Gaume and  Paolo Di Vecchia
for useful discussions on related topics.
}

\bigskip

\appendix

\setcounter{equation}0
\def\theequation{A.\arabic{equation}}
\section{On D=4 spinor helicity formalism}

In D=4 $Spin(1,3)=SL(2,{\bb C})$ and  the spinor frame or Lorentz harmonic variables $v_\alpha^{\pm}=(v_{\dot\alpha}^{\pm})^*$ \cite{Bandos:1990ji}
are restricted by the only condition $v^{-\alpha}v_\alpha^+=1$,
\begin{eqnarray}\label{VinSL2}
(v_\alpha^+, v_\alpha^-) \; \in\; SL(2,{\bb C}) \qquad \Leftrightarrow \qquad v^{-\alpha}v_\alpha^+=1 \; .
 \end{eqnarray}

In a theory invariant under $[SO(1,1)\otimes SO(2)]\subset\!\!\!\!\!\!\times {} {\bb K}_2$ transformations
\begin{eqnarray}\label{v+'=}
& v_\alpha^+ \mapsto e^{a +ib} (v^+_\alpha + k^{\#} v^-_\alpha )\; , \qquad & \bar{v}_{\dot\alpha}^+ \mapsto e^{a -ib} (\bar{v}_{\dot\alpha}^+ + \bar{k}{}^{\#} \bar{v}_{\dot\alpha}^-)\; , \qquad \\
\label{v-'=}
& v_\alpha^-  \mapsto e^{-a -ib} v^-_\alpha  \; , \qquad & \bar{v}_{\dot\alpha}^-  \mapsto e^{-a +ib} \bar{v}_{\dot\alpha}^- \; , \qquad
 \end{eqnarray}
 the set of such harmonic variables parametrize the sphere ${\bb S}^2$ \cite{Galperin:1991gk,Delduc:1991ir},
\begin{eqnarray}\label{v-pm=S2}
\{ (v_\alpha^+, v_\alpha^-) \} = \frac {Spin(1,3)} {[SO(1,1)\otimes Spin(2)]\subset\!\!\!\!\!\!\times {} {\bb K}_2} =\frac {SL(2,{\bb C})} {[SO(1,1)\otimes U(1)]\subset\!\!\!\!\!\!\times {} {\bb K}_2} ={\bb S}^2 \; . \qquad
 \end{eqnarray}
When the spinor frame is associated with a light-like momenta by the generalized Cartan-Penrose relation  \begin{eqnarray}\label{p=rvbv}p_{\alpha\dot{\alpha}}=\rho^{\#}  v^-_{\alpha }\bar{v}_{\dot{\alpha}}^{\, -}
 \end{eqnarray}
 ({\it cf.} (\ref{p=ll=4D})), ${\bb S}^2$ in (\ref{v-pm=S2}) is the celestial sphere.

In the scattering problem we can associate the spinor frame to each of $n$ light-like momenta and to express the corresponding helicity spinors of (\ref{p=ll=4D}) in terms of the spinor harmonics
 \begin{eqnarray}\label{l=v-4D}
\lambda_{\alpha (i)} = \sqrt{\rho^{\#}_{(i)}}  v^{\; -}_{\alpha (i)} \; , \qquad \bar{\lambda}_{\dot{\alpha}(i)}= \sqrt{\rho^{\#}_{(i)}} \bar{v}_{\dot{\alpha}(i)}^{\; -} \; , \qquad p_{\alpha\dot{\alpha} (i)}= \rho^{\#}_{(i)}  v^{\; -}_{\alpha (i)}\bar{v}_{\dot{\alpha}(i)}^{\; -}
 \; . \qquad
 \end{eqnarray}
As we have used only $v^{-}_{\alpha (i)}$, the complementary spinor harmonic $v^{+}_{\alpha (i)}$ remains arbitrary up to the constraint (\ref{VinSL2}),
\begin{eqnarray}\label{v-iv+i=1}
 v^{-\alpha}_{(i)}v^{+}_{\alpha (i)}=1 \; . \qquad
 \end{eqnarray}
Actually, this is the statement of ${\bb K}_2$ symmetry (parametrized by  $k^\#$ and $\bar{k}{}^\#$  in (\ref{v+'=}), (\ref{v-'=})),  which can be used as an identification relation on the set of harmonic variables (as indicated in (\ref{v-pm=S2})), and in this sense is the gauge symmetry. We can fix these ${\bb K}_{2(i)}$ gauge symmetries by identifying (up to a complex multipliers) all the complementary spinors of the spinor frames associated to the momenta of the scattered particles
\begin{eqnarray}\label{v+i=v+j}
 (v^{+}_{ (i)}v^{+}_{(j)})\equiv v^{+\alpha}_{ (i)}v^{+}_{\alpha (j)}=0 \qquad \Leftrightarrow \qquad v^{+}_{\alpha (i)}\; \propto v^{+}_{\alpha (j)}\qquad  \forall \, i,j=1,...,n  \; . \qquad
 \end{eqnarray}

It is convenient to  reformulate this statement by introducing an auxiliary spinor frame $(v_\alpha^\pm)$, which is not associated to any of the scattered particles, and to state that any of the spinor frames $(v^{\pm}_{\alpha (i)})$ is related to that by ({\it cf.} (\ref{v+'=}), (\ref{v-'=}))
\begin{eqnarray}
\label{v+i=v++}
& v_{\alpha (i)}^{\, +}  = e^{\alpha_i +i\beta_i} v^+_\alpha  \; , \qquad & \bar{v}_{\dot{\alpha}(i)}^{\; +}  =  e^{\alpha_i -i\beta_i} \bar{v}_{\dot\alpha}^+ \; , \qquad
\\ \label{v-i=v-+}
& v^{-}_{\alpha (i)} = e^{-\alpha_i -i\beta_i} ( v^{-}_{\alpha}  + {\bb K}_{i}^{=} v^+_\alpha )\; , \qquad  & \bar{v}_{\dot{\alpha}(i)}^{\; -}= e^{-\alpha_i +i\beta_i} (\bar{v}_{\dot\alpha}^{\, -} + \bar{{\bb K}}{}^{=}_{i} \bar{v}_{\dot\alpha}^+)\; . \qquad
 \end{eqnarray}
 In this gauge the contractions of the spinors from different frames read
  \begin{eqnarray}\label{v-iv-j=4D}
{} <v^{-}_{(i)}v^{-}_{(j)}> \equiv v^{-\alpha }_{(i)}v^{-}_{\alpha (j)}  =
 e^{-(\alpha_{i} +\alpha_{j} )-i(\beta_i+\beta_j)} {\bb K}_{ji}^{=}\, ,   \qquad \nonumber \\   {} <v^{-}_{(i)}v^{+}_{(j)}>  =
 e^{(\alpha_{j} -\alpha_{i} )+i(\beta_j-\beta_i)} ,  \qquad \\
 \label{bv-ibv-j=}
 {}[\bar{v}{}^{-}_{(i)}\bar{v}{}^{-}_{(j)}] \equiv \bar{v}{}^{-\dot{\alpha}}_{(i)} \bar{v}_{\dot{\alpha}(j)}^{\; -}  =
 e^{-(\alpha_{i} +\alpha_{j} )+i(\beta_i+\beta_j)} \bar{{\bb K}}{}_{ji}^{=}\, ,     \nonumber \qquad \\ {} [\bar{v}{}^{-}_{(i)}\bar{v}{}^{+}_{(j)}] =
 e^{(\alpha_{j} -\alpha_{i} )-i(\beta_j-\beta_i)} ,  \qquad
 \end{eqnarray}
 where
\begin{eqnarray}\label{Kji=Kj-Ki}
 {\bb K}{}^=_{ji}:={\bb K}{}^=_{j}-{\bb K}{}^=_{i} \; .
  \end{eqnarray}

Of course, we can  use the $SO(1,1)_i\times SO(2)_i$ gauge symmetries to fix also
$\alpha_i=0$ and $\beta_i=0$ $\forall i=1,...,n$, but the multipliers $\beta_i$ might be useful as they actually indicate the helicity of the field or amplitude, while $\alpha_i$ can be 'eaten' by the 'energy' variables $\rho^\#_i$. Indeed,  the $i$-th light-like momentum (\ref{l=v-4D}) can be now written as
 \begin{eqnarray}\label{pi=triu}
 p_{\alpha\dot{\alpha} (i)}&=& \widetilde{\rho}{}^{\#}_{(i)}
 ( v^{-}_{\alpha}  + {\bb K}_{(i)}^{=} v^+_\alpha )\,  (\bar{v}_{\dot\alpha}^{\, -} + \bar{{\bb K}}{}^{=}_{(i)} \bar{v}_{\dot\alpha}^+) \nonumber \\ &=& \widetilde{\rho}{}^{\#}_{(i)} u^=_{\alpha\dot\alpha} +  \widetilde{\rho}{}^{\#}_{(i)} {\bb K}^=_{(i)}  u^{+-}_{\alpha\dot\alpha} +  \widetilde{\rho}{}^{\#}_{(i)} \bar{{\bb K}}{}^=_{(i)}  u^{-+}_{\alpha\dot\alpha}+  \widetilde{\rho}{}^{\#}_{(i)} {\bb K}^=_{(i)} \bar{{\bb K}}{}^=_{(i)}
u^\#_{\alpha\dot\alpha}
 \; , \qquad
 \end{eqnarray}
where $\tilde{\rho}{}^\#_i= e^{-2\alpha_i}\rho^\#_i$ (\ref{r--tr}) and
\begin{eqnarray}\label{u=NP}
  u^=_{\alpha\dot\alpha}= v^{-}_{\alpha} \bar{v}_{\dot\alpha}^{\, -}\; , \qquad
u^\#_{\alpha\dot\alpha} = v^{\, +}_{\alpha} \bar{v}_{\dot\alpha}^{\, +}\; , \qquad
 u^{\pm\mp}_{\alpha\dot\alpha} = v^{\,\pm}_{\alpha} \bar{v}_{\dot\alpha}^{\, \mp}\; , \qquad
  \qquad
 \end{eqnarray}
are two real and two complex conjugate ($u_a^{+-}=(u_a^{-+})^* $) vectors of Newman-Penrose light-like tetrade (see \cite{Penrose:1986ca} and refs. therein).

Using the complementary harmonics $v_{\alpha i}^{+}$, $\bar{v}_{\dot{\alpha} i}^{+}$ of the auxiliary frame as reference spinors, we can identify  polarization vectors with the $i$-th frame counterparts of  the above described complex
null-vectors $u_a^{-+}$ and $u_a^{+-}=(u_a^{-+})^* $:
 \begin{eqnarray}\label{polar=u}
\varepsilon^{(+)}_{{\alpha\dot{\alpha}(i)}}= u^{-+}_{{\alpha\dot{\alpha}(i)}}\equiv  v^{\, -}_{\alpha (i)} \bar{v}_{\dot{\alpha} (i)}^{\, + } \; , \qquad \varepsilon^{(-)}_{{\alpha\dot{\alpha}(i)}}= u^{+-}_{{\alpha\dot{\alpha}(i)}} \equiv  v^{\, +}_{\alpha (i)} \bar{v}_{\dot{\alpha} (i)}^{\, - } \; . \qquad
  \end{eqnarray}
In the gauge (\ref{v+i=v+j}) these identification implies that
\begin{eqnarray}\label{e+e+=0}
\varepsilon^{(+)}_{(i)}\cdot \varepsilon^{(+)}_{(j)}:= \frac 1 2 \varepsilon^{(+)}_{{\alpha\dot{\alpha}(i)}} \varepsilon^{(+)\alpha\dot{\alpha}}_{(j)}=0 \; . \qquad
  \end{eqnarray}
 Using (\ref{pi=triu}) and (\ref{polar=u}) we can easily find
  \begin{eqnarray}\label{ek=}
\varepsilon^{(+)}_{(i)}k_{(j)}=  \frac {\rho^{\#}_{(i)}} 2
 ({v}{}^{-}_{(j)}{v}{}^{-}_{(i)})(\bar{v}{}^{-}_{(j)}\bar{v}{}^{+}_{(i)})\; , \qquad
\varepsilon^{(+)}_{(i)}\varepsilon^{(-)}_{(j)}=  -\frac 1 2
 ({v}{}^{-}_{(i)}{v}{}^{+}_{(j)})(\bar{v}{}^{-}_{(j)}\bar{v}{}^{+}_{(i)})\; , \qquad
  \end{eqnarray}
  and then, for instance,
  \begin{eqnarray}\label{ekee=}
 (\varepsilon^{(+)}_{(1)} k_{(2)})\;  (\varepsilon^{(+)}_{(2)}\varepsilon^{(-)}_{(3)})  &=&  -\frac {\rho^{\#}_{(2)}} 4
 ({v}{}^{-}_{(2)}{v}{}^{-}_{(1)})\; ({v}{}^{-}_{(2)}{v}{}^{+}_{(3)})\; (\bar{v}{}^{-}_{(2)}\bar{v}{}^{+}_{(1)})\;
 (\bar{v}{}^{-}_{(3)}\bar{v}{}^{+}_{(2)})\; \nonumber \\
 &=&  -\frac {\tilde{\rho}{}^{\#}_{(2)}} 4\; K^=_{21}\; e^{2i(\beta_3-\beta_2-\beta_1)} , \qquad
  \end{eqnarray}
This allows us to calculate  3-gluon  amplitude of ${\cal N}=4$ 4D SYM,
 \begin{eqnarray}\label{cA3g=}
 {\cal M}(1^+,2^+, 3^-)&=& g \epsilon_{(1)}^{(+)a}\epsilon_{(2)}^{(+)b} \epsilon_{(3)}^{(-)c} t_{abc}(k_1,k_2,k_3) \qquad \nonumber  \\ &=& g (\varepsilon^{(+)}_{(1)}k_{(2)}\;  \varepsilon^{(+)}_{(2)}\varepsilon^{(-)}_{(3)} + \varepsilon^{(+)}_{(2)}k_{(3)}\;  \varepsilon^{(-)}_{(3)}\varepsilon^{(+)}_{(1)}+\varepsilon^{(-)}_{(3)}k_{(1)}\;  \varepsilon^{(+)}_{(1)}\varepsilon^{(+)}_{(2)}) = \nonumber \\ &=&  -\frac {g} 4\,  e^{2i(\beta_3-\beta_2-\beta_1)} \left( \tilde{\rho}{}^{\#}_{(2)} K^=_{21} + \tilde{\rho}{}^{\#}_{(3)} K^=_{32}\right) \; \nonumber \\
&=&   \; \frac {g} 4\; \tilde{\rho}{}^{\#}_{(3)}K^=_{21}\; e^{2i(\beta_3-\beta_2-\beta_1)}  \qquad
  \end{eqnarray}
(see \cite{Green:1981ya,Schwarz:1982jn} for the definition of $t_{abc}$ tensor).  Notice that the last term in the second line of this equation  vanishes as a result of (\ref{e+e+=0}) and that at the last stage of transformations  of this equation we have used the consequence of the momentum conservation in 3-particle process which we are going to discuss now.

  \subsection{Momentum conservation in a 3-point 4D amplitude}

In our notation the momentum conservation in the 3-particle process is expressed by
 \begin{eqnarray}\label{p1+p2+p3=0-4D}
{\rho}{}^{\#}_{(1)} v^{\, -}_{\alpha (1)} \bar{v}_{\dot{\alpha} (1)}^{\,- }+{\rho}{}^{\#}_{(2)} v^{\, -}_{\alpha (2)} \bar{v}_{\dot{\alpha} (2)}^{\, - }+{\rho}{}^{\#}_{(3)} v^{\, -}_{\alpha (3)} \bar{v}_{\dot{\alpha} (3)}^{\, - }=0   \qquad
  \end{eqnarray}
  This implies
  \begin{eqnarray}\label{Sr123=0}
\tilde{\rho}{}^{\#}_{1}+ \tilde{\rho}{}^{\#}_{2} + \tilde{\rho}{}^{\#}_{3}=0\; , \qquad
\\ \label{SrK123=0}
\tilde{\rho}{}^{\#}_{1} K^=_{1}+ \tilde{\rho}{}^{\#}_{2} K^=_{2} + \tilde{\rho}{}^{\#}_{3} K^=_{3}=0\; , \qquad \nonumber \\
\Rightarrow \qquad   K^=_{32}=\frac {\tilde{\rho}{}^{\#}_{1}}{\tilde{\rho}{}^{\#}_{3}}
K^=_{21}= \frac {\tilde{\rho}{}^{\#}_{1}}{\tilde{\rho}{}^{\#}_{2}}
K^=_{13} \quad \Rightarrow \quad K^=_{13}= \frac {\tilde{\rho}{}^{\#}_{2}}{\tilde{\rho}{}^{\#}_{3}}
K^=_{21}
  \end{eqnarray}
  as well as
   \begin{eqnarray}\label{K12bK12=0}
 K^=_{32}\bar{K}^=_{32}:=(K^=_{3}-K^=_{2})(\bar{K}^=_{3}-\bar{K}^=_{2})=0 \; .
  \end{eqnarray}
 Here we have used the notation (\ref{Kji=Kj-Ki}) and
\begin{eqnarray} \label{tilrho=}
& \tilde{\rho}{}^{\#}_{i}:=e^{-2\alpha_i } {\rho}{}^{\#}_{i}  \; .  \qquad
 \end{eqnarray}

The solution of Eq. (\ref{K12bK12=0}) is nontrivial only if $(\bar{K}^=_{(32)})^*\not= {K}^=_{(32)}$.  In this case one of two branches of the general solution is described by
   \begin{eqnarray}\label{SbK1=2=3}
\bar{K}^=_{3}=\bar{K}^=_{2}=\bar{K}^=_{1} \;
  \end{eqnarray}
  while ${K}^=_{(1,2,3)}$ can be different but obeying (\ref{SrK123=0}) with $\tilde{\rho}{}^{\#}_{(1,2,3)}$ restricted by (\ref{Sr123=0}).  From now on we will denote these complex nonvanishing ${K}^=_{(1,2,3)}$ restricted by 3-particle kinematics by ${\bb K}^=_{(1,2,3)}$. We will also use the solution of (\ref{SrK123=0}) in terms of complex non-vanishing ${\bb K}^{==}$
 \begin{eqnarray}\label{bbK==:=}
  \frac {{\bb K}^=_{32}}{\tilde{\rho}{}^{\#}_{1}}=   \frac {{\bb K}^=_{21}}{\tilde{\rho}{}^{\#}_{3}}=  \frac {{\bb K}^=_{13}}{\tilde{\rho}{}^{\#}_{2}}=: {\bb K}^{==}\; , \qquad
  \end{eqnarray}

   Eq. (\ref{SbK1=2=3}) implies
   \begin{eqnarray}\label{bv1=2=3}
\bar{v}_{\dot{\alpha} (1)}^{\, - }\propto  \bar{v}_{\dot{\alpha} (2)}^{\, -}\propto  \bar{v}_{\dot{\alpha} (3)}^{\, -}  \qquad
  \end{eqnarray}
  while  $ v^{\, -}_{\alpha (1)}$,   $ v^{\, -}_{\alpha (2)}$ and   $ v^{\, -}_{\alpha (3)}$ are  different.

  \subsection{ 3-gluon amplitude and superamplitude in maximal D=4 SYM }
  The standard expression for  the 3-point amplitude in D=4 SYM is written in terms of  \begin{eqnarray}\label{ij=lili=}
  <ij> =<\lambda_i\lambda_j>=\lambda^\alpha_{i}\lambda_{\alpha j} &=& \sqrt{\rho^{\#}_{i}\rho^{\#}_j}<v^-_iv^-_{j}> \qquad \nonumber \\ &=&  \sqrt{\tilde{\rho}{}^{\#}_{i}\tilde{\rho}{}^{\#}_{j}} e^{-i(\beta_i+\beta_j)} {\bb K}_{ji}^{=}\, . \end{eqnarray}
If we were trying to guess the corresponding expression starting from (\ref{cA3g=}), the $\beta_{i}$ dependence indicates  that this should be (up to a coefficient)
  \begin{eqnarray}\label{cM3g=stand}
 {\cal M}(1^+,2^+, 3^-)&=&  \frac {<12>^3} {<23><31>} \equiv  \frac {<12>^4} {<12> <23><31>}\; .
  \end{eqnarray}
Using (\ref{ij=lili=}) and (\ref{SrK123=0}) one can easily check that this expression indeed reproduce  (\ref{cA3g=}),
 \begin{eqnarray}\label{PT3=rhoK}
\frac {<12>^4} {<12> <23><31>}= \tilde{\rho}{}^{\#}_{3}{\bb K}^=_{21}\, e^{2i(\beta_3-\beta_2-\beta_1)}=
(\tilde{\rho}{}^{\#}_{3})^2 {\bb K}^{==}\, e^{2i(\beta_3-\beta_2-\beta_1)}
\; .   \end{eqnarray}

In our notation the anti-MHV ($\overline{{\rm MHV}}$) type superamplitude reads (see (\ref{cA3g=bMHV}))
   \begin{eqnarray}\label{cA3g=stand-A}
 {\cal A}^{\overline{{\rm MHV}}}(1,2, 3)=  \frac {1} {<12> <23><31>}\; \delta^4 \left( \eta_{1}<23> +  \eta_{2}<31>+  \eta_{3} <12> \right) ,  \qquad
  \end{eqnarray}
  while the MHV amplitude is
  \begin{eqnarray}\label{cA3g=MHV-A}
 {\cal A}^{MHV}(1,2, 3) &= &  \frac {1} {[12] \,[23]\, [31]}\; \delta^8 \left( \bar{\lambda}_{ \dot{\alpha}1}\eta_{A1}+  \bar{\lambda}_{ \dot{\alpha}2}\eta_{A2}+ \bar{\lambda}_{ \dot{\alpha}3} {\eta}_{A3} \right) = \qquad\nonumber \\
 & =& \frac {1} {[12] \,[23]\, [31]}\; \frac 1 {2^4} \; \prod\limits_{A=1}^4 \sum\limits_{i,j=1}^3 \;[ij] \; \eta_{Ai}\eta_{Aj}\;
 .  \qquad
  \end{eqnarray}

The covariance of $\delta$ function under the phase transformations of the bosonic spinors holds when
the fermionic variables $\eta_{Ai}$ have the same phase transformation property as $\lambda_{\alpha i}$. This reflects its origin in Penrose-Ferber  incidence type relation  $\eta_{Ai}=\theta^{\alpha}_{A i}\lambda_{\alpha i}$ \cite{Ferber:1977qx}
which in terms of our Lorentz harmonic notation reads  $\eta_{Ai}=  \sqrt{\rho^{\#}_{i}} \eta^-_{Ai} := \sqrt{\rho^{\#}_{i}} \theta^{\alpha}_{A i}v^-_{\alpha i}$.

Notice also that the indices $A$ of all the fermionic coordinates are transformed by the same $SU(4)$, which is the R-symmetry group of ${\cal N}=4$ D=4 SYM.

Using (\ref{v-i=v-+}), (\ref{v+i=v++}) and (\ref{Sr123=0}), (\ref{bbK==:=}),  we can write the
Grassmann delta function  of (\ref{cA3g=MHV-A}) in the form
\begin{eqnarray}\label{delta4=K}
   &&  \delta^4 \left( \eta_{1}<23>  +  \eta_{2}<31>+  \eta_{3} <12> \right) = \nonumber
   \\ &&{}\qquad =
{ \left(\tilde{\rho}{}^{\#}_{1} \tilde{\rho}{}^{\#}_{2} \tilde{\rho}{}^{\#}_{3} \right)^{2}} \, e^{-4i(\beta_1+\beta_2+\beta_3)} \,  \delta^4 \left( \tilde{\eta}^-_{1A} {\bb K}{}^=_{32}     + \tilde{\eta}^-_{2A} {\bb K}{}^=_{13} + \tilde{\eta}^-_{3A} {\bb K}{}^=_{21}  \right)\; \nonumber
\\
&&{}\qquad   =
{ \left(\tilde{\rho}{}^{\#}_{1} \tilde{\rho}{}^{\#}_{2} \tilde{\rho}{}^{\#}_{3} \right)^{2}} \,  ( {\bb K}{}^{==} )^{4}\, e^{-4i(\beta_1+\beta_2+\beta_3)} \, \,  \delta^4 \left( \tilde{\rho}{}^{\#}_{1} \eta^-_{1A}   + \tilde{\rho}{}^{\#}_2 \tilde{\eta}^-_{2A}  + \tilde{\rho}{}^{\#}_3\tilde{\eta}^-_{3A}  \right)\; \nonumber
\\
&& {}\qquad =
\frac { \left(\tilde{\rho}{}^{\#}_{1} \tilde{\rho}{}^{\#}_{2} \right)^{2} \, ( {\bb K}{}^{=}_{21} )^{4}} {(\tilde{\rho}{}^{\#}_{3})^2  }\, e^{-4i(\beta_1+\beta_2+\beta_3)}    \delta^4 \left( \tilde{\rho}{}^{\#}_{1} \eta^-_{[13]A}   + \tilde{\rho}{}^{\#}_2 \tilde{\eta}^-_{[23]A}    \right)\; , \nonumber \\ && {}
\end{eqnarray}
where  \footnote{One can check that $\tilde{\eta}{}^-_{Ai} =\theta^{\alpha}_{A i}(v^-_{\alpha}+ {\bb K}_{i}^= v^+_{\alpha})$ which makes transparent that all $\tilde{\eta}{}^-_{Ai}$ are transformed by the common $U(1)\otimes SO(1,1)$ group, but are inert under all the $U(1)_j\otimes SO(1,1)_j$ gauge symmetries, including the one with $i=j$.}
\begin{eqnarray}\label{tileta=}
& \tilde{\eta}{}^-_{Ai}:=e^{\alpha_i +i\beta_i}   {\eta}^-_{Ai} \; , \qquad  &  \tilde{\eta}{}^-_{A[ji]}=
\tilde{\eta}{}^-_{Aj}-\tilde{\eta}{}^-_{Ai}\; , \qquad
 \end{eqnarray} $\tilde{\rho}{}^{\#}_i$ is defined in (\ref{tilrho=}).

 Similarly, the fermionic delta function in (\ref{cA3g=MHV}) can be written as
\begin{eqnarray} \label{delta8=}
&& \hspace{-0.5cm }\delta^8 \left( \bar{\lambda}_{ \dot{\alpha}1}\eta_{A1}+  \bar{\lambda}_{ \dot{\alpha}2}\eta_{A2}+ \bar{\lambda}_{ \dot{\alpha}3} {\eta}_{A3} \right) = \delta^8 \left( {\rho}{}^{\#}_{(1)}\bar{v}_{ \dot{\alpha}(1)}^- \eta^-_{A(1)}+  {\rho}{}^{\#}_{(2)}\bar{v}_{ \dot{\alpha}2}^- \eta^-_{A2}+{\rho}{}^{\#}_{3}\bar{v}_{ \dot{\alpha}3}^- \eta^-_{A3}\right) = \nonumber \\ && \delta^8\left(\bar{v}{}_{\dot{\alpha}}^- \left(\tilde{\rho}{}^{\#}_{1} \tilde{\eta}{}^-_{A[13]}  +\tilde{\rho}{}^{\#}_{2} \tilde{\eta}{}^-_{A[23]} \right)+\bar{v}{}_{\dot{\alpha}}^+\left(\bar{{\bb K}}^=_{1}\tilde{\rho}{}^{\#}_{1} \tilde{\eta}{}^-_{A[13]}  +\bar{{\bb K}}{}^=_{2} \tilde{\rho}{}^{\#}_{2} \tilde{\eta}{}^-_{A[23]}  \right)\right) = \qquad \nonumber \\ &&{}\qquad \delta^8\left(\left( \bar{v}{}_{\dot{\alpha}}^- + \bar{{\bb K}}^=_{1} \bar{v}{}_{\dot{\alpha}}^+\right) \, \tilde{\rho}{}^{\#}_{1} \tilde{\eta}{}^-_{A[13]}+ \left( \bar{v}{}_{\dot{\alpha}}^- + \bar{{\bb K}}^=_{2} \bar{v}{}_{\dot{\alpha}}^+\right)\, \tilde{\rho}{}^{\#}_{2} \tilde{\eta}{}^-_{A[23]}  \right) \; . \qquad
\end{eqnarray}

In this notation, the multiplier  in the $\overline{{\rm MHV}}$ superamplitude (\ref{cA3g=bMHV}) reads
 \begin{eqnarray}\label{12-23-34}
 \frac {1} {<12> <23><31>} =
  \frac {e^{2i(\beta_1+\beta_2+\beta_3)}} {\tilde{\rho}{}^{\#}_{(1)} \tilde{\rho}{}^{\#}_{(2)} \tilde{\rho}{}^{\#}_{(3)} }\;  \frac {1}{{\bb K}{}^=_{21}{\bb K}{}^=_{32} {\bb K}{}^=_{13}}=
  \frac {e^{2i(\beta_1+\beta_2+\beta_3)}} {(\tilde{\rho}{}^{\#}_{1} \tilde{\rho}{}^{\#}_{2} )^2 }\;  \frac { \tilde{\rho}{}^{\#}_{3}}{({\bb K}{}^=_{21})^3} \nonumber \\
  =
  \frac {e^{2i(\beta_1+\beta_2+\beta_3)}} {(\tilde{\rho}{}^{\#}_{1} \tilde{\rho}{}^{\#}_{2}  \tilde{\rho}{}^{\#}_{3})^2 }\;  \frac { 1}{({\bb K}{}^{==})^3} \; .  \qquad
  \end{eqnarray}

Using (\ref{12-23-34}) and (\ref{delta4=K}), we can write the 3-point anti-MHV superamplitude (\ref{cA3g=bMHV}) in the form
\begin{eqnarray}\label{cA3g=K21}
 && {\cal A}^{\overline{{\rm MHV}}}(1,2,3)= ({\bb K}{}^{==})\; e^{-2i(\beta_1+\beta_2+\beta_3)}\, \delta^4 \left( \tilde{\rho}{}^{\#}_{1} \tilde{\eta}^-_{1A}   + \tilde{\rho}{}^{\#}_2 \tilde{\eta}^-_{2A}  + \tilde{\rho}{}^{\#}_3\tilde{\eta}^-_{3A}  \right)\; ,
 \qquad \nonumber \\  &&{} \qquad =
 -\frac { {\bb K}{}^=_{21}} { (\tilde{\rho}{}^{\#}_{1} +\tilde{\rho}{}^{\#}_{2} )} e^{-2i(\beta_1+\beta_2+\beta_3)}\; \delta^4 \left( \tilde{\rho}{}^{\#}_{1} \tilde{\eta}{}^-_{[13]A}   + \tilde{\rho}{}^{\#}_2 \tilde{\eta}^-_{[23]A}   \right)\; .
 \qquad
  \end{eqnarray}

\setcounter{equation}0
\def\theequation{B.\arabic{equation}}

\section{BCFW-like deformations of complex frame and complex fermionic variables}

An important tool to reconstruct tree $D=4$ (super)amplitudes from the basic 3-point (super)amplitude is given by
BCFW recurrent relation \cite{Britto:2005fq} and their superfield generalization \cite{ArkaniHamed:2008gz}.
The counterparts of these latter 4D relations for constrained superamplitudes of 11D SUGRA and 10D SYM   have been presented in
\cite{Bandos:2016tsm,Bandos:2017eof}. They use the real BCFW deformations of real bosonic and fermionic variables of the constrained superamplitude formalism. In contrast, in the case of the BCFW-type recurrent relations for analytic superamplitudes (which are still to be derived), one expects the BCFW deformations used in such recurrent relations to have an intrinsic complex structure, similar to the one of the D=4 relations \cite{Britto:2005fq,ArkaniHamed:2008gz}
\begin{eqnarray}\label{BCFWln=4D}
  & \lambda^{A}_{(n)}\mapsto \widehat{\lambda^{A}_{(n)}} = \lambda^{A}_{(n)} + z \lambda^{A}_{(1)} , \qquad  & \qquad \bar{\lambda}{}^{\dot A}_{(n)}\mapsto \widehat{\bar{\lambda}{}^{\dot A}_{(n)}}=\bar{\lambda}{}^{\dot A}_{(n)} , \qquad
  \\
  \label{BCFWl1=4D}
 & \lambda^{A}_{(1)}\mapsto \widehat{\lambda^{A}_{(1)}} = \lambda^{A}_{(1)}\; , \qquad
 & \bar{\lambda}{}^{\dot A}_{(1)} \mapsto  \widehat{\bar{\lambda}{}^{\dot A}_{(1)}}=\bar{\lambda}{}^{\dot A}_{(1)} -z \bar{\lambda}{}^{\dot A}_{(n)} , \qquad \\ \label{BCWF=eta4D}
& \widehat{\eta_{A\,n}}= \eta_{A\,n}  + z \;
\eta^-_{A\,1}\;
\; , \qquad & \qquad  \widehat{\eta_{A\,1}}= \eta_{A\,1}
\;   \qquad \end{eqnarray}

Let us  show how this can be reached starting from the
BCFW deformations of real spinor frame variables \cite{Bandos:2016tsm,Bandos:2017eof}   \begin{eqnarray}\label{BCWF=vnM}
\widehat{v_{\alpha {q}(n)}^{\; -}}= v_{\alpha {q}(n)}^{\; -} + z \;
\sqrt{\frac {\rho^{\#}_{(1)}}{\rho^{\#}_{(n)}}} v_{\alpha {p}(1)}^{\; -} \; {\bb M}_{ {p}{q}}
\; , \qquad \\ \label{BCWF=v1M} \widehat{v_{\alpha {q}(1)}^{\; -}}= v_{\alpha {q}(1)}^{\; -} - z \;
 \sqrt{\frac {\rho^{\#}_{(n)}}{\rho^{\#}_{(1)}}}\; {\bb M}_{{q}{p}}\;  v_{\alpha {p}(n)}^{\; -}
\; , \qquad \end{eqnarray}
and of the real fermionic variables
\begin{eqnarray}\label{BCFW=thn}
\widehat{ \theta^-_{{p}(n)}}= \theta^-_{{p}(n)}+ z \,\theta^-_{{q}(1)} \, {\bb M}_{{q}{p}} \, \sqrt{\frac {\rho^{\#}_{(1)}}{\rho^{\#}_{(n)}}} \; , \qquad \\ \label{BCFW=th1}
\widehat{ \theta^-_{{q}(1)}}= \theta^-_{{q}(1)}- z \, \sqrt{\frac {\rho^{\#}_{(n)}}{\rho^{\#}_{(1)}}}  \, {\bb M}_{{q}{p}} \, \theta^-_{{p}(n)}\;  . \qquad \end{eqnarray}
Here  $\alpha=1,..,4{\cal N}$ and $q,p=1,...,4{\cal N}$
(we should set ${\cal N}=8$ and $4$ for 11D SUGRA and 10D SYM, respectively) and $z$ is an arbitrary number. In  principle this can be considered to be real $z\in {\bb R}$ \cite{CaronHuot:2010rj}, although
 $z\in {\bb C}$ is neither forbidden and actually more convenient in amplitude calculations.

The above shift of spinor moving frame variables
 results in shifting the momentum of the first and of the $n$--th particle,
\begin{eqnarray}\label{BCWF=hp}
\widehat{k_{(1)}^{a}}= k_{(1)}^{a} -z q^a \; , \qquad \widehat{k_{(n)}^{a}}= k_{(n)}^{a} +z q^a \; , \qquad
\end{eqnarray}
on a light-like vector $q^a$
 orthogonal to both $ k_{(1)}^{a}$ and $ k_{(n)}^{a}$,
\begin{eqnarray}\label{qq=0=}
q_aq^a=0\; , \qquad q_ak_{(1)}^{a} =0\; , \qquad q_ak_{(n)}^{a}= 0\; , \qquad
\end{eqnarray}
provided we choose
\begin{eqnarray}\label{Mpq=}
{\bb M}_{{q}{p}} = - \frac {1} { {\sqrt{\rho^{\#}_{(1)}\rho^{\#}_{(n)} }} (u_{(1)}^=u^{=}_{(n)}) }\; {(v_{{q}(1)}^{\; -} \, \tilde{q}\!\!\!/{} v_{{p}(n)}^{\; -})}  \; , \qquad
\\
\label{tq=qts}
\tilde{q}\!\!\!/{}^{ \alpha\beta}:= q^a
\tilde{\Gamma}_a^{ \alpha\beta} \; , \qquad {q}\!\!\!/{}_{ \alpha\beta}:= q_a
{\Gamma}^a_{ \alpha\beta} \; . \qquad
\end{eqnarray}
The light-likeness of $q^a$ (\ref{qq=0=}) implies the nilpotency of the
matrix  ${\bb M} $,
\begin{eqnarray}\label{MMT=0}
{\bb M}_{rp} {\bb M}_{r{q}} =0\;  ,\qquad {\bb M}_{{q}r } {\bb M}_{{p}r} =0\; .\qquad
\end{eqnarray}
We can also write the expression for light-like complex vector in terms of deformation matrix,
\begin{eqnarray}\label{q:=M}
q^a= \frac 1 {{\cal N}}\,
{\sqrt{\rho^{\#}_1\rho^{\#}_n}} \;  v_{{q}(1)}^{\; -}\tilde{\Gamma}^a{\bb M}_{{q}{p}}v_{{p}(n)}^{\; -}
\; . \qquad
\end{eqnarray}

The nilpotency condition (\ref{MMT=0}) guarantees that the shifted spinor moving frame variables obey the characteristic constraints, Eqs. (\ref{k=pv-v-})  with shifted light-like momenta $k_{(1)}$ and $k_{(n)}$ (\ref{BCWF=hp}) or, equivalently, (\ref{u==v-v-}) with shifted light-like $u^{=a}_{(1)}$ and
$u^{=a}_{(n)}$,
\begin{eqnarray}\label{BCWF=hu--}
\widehat{u_{(1)}^{=a}}= u_{(1)}^{=a} -\frac {z q^a} {\rho^\#_{(1)}} \; , \qquad \widehat{u_{(n)}^{=a}}= u_{(n)}^{=a} +\frac {z q^a} {\rho^\#_{(n)}} \; . \qquad
\end{eqnarray}

Notice that (\ref{BCWF=vnM}) and (\ref{BCWF=v1M}) imply
\begin{eqnarray}\label{hp1+hpn=}
\widehat{k^a_1}+\widehat{k^a_n}= k^a_1+k^a_n\; .
\qquad
\end{eqnarray}

The complex structure similar to the one of D=4 BCFW deformations can be reproduced after passing to the complex spinor harmonics (\ref{v-Ai:=})---(\ref{bv-Ai-1=}) composed  from the spinor harmonics  and the  internal harmonic variables.
The internal harmonics can be used to solve the nilpotency conditions (\ref{MMT=0}) for the matrix
${\bb M}_{{q}{p}}$ in (\ref{BCWF=vnM})--(\ref{BCFW=th1}). The solution
\begin{eqnarray}\label{M=bwMw}
{\bb M}_{{q}{p}}= \bar{w}_{qA\, 1} \frak{M}^A{}_B w_{p\, n}^{B}\; ,
\qquad
\end{eqnarray}
with an arbitrary hermitian ${\cal N}\times {\cal N}$ matrix  $\frak{M}^A{}_B $, results in
the following deformation of the complex spinor frame variables (\ref{v-A:=}) and of the complex fermionic  variables:
\begin{eqnarray}\label{BCWF=vnfMM}
\widehat{v_{\alpha {A}(n)}^{\; -}}= v_{\alpha {q}(A)}^{\; -} + z \;
v_{\alpha {B}(1)}^{\; -} \; \frak{M}^B{}_A \; \sqrt{{\rho^{\#}_{(1)}}/{\rho^{\#}_{(n)}}}
\; , \qquad \widehat{\bar{v}_{\alpha(n)}^{\; A -} }= \bar{v}{}_{\alpha(n)}^{\; A -}
\; , \qquad \\ \label{BCWF=v1fMM} \widehat{v_{\alpha {A}(1)}^{\; -}}= v_{\alpha {A}(1)}^{\; -}
\; , \qquad  \widehat{\bar{v}_{\alpha(1)}^{\; A -} }= \bar{v}{}_{\alpha(1)}^{\; A -}
- z \; \frak{M}^A{}_B  \; \bar{v}{}_{\alpha(n)}^{\; B -} \;
 \sqrt{{\rho^{\#}_{(n)}}/{\rho^{\#}_{(1)}}}
\;  \qquad \end{eqnarray}
and
\begin{eqnarray}\label{BCWF=etaM}
\widehat{\eta^-_{A\,n}}= \eta^-_{A\,n}  + z \;
\eta^-_{B\,1}\; \frak{M}^B{}_A \; \sqrt{{\rho^{\#}_{(1)}}/{\rho^{\#}_{(n)}}}
\; , \qquad \widehat{\eta^-_{A\,1}}= \eta^-_{A\,1}
\; .  \qquad \end{eqnarray}
These are already quite similar to the  4D super-BCFW transformations (\ref{BCFWln=4D}), (\ref{BCFWl1=4D}), (\ref{BCWF=eta4D}). To make the similarity even closer, we can choose $\frak{M}^B{}_A =\delta^B{}_A$.
In such a way we arrive at (\ref{BCWF=vnfM}),  (\ref{BCWF=v1fM}), and (\ref{BCWF=eta}).


\begin{thebibliography}{99}
\renewcommand{\theequation}{R.\arabic{equation}}


\bibitem{Bern:2011qn}
  Z.~Bern, J.~J.~Carrasco, L.~J.~Dixon, H.~Johansson and R.~Roiban,
  ``Amplitudes and Ultraviolet Behavior of N = 8 Supergravity,''
  Fortsch.\ Phys.\  {\bf 59} (2011) 561
  doi:10.1002/prop.201100037
  [arXiv:1103.1848 [hep-th]].

\bibitem{Drummond:2008vq}
  J.~M.~Drummond, J.~Henn, G.~P.~Korchemsky and E.~Sokatchev,
  ``Dual superconformal symmetry of scattering amplitudes in N=4 super-Yang-Mills theory,''
  Nucl.\ Phys.\ B {\bf 828} (2010) 317
  doi:10.1016/j.nuclphysb.2009.11.022
  [arXiv:0807.1095 [hep-th]].

\bibitem{Drummond:2009fd}
  J.~M.~Drummond, J.~M.~Henn and J.~Plefka,
  JHEP {\bf 0905} (2009) 046
  doi:10.1088/1126-6708/2009/05/046
  [arXiv:0902.2987 [hep-th]].


\bibitem{Eden:2011ku}
  B.~Eden, P.~Heslop, G.~P.~Korchemsky and E.~Sokatchev,
  ``The super-correlator/super-amplitude duality: Part II,''
  Nucl.\ Phys.\ B {\bf 869} (2013) 378
  doi:10.1016/j.nuclphysb.2012.12.014
  [arXiv:1103.4353 [hep-th]].

\bibitem{Kallosh:2012yy}
  R.~Kallosh and T.~Ortin,
  ``New E77 invariants and amplitudes,''
  JHEP {\bf 1209} (2012) 137
  doi:10.1007/JHEP09(2012)137
  [arXiv:1205.4437 [hep-th]].

\bibitem{Witten:2003nn}
  E.~Witten,
  ``Perturbative gauge theory as a string theory in twistor space,''
  Commun.\ Math.\ Phys.\  {\bf 252} (2004) 189
  doi:10.1007/s00220-004-1187-3
  [hep-th/0312171].

\bibitem{Britto:2005fq}
  R.~Britto, F.~Cachazo, B.~Feng and E.~Witten,
  ``Direct proof of tree-level recursion relation in Yang-Mills theory,''
  Phys.\ Rev.\ Lett.\  {\bf 94} (2005) 181602
  [hep-th/0501052].




\bibitem{Britto:2004ap}
  R.~Britto, F.~Cachazo and B.~Feng,
  ``New recursion relations for tree amplitudes of gluons,''
  Nucl.\ Phys.\ B {\bf 715} (2005) 499
  [hep-th/0412308].

\bibitem{Bianchi:2008pu}
  M.~Bianchi, H.~Elvang and D.~Z.~Freedman,
  ``Generating Tree Amplitudes in N=4 SYM and N = 8 SG,''
  JHEP {\bf 0809} (2008) 063
  [arXiv:0805.0757 [hep-th]].


\bibitem{Brandhuber:2008pf}
  A.~Brandhuber, P.~Heslop and G.~Travaglini,
  ``A Note on dual superconformal symmetry of the N=4 super Yang-Mills S-matrix,''
  Phys.\ Rev.\ D {\bf 78} (2008) 125005
  [arXiv:0807.4097 [hep-th]].

\bibitem{ArkaniHamed:2008gz}
  N.~Arkani-Hamed, F.~Cachazo and J.~Kaplan,
  ``What is the Simplest Quantum Field Theory?,''
  JHEP {\bf 1009} (2010) 016
  [arXiv:0808.1446 [hep-th]].


\bibitem{Heslop:2016plj}
  P.~Heslop and A.~E.~Lipstein,
  ``On-shell diagrams for $ \mathcal{N} $ = 8 supergravity amplitudes,''
  JHEP {\bf 1606} (2016) 069
  [arXiv:1604.03046 [hep-th]].

\bibitem{Herrmann:2016qea}
  E.~Herrmann and J.~Trnka,
  ``Gravity On-shell Diagrams,''
  JHEP {\bf 1611}, 136 (2016)
  [arXiv:1604.03479 [hep-th]].

\bibitem{CaronHuot:2010rj}
  S.~Caron-Huot and D.~O'Connell,
  ``Spinor Helicity and Dual Conformal Symmetry in Ten Dimensions,''
  JHEP {\bf 1108} (2011) 014
  [arXiv:1010.5487 [hep-th]].


\bibitem{Bandos:2016tsm}
  I.~Bandos,
  ``BCFW-Type Recurrent Relations for Tree Superamplitudes of D=11 Supergravity,''
  Phys.\ Rev.\ Lett.\  {\bf 118} (2017) no.3,  031601
  [arXiv:1605.00036 [hep-th]].



\bibitem{Bandos:2017eof}
  I.~Bandos,
  ``Spinor frame formalism for amplitudes and constrained superamplitudes of 10D SYM and 11D supergravity,''
  arXiv:1711.00914 [hep-th].


\bibitem{Galperin:1991gk}
  A.~S.~Galperin, P.~S.~Howe and K.~S.~Stelle,
  ``The Superparticle and the Lorentz group,''
  Nucl.\ Phys.\ B {\bf 368} (1992) 248
  [hep-th/9201020].

\bibitem{Delduc:1991ir}
  F.~Delduc, A.~Galperin and E.~Sokatchev,
  ``Lorentz harmonic (super)fields and (super)particles,''
  Nucl.\ Phys.\ B {\bf 368} (1992) 143.

\bibitem{Bandos:1996ju}
  I.~A.~Bandos and A.~Y.~Nurmagambetov,
  ``Generalized action principle and extrinsic geometry for N=1 superparticle,''
  Class.\ Quant.\ Grav.\  {\bf 14} (1997) 1597
  doi:10.1088/0264-9381/14/7/004
  [hep-th/9610098].



\bibitem{Uvarov:2015rxa}
  D.~V.~Uvarov,
  ``Spinor description of D = 5 massless low-spin gauge fields,''
  Class.\ Quant.\ Grav.\  {\bf 33} (2016) no.13,  135010
  doi:10.1088/0264-9381/33/13/135010
  [arXiv:1506.01881 [hep-th]].



\bibitem{Galperin:1992pz}
  A.~S.~Galperin, P.~S.~Howe and P.~K.~Townsend,
  ``Twistor transform for superfields,''
  Nucl.\ Phys.\ B {\bf 402} (1993) 531.



\bibitem{Bandos:2006nr}
  I.~A.~Bandos, J.~A.~de Azcarraga and D.~P.~Sorokin,
  ``On D=11 supertwistors, superparticle quantization and a hidden SO(16) symmetry of supergravity,''  in: "Quantum, Super and Twistors, Proc. XXII Max Born Symposium, Wroclaw (Poland) 2006", Eds: J. Kowalski-Glikman and Ludwik Turko,  Wroclaw University Press  2008, pp. 25-32 [hep-th/0612252].

\bibitem{Bandos:2007mi}
  I.~A.~Bandos,
  ``Spinor moving frame, M0-brane covariant BRST quantization and intrinsic complexity of the pure spinor approach,''
  Phys.\ Lett.\ B {\bf 659} (2008) 388
  [arXiv:0707.2336 [hep-th]].

\bibitem{Bandos:2007wm}
  I.~A.~Bandos,
  ``D=11 massless superparticle covariant quantization, pure spinor BRST charge and hidden symmetries,''
  Nucl.\ Phys.\ B {\bf 796} (2008) 360
  [arXiv:0710.4342 [hep-th]].


\bibitem{Boels:2012ie}
  R.~H.~Boels and D.~O'Connell,
  ``Simple superamplitudes in higher dimensions,''
  JHEP {\bf 1206} (2012) 163
  [arXiv:1201.2653 [hep-th]].


\bibitem{Boels:2012zr}
  R.~H.~Boels,
  ``Maximal R-symmetry violating amplitudes in type IIB superstring theory,''
  Phys.\ Rev.\ Lett.\  {\bf 109} (2012) 081602
  [arXiv:1204.4208 [hep-th]].



\bibitem{Wang:2015jna}
  Y.~Wang and X.~Yin,
  ``Constraining Higher Derivative Supergravity with Scattering Amplitudes,''
  Phys.\ Rev.\ D {\bf 92} (2015) no.4,  041701
  doi:10.1103/PhysRevD.92.041701
  [arXiv:1502.03810 [hep-th]].

\bibitem{Wang:2015aua}
  Y.~Wang and X.~Yin,
  ``Supervertices and Non-renormalization Conditions in Maximal Supergravity Theories,''
  arXiv:1505.05861 [hep-th].

\bibitem{Bork:2015zaa}
  L.~V.~Bork, D.~I.~Kazakov, M.~V.~Kompaniets, D.~M.~Tolkachev and D.~E.~Vlasenko,
  ``Divergences in maximal supersymmetric Yang-Mills theories in diverse dimensions,''
  JHEP {\bf 1511} (2015) 059
  doi:10.1007/JHEP11(2015)059
  [arXiv:1508.05570 [hep-th]].

\bibitem{Borlakov:2016mwp}
  A.~T.~Borlakov, D.~I.~Kazakov, D.~M.~Tolkachev and D.~E.~Vlasenko,
  ``Summation of all-loop UV Divergences in Maximally Supersymmetric Gauge Theories,''
  JHEP {\bf 1612} (2016) 154
  doi:10.1007/JHEP12(2016)154
  [arXiv:1610.05549 [hep-th]].

\bibitem{Bandos:2017kdq}
  I.~Bandos,
  ``On 10D SYM superamplitudes,''
  arXiv:1712.02857 [hep-th].  To appear in SQS17 proceedings, JINR, Dubna, 2018.


\bibitem{Galperin:1984av}
  A.~Galperin, E.~Ivanov, S.~Kalitsyn, V.~Ogievetsky and E.~Sokatchev,
  ``Unconstrained N=2 Matter, Yang-Mills and Supergravity Theories in Harmonic Superspace,''
  Class.\ Quant.\ Grav.\  {\bf 1} (1984) 469
   Erratum: [Class.\ Quant.\ Grav.\  {\bf 2} (1985) 127].
  doi:10.1088/0264-9381/1/5/004

\bibitem{Galperin:1984bu}
  A.~Galperin, E.~Ivanov, S.~Kalitsyn, V.~Ogievetsky and E.~Sokatchev,
  ``Unconstrained Off-Shell N=3 Supersymmetric Yang-Mills Theory,''
  Class.\ Quant.\ Grav.\  {\bf 2} (1985) 155.
  doi:10.1088/0264-9381/2/2/009

\bibitem{Galperin:2001uw}
  A.~S.~Galperin, E.~A.~Ivanov, V.~I.~Ogievetsky and E.~S.~Sokatchev,
  ``Harmonic superspace,''
  Cambridge, UK: Univ. Pr. (2001) 306 p
  doi:10.1017/CBO9780511535109


\bibitem{BZ-str}
 I.~A.~Bandos and A.~A.~Zheltukhin, {\it Spinor Cartan moving n hedron, Lorentz harmonic formulations of superstrings, and kappa symmetry},
  JETP Lett.\  {\bf 54} (1991) 421;
  {\it Green-Schwarz   superstrings in spinor moving frame formalism},
  Phys.\ Lett.\  {\bf B288}, 77-83 (1992);
{\it D = 10 superstring:
Lagrangian and Hamiltonian mechanics in twistor-like Lorentz harmonic formulation},
Phys.\ Part.\ Nucl.\ {\bf 25} (1994) 453-477 [Preprint IC-92-422, ICTP, Trieste, 1992,
81pp.]



\bibitem{BZ-M2}
I.~A.~Bandos and A.~A.~Zheltukhin, ``Generalization of Newman-Penrose dyads in
connection with the action integral for supermembranes in an eleven-dimensional
space'',   JETP Lett.\ {\bf 55} (1992) 81 [Pisma Zh.\ Eksp.\ Teor.\ Fiz.\ {\bf 55}
(1992) 81 ].
\\
I.~A.~Bandos and A.~A.~Zheltukhin, ``Eleven-dimensional supermembrane in a spinor
moving repere formalism'', Int.\ J.\ Mod.\ Phys.\ A {\bf 8}, 1081 (1993);

\bibitem{BZ-p}
I.~A.~Bandos and A.~A.~Zheltukhin, ``N=1 super-p-branes in twistor - like Lorentz
harmonic formulation'', Class.\ Quant.\ Grav.\ {\bf 12}, 609 (1995)
[arXiv:hep-th/9405113].


\bibitem{Penrose:1967wn}
  R.~Penrose,
  ``Twistor algebra,''
  J.\ Math.\ Phys.\  {\bf 8} (1967) 345.
  doi:10.1063/1.1705200

\bibitem{Penrose:1972ia}
  R.~Penrose and M.~A.~H.~MacCallum,
  ``Twistor theory: An Approach to the quantization of fields and space-time,''
  Phys.\ Rept.\  {\bf 6} (1972) 241.
  doi:10.1016/0370-1573(73)90008-2

\bibitem{Penrose:1986ca}
  R.~Penrose and W.~Rindler, ``Spinors And Space-time. Volume 1 , Two-Spinor Calculus and Relativistic Fields,'' Cambridge University Press 1984;
  ``Spinors And Space-time. Vol. 2: Spinor And Twistor Methods In Space-time Geometry,''  Cambridge University Press 1986.


\bibitem{Ferber:1977qx}
  A.~Ferber,
  ``Supertwistors and Conformal Supersymmetry,''
  Nucl.\ Phys.\ B {\bf 132} (1978) 55.
  doi:10.1016/0550-3213(78)90257-2

\bibitem{Shirafuji:1983zd}
  T.~Shirafuji,
  ``Lagrangian Mechanics of Massless Particles With Spin,''
  Prog.\ Theor.\ Phys.\  {\bf 70} (1983) 18.
  doi:10.1143/PTP.70.18


\bibitem{Gumenchuk:1990db} A.~I.~Gumenchuk and D.~P.~Sorokin,
``Relativistic superparticle dynamics and twistor correspondence. (In Russian),''
  Sov.\ J.\ Nucl.\ Phys.\ {\bf 51} (1990) 350
   [Yad.\ Fiz.\ {\bf 51} (1990) 549].

\bibitem{Sorokin:1989jj}
  D.~P.~Sorokin,
  ``Double Supersymmetric Particle Theories,''
  Fortsch.\ Phys.\  {\bf 38} (1990) 923.

\bibitem{Bandos:1990ji}
  I.~A.~Bandos,
  ``Superparticle in Lorentz harmonic superspace,''
  Sov.\ J.\ Nucl.\ Phys.\  {\bf 51} (1990) 906

\bibitem{Bandos:1991my}
  I.~A.~Bandos and A.~A.~Zheltukhin,
  ``Null super p-brane: Hamiltonian dynamics and quantization,''
  Phys.\ Lett.\ B {\bf 261} (1991) 245.
  doi:10.1016/0370-2693(91)90322-H



\bibitem{Sokatchev:1985tc}
  E.~Sokatchev,
  ``Light Cone Harmonic Superspace and Its Applications,''
  Phys.\ Lett.\ B {\bf 169} (1986) 209.
  doi:10.1016/0370-2693(86)90652-0

\bibitem{Sokatchev:1987nk}
  E.~Sokatchev,
  ``Harmonic Superparticle,''
  Class.\ Quant.\ Grav.\  {\bf 4} (1987) 237.
  doi:10.1088/0264-9381/4/2/007


\bibitem{Akulov:1988tm}
  V.~P.~Akulov, D.~P.~Sorokin and I.~A.~Bandos,
  ``Particle Mechanics in Harmonic Superspace,''
  Mod.\ Phys.\ Lett.\ A {\bf 3} (1988) 1633.
  doi:10.1142/S0217732388001951


\bibitem{Buchbinder:2008pn}
  I.~L.~Buchbinder and I.~B.~Samsonov,
  ``N=3 Superparticle model,''
  Nucl.\ Phys.\ B {\bf 802} (2008) 180
  doi:10.1016/j.nuclphysb.2008.05.014
  [arXiv:0801.4907 [hep-th]].

\bibitem{Buchbinder:2008ub}
  I.~L.~Buchbinder, O.~Lechtenfeld and I.~B.~Samsonov,
  ``N=4 superparticle and super Yang-Mills theory in USp(4) harmonic superspace,''
  Nucl.\ Phys.\ B {\bf 802} (2008) 208
  doi:10.1016/j.nuclphysb.2008.05.015
  [arXiv:0804.3063 [hep-th]].


\bibitem{Deser:2000xz}
  S.~Deser and D.~Seminara,
  ``Tree amplitudes and two loop counterterms in D = 11 supergravity,''
  Phys.\ Rev.\ D {\bf 62} (2000) 084010
  doi:10.1103/PhysRevD.62.084010
  [hep-th/0002241].


\bibitem{Green:1999by}
  M.~B.~Green, M.~Gutperle and H.~H.~Kwon,
  ``Light cone quantum mechanics of the eleven-dimensional superparticle,''
  JHEP {\bf 9908} (1999) 012
  [hep-th/9907155].




\bibitem{Green:1981ya}
  M.~B.~Green and J.~H.~Schwarz,
  ``Supersymmetrical Dual String Theory. 3. Loops and Renormalization,''
  Nucl.\ Phys.\ B {\bf 198} (1982) 441.
  doi:10.1016/0550-3213(82)90334-0

\bibitem{Schwarz:1982jn}
  J.~H.~Schwarz,
  ``Superstring Theory,''
  Phys.\ Rept.\  {\bf 89} (1982) 223.
  doi:10.1016/0370-1573(82)90087-4





\bibitem{Mason:2013sva}
  L.~Mason and D.~Skinner,
  ``Ambitwistor strings and the scattering equations,''
  JHEP {\bf 1407} (2014) 048
  doi:10.1007/JHEP07(2014)048
  [arXiv:1311.2564 [hep-th]].


\bibitem{Adamo:2013tsa}
  T.~Adamo, E.~Casali and D.~Skinner,
  ``Ambitwistor strings and the scattering equations at one loop,''
  JHEP {\bf 1404} (2014) 104
  [arXiv:1312.3828 [hep-th]].



\bibitem{Adamo:2014wea}
  T.~Adamo, E.~Casali and D.~Skinner,
  ``A Worldsheet Theory for Supergravity,''
  JHEP {\bf 1502} (2015) 116
  [arXiv:1409.5656 [hep-th]].




\bibitem{Bandos:2014lja}
  I.~Bandos,
  ``Twistor/ambitwistor strings and null-superstrings in spacetime of D=4, 10 and 11 dimensions,''
  JHEP {\bf 1409} (2014) 086
  [arXiv:1404.1299 [hep-th]].

\bibitem{Casali:2016atr}
  E.~Casali and P.~Tourkine,
  ``On the null origin of the ambitwistor string,''
  JHEP {\bf 1611} (2016) 036
  doi:10.1007/JHEP11(2016)036
  [arXiv:1606.05636 [hep-th]].

\bibitem{Casali:2017zkz}
  E.~Casali, Y.~Herfray and P.~Tourkine,
  ``The complex null string, Galilean conformal algebra and scattering equations,''
  arXiv:1707.09900 [hep-th].


\bibitem{Bandos:2006af}
  I.~A.~Bandos, J.~A.~de Azcarraga and C.~Miquel-Espanya,
  ``Superspace formulations of the (super)twistor string,''
  JHEP {\bf 0607} (2006) 005
  [hep-th/0604037].



\bibitem{Berkovits:2004hg}
  N.~Berkovits,
  ``An Alternative string theory in twistor space for N=4 superYang-Mills,''
  Phys.\ Rev.\ Lett.\  {\bf 93} (2004) 011601
  [hep-th/0402045].

\bibitem{Siegel:2004dj}
  W.~Siegel,
  ``Untwisting the twistor superstring,''
  hep-th/0404255.


\bibitem{Geyer:2014fka}
  Y.~Geyer, A.~E.~Lipstein and L.~J.~Mason,
  ``Ambitwistor Strings in Four Dimensions,''
  Phys.\ Rev.\ Lett.\  {\bf 113} (2014) 8,  081602
  [arXiv:1404.6219 [hep-th]].

\bibitem{Lipstein:2015vxa}
  A.~Lipstein and V.~Schomerus,
  ``Towards a Worldsheet Description of N=8 Supergravity,''
  arXiv:1507.02936 [hep-th].

\bibitem{Bork:2017qyh}
  L.~V.~Bork and A.~I.~Onishchenko,
  ``Four dimensional ambitwistor strings and form factors of local and Wilson line operators,''
  arXiv:1704.04758 [hep-th],
  ``Ambitwistor strings and reggeon amplitudes in N=4 SYM,''
  Phys.\ Lett.\ B {\bf 774} (2017) 403
  doi:10.1016/j.physletb.2017.08.070
  [arXiv:1704.00611 [hep-th]].

\bibitem{Farrow:2017eol}
  J.~A.~Farrow and A.~E.~Lipstein,
  ``From 4d Ambitwistor Strings to On Shell Diagrams and Back,''
  JHEP {\bf 1707} (2017) 114
  doi:10.1007/JHEP07(2017)114
  [arXiv:1705.07087 [hep-th]].



\end{thebibliography}
\end{document}